\documentclass[11pt,a4paper]{article}
\pdfoutput=1

\usepackage{jheppub} % for details on the use of the package, please
\usepackage[T1]{fontenc} % if needed
\usepackage{hyperref}

\usepackage{graphicx}%图片头文件
\usepackage{float} %指定图片位置
\usepackage{subfigure}%并排子图 共享标题 有子标题
\usepackage{caption}
\usepackage{mathrsfs}
\usepackage{amsmath}
\usepackage{extarrows}
\usepackage{appendix}

\usepackage{cancel}
\usepackage{pgfplots}
\pgfplotsset{compat=1.15}
\usepackage{mathrsfs}
\usepackage{tikz}
\usetikzlibrary{arrows}
\usetikzlibrary{arrows.meta}
\usetikzlibrary{positioning}
\definecolor{cqcqcq}{rgb}{0.7529411764705882,0.7529411764705882,0.7529411764705882}
\newcommand{\bea}{\begin{eqnarray}}
\newcommand{\eea}{\end{eqnarray}}
\newcommand{\bean}{\begin{eqnarray*}}
\newcommand{\eean}{\end{eqnarray*}}
\newcommand{\nn}{\nonumber\\}

\def\Label#1{\label{#1}%
  \smash{\hbox to0pt{\raise1ex\hbox{\tiny[#1]}\hss}}}
\def\Label#1{\label{#1}}
\renewcommand{\eqref}[1]{eq.~(\ref{#1})}
\newcommand{\figref}[1]{Fig.~\ref{#1}}

\newcommand{\secref}[1]{section~\ref{#1}}

\def\braket#1{\left\langle #1 \right\rangle}

\def\vev{\braket}

\def\Spaa{\vev}

\newcommand{\ctobedelete}[1]{}

\allowdisplaybreaks

\title{\boldmath Constructing EYM amplitudes by inverse soft limit
}

\author[a]{Shiquan Ma \footnote{The unusual ordering of authors is to let authors get proper recognition of contributions under
the practice in China.}}
\author[a,b]{Rongyu Dong}
\author[a,c]{Yi-Jian Du\footnote{Corresponding author}} 
\affiliation[a]{School of Physics and Technology, Wuhan University,No.299 Bayi Road, Wuhan 430072, China}
\affiliation[b]{Department of Physics, Brown University, Providence, RI, 02912,USA}
\affiliation[c]{Hubei Key Laboratory of Nuclear Solid Physics, Wuhan University, No.299 Bayi Road, Wuhan 430072, China}
\emailAdd{shiqma@whu.edu.cn}
\emailAdd{rongyu\_dong@brown.edu}
\emailAdd{yijian.du@whu.edu.cn}

\abstract{ It is well known that gravity amplitudes in four dimensions can be reconstructed by the inverse soft limit (ISL) method. According to ISL, a tree level $n$-graviton maximally-helicity-violating (MHV) amplitude is expressed in terms of deformed $(n-1)$-graviton amplitudes accompanied by soft graviton factors. On another hand, 
single- and double-trace tree-level Einstein-Yang-Mills (EYM) MHV amplitudes have been proven to satisfy spanning forest formulas, where each edge in a forest has the same form with a term in soft graviton factor. It is not transparent that the formulas satisfied by EYM amplitudes can be constructed with ISL. In this paper, we construct the single- and double-trace MHV amplitudes in EYM, by the ISL and show that the known formulas can be precisely reproduced. Interesting identities which are based on Schouten identity and characterized by graphs are also introduced.
 }

\keywords{ Amplitude Relations, Gauge invariance}

\begin{document}
\maketitle
\flushbottom

\section{Introduction}
\label{sec:intro}

Scattering amplitudes have been shown to possess a universal soft limit \cite{Weinberg:1965nx}, which states that an $n$-particle amplitude can factorize into a soft factor and an $(n-1)$-particle amplitude when the momentum of an external particle tends to zero. Based on the Britto-Cachazo-Feng-Witten (BCFW) recursion \cite{Britto:2004ap,Britto:2005fq}, it was further shown that one could even do the inverse soft limit (ISL) \cite{Arkani-Hamed:2009ljj,Arkani-Hamed:n=4,Arkani-Hamed:t} to reconstruct a full amplitude by associating soft factors to amplitudes with deformed momentum. The ISL method has been successfully applied to construct the MHV amplitudes in Yang-Mills (YM) and gravitational (GR) theories and further extended to situations beyond MHV \cite{Bern:1998sv,Arkani-Hamed:n=4,Boucher-Veronneau:2011rwd,Nguyen:2009jk,Bullimore_2011,Nandan:2012rk}.

In \cite{Tian:2021dzf}, a symmetric formula of double trace Einstein-Yang-Mills (EYM) amplitude\footnote{ In this paper, EYM theory means the theory that also involves antisymmetric B field and dilaton, while the external particles of amplitudes can only be gravitons and gluons.} in four dimensions was introduced through the expansion formulas \cite{Stieberger:2016lng,Nandan:2016pya,Schlotterer:2016cxa,Fu:2017uzt,Chiodaroli:2017ngp,Teng:2017tbo,Du:2017gnh}. In this formula, a double-trace MHV amplitude is expressed via summing over all possible graphs \cite{Hou:2018bwm,Du:2019vzf} obtained by the following two steps: (i). constructing a bridge, which may pass through some gravitons, between the two gluon traces, (ii). constructing a spanning forest, which roots at nodes in the previous step, by the remaining gravitons. Such a spanning forest form can be considered as a generalization of the ones for GR \cite{Nguyen:2009jk,Hodges:2012ym,Feng:2012sy} and single-trace EYM amplitudes \cite{Du:2016wkt,Fu:2017uzt,Teng:2017tbo,Chiodaroli:2017ngp}, which were investigated earlier.

A key feature of the spanning forest formulas is that each edge between two adjacent nodes occurs as a term in the soft graviton factor. Thus one may realize that the spanning forest form can result from the ISL straightforwardly. However, if one tries to reconstruct an EYM amplitude with $r$ gravitons via ISL, the soft factors accompanying a new graviton should be inserted into an EYM amplitude with $r-1$ gravitons. Graphically, this procedure can be described by adding a leaf (corresponding to the new graviton) to a graph with $r-1$ gravitons. This seems to conflict with the full spanning forest formula where a graviton can also play as an internal node. As checked in \cite{Nguyen:2009jk,Boucher-Veronneau:2011rwd} for GR amplitudes, the contributions of graphs that involves the soft gravitons are in fact absorbed by the deformation of momentum. In this note, we fill this gap for EYM amplitudes, in a systematical way. We show that Schouten identities can be braided via graphs, into more complicated ones. With this trick, we generally prove that the single- and double-trace MHV formulas of EYM amplitudes can be reconstructed by ISL.\footnote{We acknowledge the referee of the paper \cite{Tian:2021dzf} for suggesting us to recover the relationship between ISL and the spanning forest formula of EYM.} 
 
This paper is organized as follows. In \secref{Sec2}, we provide a brief review of the soft limit and inverse soft limit. A generalized identity which is obtained by braiding Schouten identities is proved in \secref{Sec3}. Using the identity supposed in \secref{Sec3}, we construct the single- and double-trace MHV amplitudes in \secref{Sec4} and \secref{Sec5}, respectively. We finally summarize this work in \secref{Sec6}.

\newpage
\section{A brief review of soft limit and inverse soft limit}
\label{Sec2}

Scattering amplitudes in various theories have been shown to satisfy a universal soft behavior: when the momentum of an external particle tends to zero, the amplitude factorizes into a soft factor times a lower-point amplitude where the soft particle has been removed \cite{Weinberg:1964ew}. For $n$-point color ordered Yang-Mills (YM) amplitude $A_n$, this factorization takes the form
\begin{equation}
\lim_{p^{\mu}_j\to0} A_n(1,\dots,i,j,k,
\dots,n)=\mathcal{S}(i,j,k)A_{n-1}(1,\dots,i,k,\dots,n),
\label{2.1}
\end{equation}
where $\mathcal{S}$ is the soft factor that depends on the helicity of the soft gluon $j$  and the momenta of its adjacent gluons $i$, $k$. The explicit expressions in spinor-helicity formalism \cite{Xu:1986xb} are given as follows
\begin{equation}
\mathcal{S}(i,j^+,k)=\frac{\langle i,k\rangle}{\langle i,j\rangle\langle j,k\rangle},\quad {\mathcal{S}(i,j^-,k)=-\frac{[ i,k]}{[ i,j][ j,k]}}.
\label{2.2}
\end{equation}
Since there is no color-order feature in gravity amplitudes, the gravity soft factor involves contributions of all external momenta of other gravitons: 
\begin{equation}
\lim_{p^{\mu}_j\to0} M_n(1,\dots,i,j,k,
\dots,n)=\left(\sum_{l\in\{1,\dots,\widehat{j},\dots,n\}}\mathcal{G}(l,j)\right)M_{n-1}(1,\dots,i,k,\dots,n).
\label{2.3}
\end{equation}
The leading soft limits of gravity amplitudes have already been mentioned in \cite{Weinberg:1965nx}. For a positive-helicity soft graviton $j^+$, its soft factor can be expressed as
\begin{equation}
\sum_{l\in\textsc{H}}\mathcal{G}(l,j)=\sum_{l\in\textsc{H}}\frac{\langle l,\xi\rangle\langle l,\eta\rangle[l,j]}{\langle j,\xi\rangle\langle j,\eta\rangle\langle l,j\rangle},
\label{2.4}
\end{equation}
where $\xi\rangle$ and $\eta\rangle$ are two arbitrary reference spinors coming from the symmetric polarization tensor of graviton $j$, H denotes the graviton sector with graviton $j$ been removed.

The soft behavior of an amplitude holds when the momentum of a particle goes to zero, one can reverse the process to reconstruct the full amplitude information,  inspired by  BCFW \cite{Britto:2004ap,Britto:2005fq} recursion.
Briefly speaking, the ISL  is to express an amplitude by multiplying the soft factor to a lower-point amplitude with shifted momenta. For example, a full tree-level MHV amplitude of  color-ordered YM theory is expressed by
\begin{equation}
A_{\text{MHV}}(1,\dots,i,j,k,\dots,n)=\mathcal{S}(i,j,k)A'_{\text{MHV}}\left(1,\dots,i',k',\dots,n\right),
\label{2.5}
\end{equation}
where $j$ is a positive-helicity gluon and the spinors corresponding to the momenta of  $i$ and $k$ (the particles adjacent to $j$) are deformed as
\begin{equation}
  \begin{aligned} 
    \langle i' & \rightarrow \langle i,\quad  &i' ] &\rightarrow i] +  \frac{\langle j,k \rangle }{\langle i, k \rangle}j],\\
    \langle k' & \rightarrow \langle k, \quad &k'] &\rightarrow k] + \frac{\langle i,j \rangle }{\langle i, k \rangle}j].
  \end{aligned}
\label{2.6}
\end{equation}
The factor $\mathcal{S}(i,j,k)$ is identical to the soft factor in \eqref{2.2}. Similarly, a tree-level MHV amplitude in GR also has an ISL construction:
\begin{equation}
M_{\text{MHV}}(1,\dots,j,\dots,n)=\sum_{l\in\{1,\dots,\widehat{j},...,\widehat{\xi},...,n\}}\mathcal{G}(l,j,\xi) M'_{\text{MHV}}(1,\dots,\widehat{j},\dots,l',\dots,\xi',\dots,n).
\label{2.7}
\end{equation}
Here the factor $\mathcal{G}(l,j,\xi)$ has the same form as that in \eqref{2.4} where $ \xi\rangle$ is a reference spinor which has been fixed as the spinor corresponding to an arbitrary particle. The $\widehat{j}$ means that $j$ has been removed from the amplitude. Since the graviton $j$ has positive helicity, the shifted spinors in \eqref{2.7} has the same form as \eqref{2.6}
\begin{equation}
  \begin{aligned} 
    \langle \xi' & \rightarrow \langle \xi,\quad  &\xi' ] &\rightarrow \xi] +  \frac{\langle j,l \rangle }{\langle \xi, l \rangle}j],\\
    \langle l' & \rightarrow \langle l, \quad & l'] &\rightarrow l] + \frac{\langle \xi,j \rangle }{\langle \xi, l \rangle}j].
  \end{aligned}
\label{2.8}
\end{equation}
%The term with $l=\xi$ has to vanish due to our choice of gauge, thus the summation in \eqref{2.7} does not involve this term.

In this paper, we construct EYM amplitudes in MHV configuration by ISL and compare the results with those proposed in \cite{Du:2016wkt,Tian:2021dzf}.

\section{Fun with Schouten identity}
\label{Sec3}

\begin{figure}
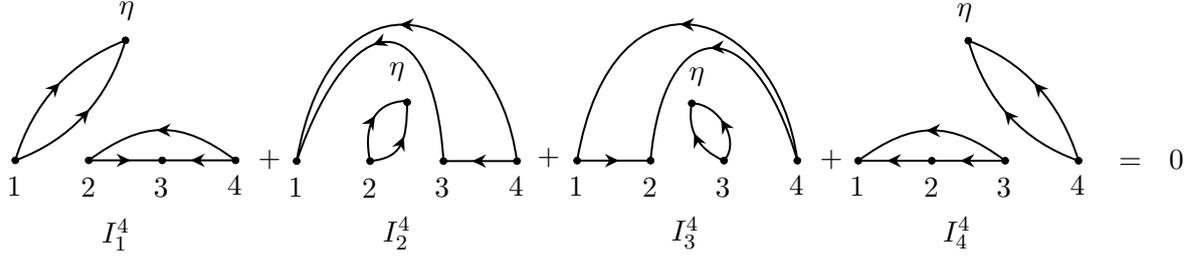

\centering

\tikzset{every picture/.style={line width=0.75pt}} %set default line width to 0.75pt        


\caption{Graph representation of \eqref{S-id-ab} with $n=4$, where the arrow line pointing from node $i$ to node $j$ denotes the spinor product $\langle i,j\rangle$.}
\label{fig-4-id}
\end{figure}

In this section, we introduce an interesting identity attributed to the Schouten identity, which is helpful in the coming discussions and may be further generalized and applied in other situations. The original Schouten identity is 
\begin{equation}
\langle a,b\rangle\langle c,d\rangle+\langle a,c\rangle\langle d,b\rangle+\langle a,d\rangle\langle b,c\rangle=0.
\label{Sc-id}
\end{equation}
The generalized Schouten identity used in this paper is given by
\begin{equation}
\sum\limits_{1\leq k\leq n}I^{n}_k\equiv\sum\limits_{1\leq k\leq n}(-1)^{n-k}\Bigg[\langle k,\eta\rangle^{n-2}\prod\limits_{\substack{1\leq j<i\leq n\\i,j\neq k}}\langle i,j\rangle\Bigg]=0,
\label{S-id-ab}
\end{equation}
where $\eta\rangle,1\rangle,2\rangle,\dots,n\rangle$ are $n+1$ spinors. Obviously, when $n=3$, what we get is just the Schouten identity \eqref{Sc-id}. In the following, we show some nontrivial examples and then prove the \eqref{S-id-ab}.

\subsection{Nontrivial examples}
  
The first nontrivial example is the $n=4$ case which has the explicit form
\begin{equation}
\begin{aligned}
\underbrace{\langle 1,\eta\rangle^2\langle 4,3\rangle\langle 4,2\rangle\langle 2,3\rangle}_{I^4_1}+\underbrace{\langle 2,\eta\rangle^2\langle 4,3\rangle\langle 4,1\rangle\langle 3,1\rangle}_{I^4_2}+\underbrace{\langle 3,\eta\rangle^2\langle 4,2\rangle\langle 4,1\rangle\langle 1,2\rangle}_{I^4_3}+\underbrace{\langle 4,\eta\rangle^2\langle 3,2\rangle\langle 3,1\rangle\langle 2,1\rangle}_{I^4_4}=0.
\end{aligned}
\label{4-id}
\end{equation}
This identity can be shown graphically (see \figref{fig-4-id}) by associating an arrow line pointing from $a$ to $b$ to each spinor product $\langle a,b\rangle$.
To prove \eqref{4-id}, we express the factors $\langle 1,\eta\rangle\langle 4,3\rangle $  and   $\langle 2,\eta\rangle\langle 4,3\rangle$ by Schouten identity \eqref{Sc-id}, then $I^4_1$ and $I^4_2$ read
\begin{equation}
\begin{aligned}
I^4_1&=\underbrace{\langle 1,\eta\rangle\langle 3,\eta\rangle\langle 4,1\rangle\langle 4,2\rangle\langle 2,3\rangle}_{I^4_{1_1}}+
\underbrace{\langle 1,\eta\rangle\langle 4,\eta\rangle\langle 1,3\rangle\langle 4,2\rangle\langle 2,3\rangle}_{I^4_{1_2}},\\
I^4_2&=\underbrace{\langle 2,\eta\rangle\langle 3,\eta\rangle\langle 4,2\rangle\langle 4,1\rangle\langle 3,1\rangle}_{I^4_{2_1}}+
\underbrace{\langle 2,\eta\rangle\langle 4,\eta\rangle\langle 2,3\rangle\langle 4,1\rangle\langle 3,1\rangle}_{I^4_{2_2}}.
\end{aligned}
\label{4-split}
\end{equation}
Terms in \eqref{4-split} are recombined as
\bea
I^4_{1_1}+I^4_{2_1}&=&\langle3,\eta\rangle\langle4,1\rangle\langle 4,2\rangle\Bigl[\langle 1,\eta\rangle\langle 2,3\rangle+\langle 2,\eta\rangle\langle 3,1\rangle\Bigr]=\langle 3,\eta\rangle^2\langle 4,1\rangle\langle 4,2\rangle\langle 2,1\rangle=-I^4_3,\\
I^4_{1_2}+I^4_{2_2}&=&\langle4,\eta\rangle\langle3,1\rangle\langle 3,2\rangle\Bigl[\langle 1,\eta\rangle\langle 4,2\rangle+\langle 2,\eta\rangle\langle 1,4\rangle\Bigr]=\langle 4,\eta\rangle^2\langle 3,1\rangle\langle 2,3\rangle\langle 1,2\rangle=-I^4_4,
\eea
which cancel with the last two terms $I_3^4$ and $I_4^4$, thus \eqref{4-id} has been proven.

\begin{figure}
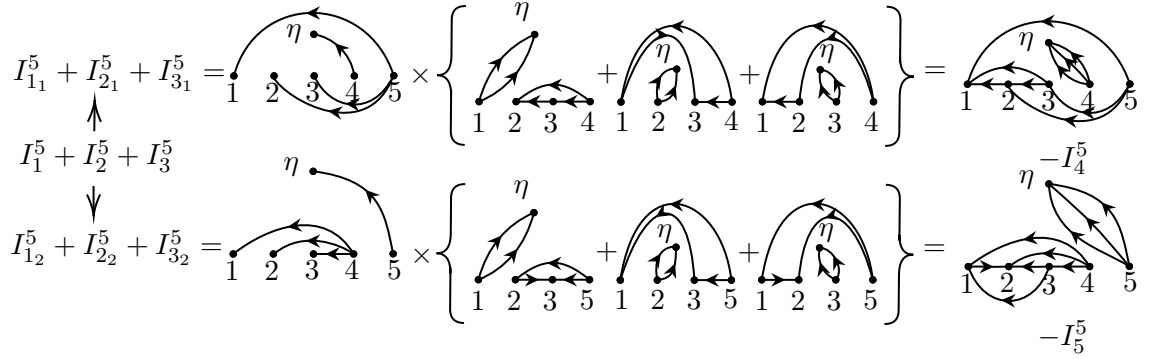

  \centering

  \tikzset{every picture/.style={line width=0.75pt}} %set default line width to 0.75pt        
  

  
  \caption{The $I^5_1+I^5_2+I^5_3$  is decomposed into two terms $I^5_{1_1}+I^5_{2_1}+I^5_{3_1}$ and  $I^5_{1_2}+I^5_{2_2}+I^5_{3_2}$ with Schouten identity. Each term, when expressed by graphs, contains a factor (the expression inside the braces) which has the same pattern with the identity (\ref{4-id}).}
  \label{fig-5-id}
  \end{figure}

For $n=5$ case, the identity (\ref{S-id-ab}) becomes more bloated, 
but the proof can actually be done using the $n=4$ example to circumvent tedious algebraic calculations. The identity with $n=5$ is explicitly displayed as
\begin{equation}
\begin{aligned}
&\underbrace{\langle 1,\eta\rangle^3\langle 5,4\rangle\langle 5,3\rangle\langle 5,2\rangle\langle 4,3\rangle\langle 4,2\rangle\langle 3,2\rangle}_{I^5_1}+\underbrace{\langle 2,\eta\rangle^3\langle 5,4\rangle\langle 5,3\rangle\langle 5,1\rangle\langle 4,3\rangle\langle 4,1\rangle\langle 1,3\rangle}_{I^5_2}\\
&\underbrace{\langle 3,\eta\rangle^3\langle 5,4\rangle\langle 5,2\rangle\langle 5,1\rangle\langle 4,2\rangle\langle 4,1\rangle\langle 2,1\rangle}_{I^5_3}+\underbrace{\langle 4,\eta\rangle^3\langle 5,3\rangle\langle 5,2\rangle\langle 5,1\rangle\langle 3,2\rangle\langle 3,1\rangle\langle 1,2\rangle}_{I^5_4}\\
&+\underbrace{\langle 5,\eta\rangle^3\langle 4,3\rangle\langle 4,2\rangle\langle 4,1\rangle\langle 3, 2\rangle\langle 3,1\rangle\langle 2,1\rangle}_{I^5_5}=0.
\end{aligned}
\label{5-id}
\end{equation}
Using Schouten identity, we decompose the factors $\langle 1,\eta\rangle\langle 5,4\rangle $, $\langle 2,\eta\rangle\langle 5,4\rangle $ and $\langle 3,\eta\rangle\langle 5,4\rangle $ inside terms $I^5_1$, $I^5_2$ and $I^5_3$ respectively, then we get
\begin{equation}
\begin{aligned}
I^5_1&=\underbrace{\langle 1,\eta\rangle^2\langle 4,\eta\rangle\langle5,1\rangle\langle 5,3\rangle\langle 5,2\rangle\langle 4,3\rangle\langle 4,2\rangle\langle 3,2\rangle}_{I^5_{1_1}}+\underbrace{\langle 1,\eta\rangle^2\langle 5,\eta\rangle\langle 1,4\rangle\langle 5,3\rangle\langle 5,2\rangle\langle 4,3\rangle\langle 4,2\rangle\langle 3,2\rangle}_{I^5_{1_2}},\\
I^5_2&=\underbrace{\langle 2,\eta\rangle^2\langle 4,\eta\rangle\langle 5,2\rangle\langle 5,3\rangle\langle 5,1\rangle\langle 4,3\rangle\langle4,1\rangle\langle1,3\rangle}_{I^5_{2_1}}+\underbrace{\langle 2,\eta\rangle^2\langle 5,\eta\rangle\langle 2,4\rangle\langle 5,3\rangle\langle 5,1\rangle\langle 4,3\rangle\langle4,1\rangle\langle1,3\rangle}_{I^5_{2_2}},\\
I^5_3&=\underbrace{\langle 3,\eta\rangle^2\langle 4,\eta\rangle\langle 5,3\rangle\langle 5,2\rangle\langle 5,1\rangle\langle 4,2\rangle\langle 4,1\rangle\langle2,1\rangle}_{I^5_{3_1}}+\underbrace{\langle 3,\eta\rangle^2\langle 5,\eta\rangle\langle 3,4\rangle\langle 5,2\rangle\langle 5,1\rangle\langle 4,2\rangle\langle 4,1\rangle\langle2,1\rangle}_{I^5_{3_2}}.
\end{aligned}
\label{5-split}
\end{equation}
Applying \eqref{4-id}, we find 

\bea
I^5_{1_1}+I^5_{2_1}+I^5_{3_1}=\langle 4,\eta\rangle\langle5,3\rangle\langle5,2\rangle\langle5,1\rangle&\times&\Bigl[\langle 1,\eta\rangle^2\langle 4,3\rangle\langle 4,2\rangle\langle 3,2\rangle\label{5-id-dt1}\\
&&+\langle 2,\eta\rangle^2\langle 4,3\rangle\langle 4,1\rangle\langle 1,3\rangle+\langle 3,\eta\rangle^2\langle 4,2\rangle\langle 4,1\rangle\langle 2,1
\rangle\Bigr]=-I^5_4,\nonumber
\eea
and 
\bea
I^5_{1_2}+I^5_{2_2}+I^5_{3_2}=\langle 5,\eta\rangle\langle4,3\rangle\langle4,2\rangle\langle4,1\rangle&\times&\Bigl[\langle 1,\eta\rangle^2\langle 5,3\rangle\langle 5,2\rangle\langle 2,3\rangle\label{5-id-dt2}\\
&&+\langle 2,\eta\rangle^2\langle 5,3\rangle\langle 5,1\rangle\langle 3,1\rangle+\langle 3,\eta\rangle^2\langle 5,2\rangle\langle 5,1\rangle\langle 1,2\rangle
\Bigr]=-I^5_5,\nonumber
\eea
which respectively cancel with the last two terms in \eqref{5-id}. Thus  \eqref{5-id} is proven. The process of this proof is graphically shown by \figref{fig-5-id}, as follows. 

\subsection{General proof of the identity}
From the discussion in the last subsection, we assert that the identity (\ref{S-id-ab}) with an arbitrary $n$ is formed by nesting Schouten identities with a level-$(n-1)$ identity. Based on this, we give an inductive proof of \eqref{S-id-ab}. Supposing it holds at level-($n-1$), so that
\begin{equation}
\sum\limits_{1\leq k\leq n-1}(-1)^{n-k}\Bigg[\langle k,\eta\rangle^{n-3}\prod\limits_{\substack{1\leq j<i\leq n-1\\i,j\neq k}}\langle i,j\rangle\Bigg]=\sum\limits_{1\leq  k\leq n-1} I^{n-1}_k=0.
\label{S-id-n-1}
\end{equation}
 For level-$n$ identity, we apply Schouten identity 
\begin{equation}
\langle k,\eta\rangle\langle n,n-1\rangle=\langle n-1,\eta\rangle\langle n,k\rangle+\langle n,\eta\rangle\langle k,n-1\rangle
\end{equation}
to terms $I^{n}_k$ ($k\in\{1,2,\dots,n-2\}$)  in \eqref{S-id-ab}. Each term $I^{n}_k$ then splits into $I^{n}_{k_1}$ and $I^{n}_{k_2}$:
\begin{equation}
\begin{aligned}
I^{n}_{k_1}&=(-1)^{n-k}\langle n-1,\eta\rangle\langle n,k\rangle\langle k,\eta\rangle^{n-3}\prod\limits_{\substack{1\leq m\leq n-2\\m\neq k}}\langle n,m\rangle \prod\limits_{\substack{1\leq j<i\leq n-1\\i,j\neq k}}\langle i,j\rangle\\
&=\langle n-1,\eta\rangle\prod\limits_{\substack{1\leq m\leq n-2}}\langle n,m\rangle\,\ (-1)^{n-1}I^{n-1}_k,
\end{aligned}
\label{Ink1}
\end{equation}
\begin{equation}
\begin{aligned}
I^{n}_{k_2}&=(-1)^{n-k+1}\langle n,\eta\rangle\langle n-1,k\rangle\langle k,\eta\rangle^{n-3}\prod\limits_{\substack{1\leq m\leq n-2\\m\neq k}}\langle n-1,m\rangle\prod\limits_{\substack{1\leq j<i\leq n\\i,j\neq n-1}}\langle i,j\rangle \\
&=\langle n,\eta\rangle\prod\limits_{\substack{1\leq m\leq n-2}}\langle n-1,m\rangle \,\ (-1)^n \tilde{I}^{n-1}_k,
\end{aligned}
\label{Ink2}
\end{equation}
where $\tilde{I}^{n-1}_k\equiv I^{n-1}_k|_{(n-1)\to n}$. When  \eqref{Ink1} and \eqref{Ink2} for all $k\in\{1,2,\dots,n-2\}$ are summed up, we respectively get 
\begin{equation}
\sum\limits_{1\leq k \leq n-2} I^{n}_{k_{1}}=\langle n-1,\eta\rangle\prod\limits_{\substack{1\leq m\leq n-2}}\langle n,m\rangle  \sum\limits_{1\leq k \leq n-2} (-1)^{n-1}I^{n-1}_k=-I^n_{n-1},
\label{In-1}
\end{equation}
and 
\begin{equation}
\sum\limits_{1\leq k \leq n-2} I^{n}_{k_{2}}=\langle n,\eta\rangle\prod\limits_{\substack{1\leq m\leq n-2}}\langle n-1,m\rangle  \sum\limits_{1\leq k \leq n-2}(-1)^n \tilde{I}^{n-1}_k=-I^n_{n},
\label{Inn}
\end{equation}
where \eqref{S-id-n-1} has been applied in the last equality in each equation. Therefore, \eqref{In-1} and \eqref{Inn} cancel out with the $k=n-1$ and $k=n$ terms of \eqref{S-id-ab} and the proof of the identity (\ref{S-id-ab}) has been completed. It seems that there may be more generic extensions and studies of the identity (\ref{S-id-ab}), especially on (i). how to reflect more properties of spinor products by properties of graphs, and (ii). how to find a more fundamental way to solve the relations between spinor products. Nevertheless, \eqref{S-id-ab} is sufficient in this paper, for the study of the ISL.
\begin{figure}
  \centering

  \tikzset{every picture/.style={line width=0.75pt}} %set default line width to 0.75pt        

  \begin{tikzpicture}[x=0.75pt,y=0.75pt,yscale=-1,xscale=1]
  %uncomment if require: \path (0,208); %set diagram left start at 0, and has height of 208
  
  %Straight Lines [id:da7458920377227638] 
  \draw  [dash pattern={on 4.5pt off 4.5pt}]  (269.58,140.66) -- (233.1,140.54) ;
  \draw [shift={(230.1,140.53)}, rotate = 0.19] [fill={rgb, 255:red, 0; green, 0; blue, 0 }  ][line width=0.08]  [draw opacity=0] (7.14,-3.43) -- (0,0) -- (7.14,3.43) -- (4.74,0) -- cycle    ;
  %Straight Lines [id:da43421945388735206] 
  \draw  [dash pattern={on 0.84pt off 2.51pt}]  (270.33,140.69) -- (429.87,140.66) ;
  \draw [shift={(429.87,140.66)}, rotate = 359.99] [color={rgb, 255:red, 0; green, 0; blue, 0 }  ][fill={rgb, 255:red, 0; green, 0; blue, 0 }  ][line width=0.75]      (0, 0) circle [x radius= 1.34, y radius= 1.34]   ;
  \draw [shift={(270.33,140.69)}, rotate = 359.99] [color={rgb, 255:red, 0; green, 0; blue, 0 }  ][fill={rgb, 255:red, 0; green, 0; blue, 0 }  ][line width=0.75]      (0, 0) circle [x radius= 1.34, y radius= 1.34]   ;
  %Straight Lines [id:da30385406132483816] 
  \draw  [dash pattern={on 4.5pt off 4.5pt}]  (470.1,140.56) -- (433.15,140.92) ;
  \draw [shift={(430.15,140.95)}, rotate = 359.45] [fill={rgb, 255:red, 0; green, 0; blue, 0 }  ][line width=0.08]  [draw opacity=0] (7.14,-3.43) -- (0,0) -- (7.14,3.43) -- (4.74,0) -- cycle    ;
  %Straight Lines [id:da2727504257100648] 
  \draw  [dash pattern={on 0.84pt off 2.51pt}]  (470.1,140.56) ;
  \draw [shift={(470.1,140.56)}, rotate = 0] [color={rgb, 255:red, 0; green, 0; blue, 0 }  ][fill={rgb, 255:red, 0; green, 0; blue, 0 }  ][line width=0.75]      (0, 0) circle [x radius= 1.34, y radius= 1.34]   ;
  \draw [shift={(470.1,140.56)}, rotate = 0] [color={rgb, 255:red, 0; green, 0; blue, 0 }  ][fill={rgb, 255:red, 0; green, 0; blue, 0 }  ][line width=0.75]      (0, 0) circle [x radius= 1.34, y radius= 1.34]   ;
  %Straight Lines [id:da41186516054656264] 
  \draw  [dash pattern={on 0.84pt off 2.51pt}]  (229.9,140.97) -- (230.04,140.84) ;
  \draw [shift={(230.04,140.84)}, rotate = 316.62] [color={rgb, 255:red, 0; green, 0; blue, 0 }  ][fill={rgb, 255:red, 0; green, 0; blue, 0 }  ][line width=0.75]      (0, 0) circle [x radius= 1.34, y radius= 1.34]   ;
  \draw [shift={(229.9,140.97)}, rotate = 316.62] [color={rgb, 255:red, 0; green, 0; blue, 0 }  ][fill={rgb, 255:red, 0; green, 0; blue, 0 }  ][line width=0.75]      (0, 0) circle [x radius= 1.34, y radius= 1.34]   ;
  %Straight Lines [id:da9733366768660394] 
  \draw  [dash pattern={on 0.84pt off 2.51pt}]  (349.98,140.58) ;
  \draw [shift={(349.98,140.58)}, rotate = 0] [color={rgb, 255:red, 0; green, 0; blue, 0 }  ][fill={rgb, 255:red, 0; green, 0; blue, 0 }  ][line width=0.75]      (0, 0) circle [x radius= 1.34, y radius= 1.34]   ;
  \draw [shift={(349.98,140.58)}, rotate = 0] [color={rgb, 255:red, 0; green, 0; blue, 0 }  ][fill={rgb, 255:red, 0; green, 0; blue, 0 }  ][line width=0.75]      (0, 0) circle [x radius= 1.34, y radius= 1.34]   ;
  %Straight Lines [id:da9350597583365086] 
  \draw  [dash pattern={on 0.84pt off 2.51pt}]  (310.27,140.87) ;
  \draw [shift={(310.27,140.87)}, rotate = 0] [color={rgb, 255:red, 0; green, 0; blue, 0 }  ][fill={rgb, 255:red, 0; green, 0; blue, 0 }  ][line width=0.75]      (0, 0) circle [x radius= 1.34, y radius= 1.34]   ;
  \draw [shift={(310.27,140.87)}, rotate = 0] [color={rgb, 255:red, 0; green, 0; blue, 0 }  ][fill={rgb, 255:red, 0; green, 0; blue, 0 }  ][line width=0.75]      (0, 0) circle [x radius= 1.34, y radius= 1.34]   ;
  %Straight Lines [id:da23504513102307678] 
  \draw  [dash pattern={on 0.84pt off 2.51pt}]  (390.27,140.87) ;
  \draw [shift={(390.27,140.87)}, rotate = 0] [color={rgb, 255:red, 0; green, 0; blue, 0 }  ][fill={rgb, 255:red, 0; green, 0; blue, 0 }  ][line width=0.75]      (0, 0) circle [x radius= 1.34, y radius= 1.34]   ;
  \draw [shift={(390.27,140.87)}, rotate = 0] [color={rgb, 255:red, 0; green, 0; blue, 0 }  ][fill={rgb, 255:red, 0; green, 0; blue, 0 }  ][line width=0.75]      (0, 0) circle [x radius= 1.34, y radius= 1.34]   ;
  %Straight Lines [id:da5081381724420155] 
  \draw [color={rgb, 255:red, 0; green, 0; blue, 0 }  ,draw opacity=1 ]   (337.65,112.39) -- (348.91,138.3) ;
  \draw [shift={(350.1,141.05)}, rotate = 246.52] [fill={rgb, 255:red, 0; green, 0; blue, 0 }  ,fill opacity=1 ][line width=0.08]  [draw opacity=0] (7.14,-3.43) -- (0,0) -- (7.14,3.43) -- (4.74,0) -- cycle    ;
  %Straight Lines [id:da855116955802165] 
  \draw  [dash pattern={on 0.84pt off 2.51pt}]  (337.65,112.39) ;
  \draw [shift={(337.65,112.39)}, rotate = 0] [color={rgb, 255:red, 0; green, 0; blue, 0 }  ][fill={rgb, 255:red, 0; green, 0; blue, 0 }  ][line width=0.75]      (0, 0) circle [x radius= 1.34, y radius= 1.34]   ;
  \draw [shift={(337.65,112.39)}, rotate = 0] [color={rgb, 255:red, 0; green, 0; blue, 0 }  ][fill={rgb, 255:red, 0; green, 0; blue, 0 }  ][line width=0.75]      (0, 0) circle [x radius= 1.34, y radius= 1.34]   ;
  %Straight Lines [id:da1505518548494431] 
  \draw [color={rgb, 255:red, 0; green, 0; blue, 0 }  ,draw opacity=1 ]   (362.72,112.87) -- (350.52,138.35) ;
  \draw [shift={(349.23,141.05)}, rotate = 295.59] [fill={rgb, 255:red, 0; green, 0; blue, 0 }  ,fill opacity=1 ][line width=0.08]  [draw opacity=0] (7.14,-3.43) -- (0,0) -- (7.14,3.43) -- (4.74,0) -- cycle    ;
  %Straight Lines [id:da014978647100611475] 
  \draw  [dash pattern={on 0.84pt off 2.51pt}]  (362.72,112.87) ;
  \draw [shift={(362.72,112.87)}, rotate = 0] [color={rgb, 255:red, 0; green, 0; blue, 0 }  ][fill={rgb, 255:red, 0; green, 0; blue, 0 }  ][line width=0.75]      (0, 0) circle [x radius= 1.34, y radius= 1.34]   ;
  \draw [shift={(362.72,112.87)}, rotate = 0] [color={rgb, 255:red, 0; green, 0; blue, 0 }  ][fill={rgb, 255:red, 0; green, 0; blue, 0 }  ][line width=0.75]      (0, 0) circle [x radius= 1.34, y radius= 1.34]   ;
  %Straight Lines [id:da40486291189558554] 
  \draw [color={rgb, 255:red, 0; green, 0; blue, 0 }  ,draw opacity=1 ]   (312.74,91.08) -- (334.66,109.11) ;
  \draw [shift={(336.98,111.02)}, rotate = 219.44] [fill={rgb, 255:red, 0; green, 0; blue, 0 }  ,fill opacity=1 ][line width=0.08]  [draw opacity=0] (7.14,-3.43) -- (0,0) -- (7.14,3.43) -- (4.74,0) -- cycle    ;
  %Straight Lines [id:da7568469819092347] 
  \draw [color={rgb, 255:red, 0; green, 0; blue, 0 }  ,draw opacity=1 ] [dash pattern={on 0.84pt off 2.51pt}]  (312.91,91.16) ;
  \draw [shift={(312.91,91.16)}, rotate = 0] [color={rgb, 255:red, 0; green, 0; blue, 0 }  ,draw opacity=1 ][fill={rgb, 255:red, 0; green, 0; blue, 0 }  ,fill opacity=1 ][line width=0.75]      (0, 0) circle [x radius= 1.34, y radius= 1.34]   ;
  \draw [shift={(312.91,91.16)}, rotate = 0] [color={rgb, 255:red, 0; green, 0; blue, 0 }  ,draw opacity=1 ][fill={rgb, 255:red, 0; green, 0; blue, 0 }  ,fill opacity=1 ][line width=0.75]      (0, 0) circle [x radius= 1.34, y radius= 1.34]   ;
  %Straight Lines [id:da2962207708932074] 
  \draw [color={rgb, 255:red, 0; green, 0; blue, 0 }  ,draw opacity=1 ]   (363.05,93.81) -- (339.81,110.11) ;
  \draw [shift={(337.35,111.83)}, rotate = 324.96] [fill={rgb, 255:red, 0; green, 0; blue, 0 }  ,fill opacity=1 ][line width=0.08]  [draw opacity=0] (7.14,-3.43) -- (0,0) -- (7.14,3.43) -- (4.74,0) -- cycle    ;
  %Straight Lines [id:da18975063903528855] 
  \draw  [dash pattern={on 0.84pt off 2.51pt}]  (363.72,93.47) ;
  \draw [shift={(363.72,93.47)}, rotate = 0] [color={rgb, 255:red, 0; green, 0; blue, 0 }  ][fill={rgb, 255:red, 0; green, 0; blue, 0 }  ][line width=0.75]      (0, 0) circle [x radius= 1.34, y radius= 1.34]   ;
  \draw [shift={(363.72,93.47)}, rotate = 0] [color={rgb, 255:red, 0; green, 0; blue, 0 }  ][fill={rgb, 255:red, 0; green, 0; blue, 0 }  ][line width=0.75]      (0, 0) circle [x radius= 1.34, y radius= 1.34]   ;
  %Straight Lines [id:da8331300830729897] 
  \draw [color={rgb, 255:red, 0; green, 0; blue, 0 }  ,draw opacity=1 ]   (337.59,79.82) -- (337.03,108.02) ;
  \draw [shift={(336.98,111.02)}, rotate = 271.12] [fill={rgb, 255:red, 0; green, 0; blue, 0 }  ,fill opacity=1 ][line width=0.08]  [draw opacity=0] (7.14,-3.43) -- (0,0) -- (7.14,3.43) -- (4.74,0) -- cycle    ;
  %Straight Lines [id:da7514372046666631] 
  \draw [color={rgb, 255:red, 0; green, 0; blue, 0 }  ,draw opacity=1 ] [dash pattern={on 0.84pt off 2.51pt}]  (337.68,79.75) ;
  \draw [shift={(337.68,79.75)}, rotate = 0] [color={rgb, 255:red, 0; green, 0; blue, 0 }  ,draw opacity=1 ][fill={rgb, 255:red, 0; green, 0; blue, 0 }  ,fill opacity=1 ][line width=0.75]      (0, 0) circle [x radius= 1.34, y radius= 1.34]   ;
  \draw [shift={(337.68,79.75)}, rotate = 0] [color={rgb, 255:red, 0; green, 0; blue, 0 }  ,draw opacity=1 ][fill={rgb, 255:red, 0; green, 0; blue, 0 }  ,fill opacity=1 ][line width=0.75]      (0, 0) circle [x radius= 1.34, y radius= 1.34]   ;
  %Curve Lines [id:da2177956266519765] 
  \draw [color={rgb, 255:red, 0; green, 0; blue, 0 }  ,draw opacity=1 ] [dash pattern={on 0.84pt off 2.51pt}]  (316.93,86.49) .. controls (319.9,76.4) and (328.81,76.82) .. (333.48,82.29) ;
  %Straight Lines [id:da5342312720825848] 
  \draw [color={rgb, 255:red, 0; green, 0; blue, 0 }  ,draw opacity=1 ]   (350.81,65.46) -- (362.99,91.1) ;
  \draw [shift={(364.27,93.81)}, rotate = 244.6] [fill={rgb, 255:red, 0; green, 0; blue, 0 }  ,fill opacity=1 ][line width=0.08]  [draw opacity=0] (7.14,-3.43) -- (0,0) -- (7.14,3.43) -- (4.74,0) -- cycle    ;
  %Straight Lines [id:da7809420734670338] 
  \draw [color={rgb, 255:red, 0; green, 0; blue, 0 }  ,draw opacity=1 ] [dash pattern={on 0.84pt off 2.51pt}]  (350.6,65.61) ;
  \draw [shift={(350.6,65.61)}, rotate = 0] [color={rgb, 255:red, 0; green, 0; blue, 0 }  ,draw opacity=1 ][fill={rgb, 255:red, 0; green, 0; blue, 0 }  ,fill opacity=1 ][line width=0.75]      (0, 0) circle [x radius= 1.34, y radius= 1.34]   ;
  \draw [shift={(350.6,65.61)}, rotate = 0] [color={rgb, 255:red, 0; green, 0; blue, 0 }  ,draw opacity=1 ][fill={rgb, 255:red, 0; green, 0; blue, 0 }  ,fill opacity=1 ][line width=0.75]      (0, 0) circle [x radius= 1.34, y radius= 1.34]   ;
  %Straight Lines [id:da2987442032064609] 
  \draw [color={rgb, 255:red, 0; green, 0; blue, 0 }  ,draw opacity=1 ]   (377.76,65.83) -- (365.27,91.12) ;
  \draw [shift={(363.94,93.81)}, rotate = 296.28] [fill={rgb, 255:red, 0; green, 0; blue, 0 }  ,fill opacity=1 ][line width=0.08]  [draw opacity=0] (7.14,-3.43) -- (0,0) -- (7.14,3.43) -- (4.74,0) -- cycle    ;
  %Straight Lines [id:da6707850844423309] 
  \draw [color={rgb, 255:red, 0; green, 0; blue, 0 }  ,draw opacity=1 ] [dash pattern={on 0.84pt off 2.51pt}]  (377.87,65.81) ;
  \draw [shift={(377.87,65.81)}, rotate = 0] [color={rgb, 255:red, 0; green, 0; blue, 0 }  ,draw opacity=1 ][fill={rgb, 255:red, 0; green, 0; blue, 0 }  ,fill opacity=1 ][line width=0.75]      (0, 0) circle [x radius= 1.34, y radius= 1.34]   ;
  \draw [shift={(377.87,65.81)}, rotate = 0] [color={rgb, 255:red, 0; green, 0; blue, 0 }  ,draw opacity=1 ][fill={rgb, 255:red, 0; green, 0; blue, 0 }  ,fill opacity=1 ][line width=0.75]      (0, 0) circle [x radius= 1.34, y radius= 1.34]   ;
  %Curve Lines [id:da024852857747875712] 
  \draw [color={rgb, 255:red, 0; green, 0; blue, 0 }  ,draw opacity=1 ] [dash pattern={on 0.84pt off 2.51pt}]  (356.23,63.09) .. controls (363.2,55.22) and (371.09,59.39) .. (372.99,66.32) ;
  %Straight Lines [id:da4500326727979629] 
  \draw [color={rgb, 255:red, 0; green, 0; blue, 0 }  ,draw opacity=1 ]   (286.07,120.08) -- (307.99,138.11) ;
  \draw [shift={(310.31,140.02)}, rotate = 219.44] [fill={rgb, 255:red, 0; green, 0; blue, 0 }  ,fill opacity=1 ][line width=0.08]  [draw opacity=0] (7.14,-3.43) -- (0,0) -- (7.14,3.43) -- (4.74,0) -- cycle    ;
  %Straight Lines [id:da42985809731116853] 
  \draw [color={rgb, 255:red, 0; green, 0; blue, 0 }  ,draw opacity=1 ] [dash pattern={on 0.84pt off 2.51pt}]  (286.25,120.16) ;
  \draw [shift={(286.25,120.16)}, rotate = 0] [color={rgb, 255:red, 0; green, 0; blue, 0 }  ,draw opacity=1 ][fill={rgb, 255:red, 0; green, 0; blue, 0 }  ,fill opacity=1 ][line width=0.75]      (0, 0) circle [x radius= 1.34, y radius= 1.34]   ;
  \draw [shift={(286.25,120.16)}, rotate = 0] [color={rgb, 255:red, 0; green, 0; blue, 0 }  ,draw opacity=1 ][fill={rgb, 255:red, 0; green, 0; blue, 0 }  ,fill opacity=1 ][line width=0.75]      (0, 0) circle [x radius= 1.34, y radius= 1.34]   ;
  %Straight Lines [id:da4156755764702287] 
  \draw [color={rgb, 255:red, 0; green, 0; blue, 0 }  ,draw opacity=1 ]   (310.92,108.82) -- (310.37,137.02) ;
  \draw [shift={(310.31,140.02)}, rotate = 271.12] [fill={rgb, 255:red, 0; green, 0; blue, 0 }  ,fill opacity=1 ][line width=0.08]  [draw opacity=0] (7.14,-3.43) -- (0,0) -- (7.14,3.43) -- (4.74,0) -- cycle    ;
  %Straight Lines [id:da5033103845679101] 
  \draw [color={rgb, 255:red, 0; green, 0; blue, 0 }  ,draw opacity=1 ] [dash pattern={on 0.84pt off 2.51pt}]  (311.02,108.75) ;
  \draw [shift={(311.02,108.75)}, rotate = 0] [color={rgb, 255:red, 0; green, 0; blue, 0 }  ,draw opacity=1 ][fill={rgb, 255:red, 0; green, 0; blue, 0 }  ,fill opacity=1 ][line width=0.75]      (0, 0) circle [x radius= 1.34, y radius= 1.34]   ;
  \draw [shift={(311.02,108.75)}, rotate = 0] [color={rgb, 255:red, 0; green, 0; blue, 0 }  ,draw opacity=1 ][fill={rgb, 255:red, 0; green, 0; blue, 0 }  ,fill opacity=1 ][line width=0.75]      (0, 0) circle [x radius= 1.34, y radius= 1.34]   ;
  %Curve Lines [id:da022764317405412626] 
  \draw [color={rgb, 255:red, 0; green, 0; blue, 0 }  ,draw opacity=1 ] [dash pattern={on 0.84pt off 2.51pt}]  (290.27,115.49) .. controls (293.24,105.4) and (302.15,105.82) .. (306.82,111.29) ;
  %Straight Lines [id:da3200631770403364] 
  \draw [color={rgb, 255:red, 0; green, 0; blue, 0 }  ,draw opacity=1 ]   (295.01,79.01) -- (295.55,107.25) ;
  \draw [shift={(295.61,110.25)}, rotate = 268.9] [fill={rgb, 255:red, 0; green, 0; blue, 0 }  ,fill opacity=1 ][line width=0.08]  [draw opacity=0] (7.14,-3.43) -- (0,0) -- (7.14,3.43) -- (4.74,0) -- cycle    ;
  %Straight Lines [id:da9739104810893084] 
  \draw  [dash pattern={on 0.84pt off 2.51pt}]  (295.01,79.01) ;
  \draw [shift={(295.01,79.01)}, rotate = 0] [color={rgb, 255:red, 0; green, 0; blue, 0 }  ][fill={rgb, 255:red, 0; green, 0; blue, 0 }  ][line width=0.75]      (0, 0) circle [x radius= 1.34, y radius= 1.34]   ;
  \draw [shift={(295.01,79.01)}, rotate = 0] [color={rgb, 255:red, 0; green, 0; blue, 0 }  ][fill={rgb, 255:red, 0; green, 0; blue, 0 }  ][line width=0.75]      (0, 0) circle [x radius= 1.34, y radius= 1.34]   ;
  %Straight Lines [id:da05785655246708288] 
  \draw  [dash pattern={on 0.84pt off 2.51pt}]  (295.6,108.87) ;
  \draw [shift={(295.6,108.87)}, rotate = 0] [color={rgb, 255:red, 0; green, 0; blue, 0 }  ][fill={rgb, 255:red, 0; green, 0; blue, 0 }  ][line width=0.75]      (0, 0) circle [x radius= 1.34, y radius= 1.34]   ;
  \draw [shift={(295.6,108.87)}, rotate = 0] [color={rgb, 255:red, 0; green, 0; blue, 0 }  ][fill={rgb, 255:red, 0; green, 0; blue, 0 }  ][line width=0.75]      (0, 0) circle [x radius= 1.34, y radius= 1.34]   ;
  %Straight Lines [id:da06125574285162161] 
  \draw [color={rgb, 255:red, 0; green, 0; blue, 0 }  ,draw opacity=1 ]   (396.74,110.01) -- (391.19,138.31) ;
  \draw [shift={(390.61,141.25)}, rotate = 281.1] [fill={rgb, 255:red, 0; green, 0; blue, 0 }  ,fill opacity=1 ][line width=0.08]  [draw opacity=0] (7.14,-3.43) -- (0,0) -- (7.14,3.43) -- (4.74,0) -- cycle    ;
  %Straight Lines [id:da9447643890596567] 
  \draw  [dash pattern={on 0.84pt off 2.51pt}]  (396.74,110.01) ;
  \draw [shift={(396.74,110.01)}, rotate = 0] [color={rgb, 255:red, 0; green, 0; blue, 0 }  ][fill={rgb, 255:red, 0; green, 0; blue, 0 }  ][line width=0.75]      (0, 0) circle [x radius= 1.34, y radius= 1.34]   ;
  \draw [shift={(396.74,110.01)}, rotate = 0] [color={rgb, 255:red, 0; green, 0; blue, 0 }  ][fill={rgb, 255:red, 0; green, 0; blue, 0 }  ][line width=0.75]      (0, 0) circle [x radius= 1.34, y radius= 1.34]   ;
  
% Text Node
\draw (224.79,149.27) node [anchor=north west][inner sep=0.75pt]   [align=left] {$1$};
% Text Node
\draw (464.28,152) node [anchor=north west][inner sep=0.75pt]   [align=left] {$r$};
% Text Node
\draw (264.79,149.56) node [anchor=north west][inner sep=0.75pt]   [align=left] {$2$};
% Text Node
\draw (416.44,149.11) node [anchor=north west][inner sep=0.75pt]   [align=left] {$r-1$};
% Text Node
\draw (423.33,88.33) node [anchor=north west][inner sep=0.75pt]   [align=left] {$\mathscr{G}$};
  
  \end{tikzpicture}

   \caption{A typical spanning forest graph for single-trace MHV amplitude $A^{(g^-_i,g^-_j)}(1,\dots r||\textsc{H})$, where all gravitons and gluons play as nodes. The dashed arrow lines between gluons only reflect the relative orders in the Parke-Taylor factor. Each solid arrow line connecting two gravitons or a graviton and a gluon represents a factor (\ref{G-factor}).} 
   \label{single-t-fig1}
\end{figure}
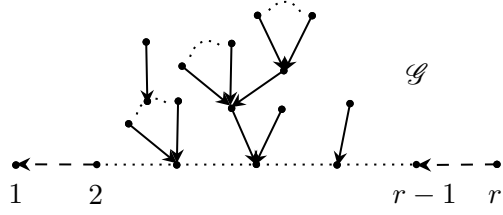

\section{Constructing single-trace MHV amplitudes}
\label{Sec4}
In this section, we construct single-trace MHV amplitudes in EYM by ISL. As pointed in \cite{Du:2016wkt,Tian:2021dzf}, EYM amplitudes in four dimensions with ($g_i^-,g_j^-$) configuration have the following form
\begin{equation}
   A_{\text{MHV}}(1,\dots,r||\textsc{H}) =\frac{\langle ij\rangle ^4}{\langle 12\rangle \langle 23\rangle \cdots\langle r1\rangle }
   \left[\sum_\mathscr{G}\prod_{e(x,y) \in \mathcal{E}(\mathscr{G})}\frac{\langle y, \xi \rangle \langle y, \eta \rangle [y,x]}{\langle x, \xi \rangle \langle x, \eta \rangle \langle y,x \rangle}
\right], 
\label{single-t-gra}
\end{equation}
where $1$, $2$, ..., $r$ are gluons, H denotes the set of gravitons. On the RHS, the color ordering of gluons is encoded into the Park-Taylor \cite{Parke:1986gb,Mangano:1987xk} factor $\frac{\langle ij\rangle ^4}{\langle 12\rangle \langle 23\rangle \cdots\langle r1\rangle }$, in which $i$ and $j$ denote the two negative-helicity gluons. The expression inside the square brackets is given by a summation over spanning forests $\mathscr{G}$. In each forest $\mathscr{G}$, the gluons $1$, ..., $r$ are considered as roots, while gravitons are connected to roots via tree structures, as shown in \figref{single-t-fig1}. The $\mathcal{E}(\mathscr{G})=\{e(x,y)\}$ in \eqref{single-t-gra} is the set of edges for a given forest $\mathscr{G}$. The edge $e(x,y)$, which is expressed by a solid arrow line pointing from node $x$ to node $y$, is accompanied by the following factor 
\begin{equation}
\frac{\langle y , \xi \rangle \langle y,\eta \rangle [y,x]  }{\langle x , \xi \rangle \langle x ,\eta \rangle \langle y,x \rangle }.
\label{G-factor}
\end{equation}

One may notice that the factor (\ref{G-factor}) has the same form as the soft factor (\ref{2.4}), which occurs in the ISL construction of gravity amplitudes (\ref{2.5}). Inspired by this, it is natural to think that the EYM amplitude (\ref{single-t-gra}) can be reconstructed through the ISL. We propose a reasonable recursive formula for the construction here: 
\begin{equation}
A_{\text{MHV}}(1,\dots,r||\textsc{H},h)=\sum_{l\in\{1,\dots,r\}\cup\textsc{H}}\mathcal{G}(h,l,\xi)A'_{\text{MHV}}(1,\dots,r||\textsc{H})\Big|_{l',\xi'},
\label{single-t-isl}
\end{equation}
where $l$ runs over all elements in $\{1,\dots,r\}\cup\textsc{H}$, $\xi$ is again an arbitrarily chosen spinor. The $A'_{\text{MHV}}(1,\dots,r||\textsc{H})\Big|_{l',\xi'}$ is an EYM amplitude which does not involve the graviton $h$. When expressed by the  formula (\ref{single-t-gra}), the expression of $A'_{\text{MHV}}(1,\dots,r||\textsc{H})\Big|_{l',\xi'}$  also involve the reference spinor $\xi\rangle$. The spinors corresponding to $l$ and $\xi$ in $A'_{\text{MHV}}$ are shifted as (\ref{2.8}). Although the factor $\mathcal{G}(h,l,\xi)$ in the above expression has the same form with an edge in \eqref{single-t-gra}, the relation between \eqref{single-t-isl} and \eqref{single-t-gra} is not transparent because  $A'_{\text{MHV}}$ in \eqref{single-t-isl} contains shifted spinors but \eqref{single-t-gra} does not. In the remaining part of this section, we show the relation between \eqref{single-t-isl} and \eqref{single-t-gra}.

\subsection{Single-trace $(g_i^-,g_j^-)$-amplitudes with one graviton}

Now we investigate the one-graviton single-trace MHV amplitude $ A(1,\dots,i^-,\dots,j^-,\dots,r||h^+_1)$ with the ($g_i^-,g_j^-$)-configuration where the negative-helicity particles are two gluons.
According to \eqref{2.8} and \eqref{single-t-isl}, the amplitude is written as
\bea
    A_{\text{MHV}}(1,\dots,r ||h^+_1) &=& \sum^r_{l=1}\frac{\langle l, \xi \rangle \langle l, \eta \rangle [l, h_1]}{\langle h_1, \xi \rangle \langle h_1, \eta \rangle \langle l,h_1\rangle} A'_{\text{MHV}}(1,\dots,l',\dots,\xi',\dots,r)\nn
&=&\sum^r_{l=1}\mathcal{G} \left(h^+_ { 1 } , l ,\xi\right)A_{\text{MHV}}(1,\dots,l,\dots,\xi,\dots,r)\nn
&=&\frac{\langle ij\rangle ^4}{\langle 12\rangle \langle 23\rangle\cdots\langle r1\rangle }\sum^r_{l=1}\mathcal{G} \left(h _ { 1 } ^ { + } , l ,\xi\right).\label{1graviton-isl}
\eea
In the above expression, the reference spinor $\xi$ is fixed as that of a gluon. The second equality follows from the fact that the amplitude $A'_{\text{MHV}}(1,\dots l',\dots,\xi',\dots,r)$ is the MHV gluon amplitude satisfying the Parke-Taylor formula and is not affected by the shifts of $l$ and $\xi$ in \eqref{2.8}. This example apparently agrees with \eqref{single-t-gra}.

\subsection{Single-trace $(g_i^-,g_j^-)$-amplitudes with two gravitons}
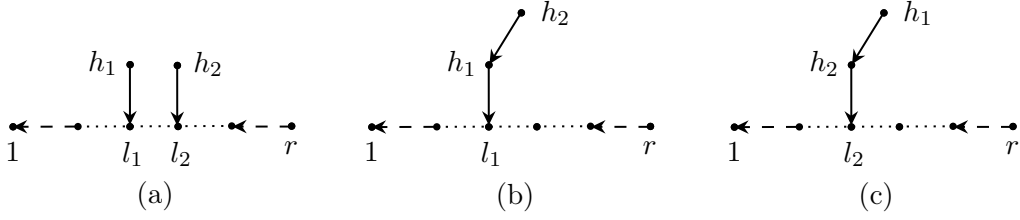
\begin{figure}
  \centering

  \tikzset{every picture/.style={line width=0.75pt}} %set default line width to 0.75pt        

  \begin{tikzpicture}[x=0.75pt,y=0.75pt,yscale=-1,xscale=1]
  %uncomment if require: \path (0,179); %set diagram left start at 0, and has height of 179
  
  %Straight Lines [id:da18656868860404652] 
  \draw  [dash pattern={on 4.5pt off 4.5pt}]  (133.38,107.2) -- (104.64,107.31) ;
  \draw [shift={(101.64,107.32)}, rotate = 359.78] [fill={rgb, 255:red, 0; green, 0; blue, 0 }  ][line width=0.08]  [draw opacity=0] (7.14,-3.43) -- (0,0) -- (7.14,3.43) -- (4.74,0) -- cycle    ;
  %Straight Lines [id:da0868630700143933] 
  \draw  [dash pattern={on 0.84pt off 2.51pt}]  (134.05,107.2) -- (210.79,106.9) ;
  \draw [shift={(210.79,106.9)}, rotate = 359.78] [color={rgb, 255:red, 0; green, 0; blue, 0 }  ][fill={rgb, 255:red, 0; green, 0; blue, 0 }  ][line width=0.75]      (0, 0) circle [x radius= 1.34, y radius= 1.34]   ;
  \draw [shift={(134.05,107.2)}, rotate = 359.78] [color={rgb, 255:red, 0; green, 0; blue, 0 }  ][fill={rgb, 255:red, 0; green, 0; blue, 0 }  ][line width=0.75]      (0, 0) circle [x radius= 1.34, y radius= 1.34]   ;
  %Straight Lines [id:da6090603629511615] 
  \draw  [dash pattern={on 4.5pt off 4.5pt}]  (240.72,107.2) -- (214.16,107.2) ;
  \draw [shift={(211.16,107.2)}, rotate = 360] [fill={rgb, 255:red, 0; green, 0; blue, 0 }  ][line width=0.08]  [draw opacity=0] (7.14,-3.43) -- (0,0) -- (7.14,3.43) -- (4.74,0) -- cycle    ;
  %Straight Lines [id:da9750279729721647] 
  \draw  [dash pattern={on 0.84pt off 2.51pt}]  (240.69,107) -- (240.72,106.98) ;
  \draw [shift={(240.72,106.98)}, rotate = 306.94] [color={rgb, 255:red, 0; green, 0; blue, 0 }  ][fill={rgb, 255:red, 0; green, 0; blue, 0 }  ][line width=0.75]      (0, 0) circle [x radius= 1.34, y radius= 1.34]   ;
  \draw [shift={(240.69,107)}, rotate = 306.94] [color={rgb, 255:red, 0; green, 0; blue, 0 }  ][fill={rgb, 255:red, 0; green, 0; blue, 0 }  ][line width=0.75]      (0, 0) circle [x radius= 1.34, y radius= 1.34]   ;
  %Straight Lines [id:da46620524914945705] 
  \draw  [dash pattern={on 0.84pt off 2.51pt}]  (101.64,107.32) ;
  \draw [shift={(101.64,107.32)}, rotate = 0] [color={rgb, 255:red, 0; green, 0; blue, 0 }  ][fill={rgb, 255:red, 0; green, 0; blue, 0 }  ][line width=0.75]      (0, 0) circle [x radius= 1.34, y radius= 1.34]   ;
  \draw [shift={(101.64,107.32)}, rotate = 0] [color={rgb, 255:red, 0; green, 0; blue, 0 }  ][fill={rgb, 255:red, 0; green, 0; blue, 0 }  ][line width=0.75]      (0, 0) circle [x radius= 1.34, y radius= 1.34]   ;
  %Straight Lines [id:da9964255624648597] 
  \draw  [dash pattern={on 0.84pt off 2.51pt}]  (160,107.23) ;
  \draw [shift={(160,107.23)}, rotate = 0] [color={rgb, 255:red, 0; green, 0; blue, 0 }  ][fill={rgb, 255:red, 0; green, 0; blue, 0 }  ][line width=0.75]      (0, 0) circle [x radius= 1.34, y radius= 1.34]   ;
  \draw [shift={(160,107.23)}, rotate = 0] [color={rgb, 255:red, 0; green, 0; blue, 0 }  ][fill={rgb, 255:red, 0; green, 0; blue, 0 }  ][line width=0.75]      (0, 0) circle [x radius= 1.34, y radius= 1.34]   ;
  %Straight Lines [id:da6509304877690671] 
  \draw  [dash pattern={on 0.84pt off 2.51pt}]  (184.01,107.17) ;
  \draw [shift={(184.01,107.17)}, rotate = 0] [color={rgb, 255:red, 0; green, 0; blue, 0 }  ][fill={rgb, 255:red, 0; green, 0; blue, 0 }  ][line width=0.75]      (0, 0) circle [x radius= 1.34, y radius= 1.34]   ;
  \draw [shift={(184.01,107.17)}, rotate = 0] [color={rgb, 255:red, 0; green, 0; blue, 0 }  ][fill={rgb, 255:red, 0; green, 0; blue, 0 }  ][line width=0.75]      (0, 0) circle [x radius= 1.34, y radius= 1.34]   ;
  %Straight Lines [id:da15543668865561067] 
  \draw [color={rgb, 255:red, 0; green, 0; blue, 0 }  ,draw opacity=1 ]   (183.57,76.53) -- (183.59,103.78) ;
  \draw [shift={(183.59,106.78)}, rotate = 269.95] [fill={rgb, 255:red, 0; green, 0; blue, 0 }  ,fill opacity=1 ][line width=0.08]  [draw opacity=0] (7.14,-3.43) -- (0,0) -- (7.14,3.43) -- (4.74,0) -- cycle    ;
  %Straight Lines [id:da8403006849171673] 
  \draw [color={rgb, 255:red, 0; green, 0; blue, 0 }  ,draw opacity=1 ] [dash pattern={on 0.84pt off 2.51pt}]  (183.67,76.47) ;
  \draw [shift={(183.67,76.47)}, rotate = 0] [color={rgb, 255:red, 0; green, 0; blue, 0 }  ,draw opacity=1 ][fill={rgb, 255:red, 0; green, 0; blue, 0 }  ,fill opacity=1 ][line width=0.75]      (0, 0) circle [x radius= 1.34, y radius= 1.34]   ;
  \draw [shift={(183.67,76.47)}, rotate = 0] [color={rgb, 255:red, 0; green, 0; blue, 0 }  ,draw opacity=1 ][fill={rgb, 255:red, 0; green, 0; blue, 0 }  ,fill opacity=1 ][line width=0.75]      (0, 0) circle [x radius= 1.34, y radius= 1.34]   ;
  %Straight Lines [id:da5430333141325867] 
  \draw [color={rgb, 255:red, 0; green, 0; blue, 0 }  ,draw opacity=1 ]   (159.97,76.13) -- (159.99,103.38) ;
  \draw [shift={(159.99,106.38)}, rotate = 269.95] [fill={rgb, 255:red, 0; green, 0; blue, 0 }  ,fill opacity=1 ][line width=0.08]  [draw opacity=0] (7.14,-3.43) -- (0,0) -- (7.14,3.43) -- (4.74,0) -- cycle    ;
  %Straight Lines [id:da03627864890684429] 
  \draw [color={rgb, 255:red, 0; green, 0; blue, 0 }  ,draw opacity=1 ] [dash pattern={on 0.84pt off 2.51pt}]  (160.07,76.07) ;
  \draw [shift={(160.07,76.07)}, rotate = 0] [color={rgb, 255:red, 0; green, 0; blue, 0 }  ,draw opacity=1 ][fill={rgb, 255:red, 0; green, 0; blue, 0 }  ,fill opacity=1 ][line width=0.75]      (0, 0) circle [x radius= 1.34, y radius= 1.34]   ;
  \draw [shift={(160.07,76.07)}, rotate = 0] [color={rgb, 255:red, 0; green, 0; blue, 0 }  ,draw opacity=1 ][fill={rgb, 255:red, 0; green, 0; blue, 0 }  ,fill opacity=1 ][line width=0.75]      (0, 0) circle [x radius= 1.34, y radius= 1.34]   ;
  %Straight Lines [id:da3481542702737992] 
  \draw  [dash pattern={on 4.5pt off 4.5pt}]  (312.49,107.34) -- (283.75,107.45) ;
  \draw [shift={(280.75,107.46)}, rotate = 359.78] [fill={rgb, 255:red, 0; green, 0; blue, 0 }  ][line width=0.08]  [draw opacity=0] (7.14,-3.43) -- (0,0) -- (7.14,3.43) -- (4.74,0) -- cycle    ;
  %Straight Lines [id:da20496517124365288] 
  \draw  [dash pattern={on 0.84pt off 2.51pt}]  (313.16,107.34) -- (389.9,107.04) ;
  \draw [shift={(389.9,107.04)}, rotate = 359.78] [color={rgb, 255:red, 0; green, 0; blue, 0 }  ][fill={rgb, 255:red, 0; green, 0; blue, 0 }  ][line width=0.75]      (0, 0) circle [x radius= 1.34, y radius= 1.34]   ;
  \draw [shift={(313.16,107.34)}, rotate = 359.78] [color={rgb, 255:red, 0; green, 0; blue, 0 }  ][fill={rgb, 255:red, 0; green, 0; blue, 0 }  ][line width=0.75]      (0, 0) circle [x radius= 1.34, y radius= 1.34]   ;
  %Straight Lines [id:da35833885841581536] 
  \draw  [dash pattern={on 4.5pt off 4.5pt}]  (419.83,107.34) -- (393.27,107.34) ;
  \draw [shift={(390.27,107.34)}, rotate = 360] [fill={rgb, 255:red, 0; green, 0; blue, 0 }  ][line width=0.08]  [draw opacity=0] (7.14,-3.43) -- (0,0) -- (7.14,3.43) -- (4.74,0) -- cycle    ;
  %Straight Lines [id:da013931563279951487] 
  \draw  [dash pattern={on 0.84pt off 2.51pt}]  (419.81,107.15) -- (419.83,107.12) ;
  \draw [shift={(419.83,107.12)}, rotate = 306.94] [color={rgb, 255:red, 0; green, 0; blue, 0 }  ][fill={rgb, 255:red, 0; green, 0; blue, 0 }  ][line width=0.75]      (0, 0) circle [x radius= 1.34, y radius= 1.34]   ;
  \draw [shift={(419.81,107.15)}, rotate = 306.94] [color={rgb, 255:red, 0; green, 0; blue, 0 }  ][fill={rgb, 255:red, 0; green, 0; blue, 0 }  ][line width=0.75]      (0, 0) circle [x radius= 1.34, y radius= 1.34]   ;
  %Straight Lines [id:da3247813037777312] 
  \draw  [dash pattern={on 0.84pt off 2.51pt}]  (280.75,107.46) ;
  \draw [shift={(280.75,107.46)}, rotate = 0] [color={rgb, 255:red, 0; green, 0; blue, 0 }  ][fill={rgb, 255:red, 0; green, 0; blue, 0 }  ][line width=0.75]      (0, 0) circle [x radius= 1.34, y radius= 1.34]   ;
  \draw [shift={(280.75,107.46)}, rotate = 0] [color={rgb, 255:red, 0; green, 0; blue, 0 }  ][fill={rgb, 255:red, 0; green, 0; blue, 0 }  ][line width=0.75]      (0, 0) circle [x radius= 1.34, y radius= 1.34]   ;
  %Straight Lines [id:da3877423051579205] 
  \draw  [dash pattern={on 0.84pt off 2.51pt}]  (339.11,107.37) ;
  \draw [shift={(339.11,107.37)}, rotate = 0] [color={rgb, 255:red, 0; green, 0; blue, 0 }  ][fill={rgb, 255:red, 0; green, 0; blue, 0 }  ][line width=0.75]      (0, 0) circle [x radius= 1.34, y radius= 1.34]   ;
  \draw [shift={(339.11,107.37)}, rotate = 0] [color={rgb, 255:red, 0; green, 0; blue, 0 }  ][fill={rgb, 255:red, 0; green, 0; blue, 0 }  ][line width=0.75]      (0, 0) circle [x radius= 1.34, y radius= 1.34]   ;
  %Straight Lines [id:da16676605997597282] 
  \draw  [dash pattern={on 0.84pt off 2.51pt}]  (363.12,107.32) ;
  \draw [shift={(363.12,107.32)}, rotate = 0] [color={rgb, 255:red, 0; green, 0; blue, 0 }  ][fill={rgb, 255:red, 0; green, 0; blue, 0 }  ][line width=0.75]      (0, 0) circle [x radius= 1.34, y radius= 1.34]   ;
  \draw [shift={(363.12,107.32)}, rotate = 0] [color={rgb, 255:red, 0; green, 0; blue, 0 }  ][fill={rgb, 255:red, 0; green, 0; blue, 0 }  ][line width=0.75]      (0, 0) circle [x radius= 1.34, y radius= 1.34]   ;
  %Straight Lines [id:da9880399636156825] 
  \draw [color={rgb, 255:red, 0; green, 0; blue, 0 }  ,draw opacity=1 ]   (339.08,76.28) -- (339.1,103.53) ;
  \draw [shift={(339.11,106.53)}, rotate = 269.95] [fill={rgb, 255:red, 0; green, 0; blue, 0 }  ,fill opacity=1 ][line width=0.08]  [draw opacity=0] (7.14,-3.43) -- (0,0) -- (7.14,3.43) -- (4.74,0) -- cycle    ;
  %Straight Lines [id:da04980510453669318] 
  \draw [color={rgb, 255:red, 0; green, 0; blue, 0 }  ,draw opacity=1 ] [dash pattern={on 0.84pt off 2.51pt}]  (339.19,76.21) ;
  \draw [shift={(339.19,76.21)}, rotate = 0] [color={rgb, 255:red, 0; green, 0; blue, 0 }  ,draw opacity=1 ][fill={rgb, 255:red, 0; green, 0; blue, 0 }  ,fill opacity=1 ][line width=0.75]      (0, 0) circle [x radius= 1.34, y radius= 1.34]   ;
  \draw [shift={(339.19,76.21)}, rotate = 0] [color={rgb, 255:red, 0; green, 0; blue, 0 }  ,draw opacity=1 ][fill={rgb, 255:red, 0; green, 0; blue, 0 }  ,fill opacity=1 ][line width=0.75]      (0, 0) circle [x radius= 1.34, y radius= 1.34]   ;
  %Straight Lines [id:da1790561981866765] 
  \draw [color={rgb, 255:red, 0; green, 0; blue, 0 }  ,draw opacity=1 ]   (355.25,49.99) -- (341.01,73.22) ;
  \draw [shift={(339.44,75.78)}, rotate = 301.51] [fill={rgb, 255:red, 0; green, 0; blue, 0 }  ,fill opacity=1 ][line width=0.08]  [draw opacity=0] (7.14,-3.43) -- (0,0) -- (7.14,3.43) -- (4.74,0) -- cycle    ;
  %Straight Lines [id:da2557377028096339] 
  \draw [color={rgb, 255:red, 0; green, 0; blue, 0 }  ,draw opacity=1 ] [dash pattern={on 0.84pt off 2.51pt}]  (355.38,50) ;
  \draw [shift={(355.38,50)}, rotate = 0] [color={rgb, 255:red, 0; green, 0; blue, 0 }  ,draw opacity=1 ][fill={rgb, 255:red, 0; green, 0; blue, 0 }  ,fill opacity=1 ][line width=0.75]      (0, 0) circle [x radius= 1.34, y radius= 1.34]   ;
  \draw [shift={(355.38,50)}, rotate = 0] [color={rgb, 255:red, 0; green, 0; blue, 0 }  ,draw opacity=1 ][fill={rgb, 255:red, 0; green, 0; blue, 0 }  ,fill opacity=1 ][line width=0.75]      (0, 0) circle [x radius= 1.34, y radius= 1.34]   ;
  %Straight Lines [id:da390202104128905] 
  \draw  [dash pattern={on 4.5pt off 4.5pt}]  (493.46,107.37) -- (464.72,107.48) ;
  \draw [shift={(461.72,107.49)}, rotate = 359.78] [fill={rgb, 255:red, 0; green, 0; blue, 0 }  ][line width=0.08]  [draw opacity=0] (7.14,-3.43) -- (0,0) -- (7.14,3.43) -- (4.74,0) -- cycle    ;
  %Straight Lines [id:da2795748279098891] 
  \draw  [dash pattern={on 0.84pt off 2.51pt}]  (494.13,107.37) -- (570.87,107.07) ;
  \draw [shift={(570.87,107.07)}, rotate = 359.78] [color={rgb, 255:red, 0; green, 0; blue, 0 }  ][fill={rgb, 255:red, 0; green, 0; blue, 0 }  ][line width=0.75]      (0, 0) circle [x radius= 1.34, y radius= 1.34]   ;
  \draw [shift={(494.13,107.37)}, rotate = 359.78] [color={rgb, 255:red, 0; green, 0; blue, 0 }  ][fill={rgb, 255:red, 0; green, 0; blue, 0 }  ][line width=0.75]      (0, 0) circle [x radius= 1.34, y radius= 1.34]   ;
  %Straight Lines [id:da8519525520676041] 
  \draw  [dash pattern={on 4.5pt off 4.5pt}]  (600.79,107.37) -- (574.24,107.37) ;
  \draw [shift={(571.24,107.37)}, rotate = 360] [fill={rgb, 255:red, 0; green, 0; blue, 0 }  ][line width=0.08]  [draw opacity=0] (7.14,-3.43) -- (0,0) -- (7.14,3.43) -- (4.74,0) -- cycle    ;
  %Straight Lines [id:da2482684413070262] 
  \draw  [dash pattern={on 0.84pt off 2.51pt}]  (600.77,107.18) -- (600.79,107.15) ;
  \draw [shift={(600.79,107.15)}, rotate = 306.94] [color={rgb, 255:red, 0; green, 0; blue, 0 }  ][fill={rgb, 255:red, 0; green, 0; blue, 0 }  ][line width=0.75]      (0, 0) circle [x radius= 1.34, y radius= 1.34]   ;
  \draw [shift={(600.77,107.18)}, rotate = 306.94] [color={rgb, 255:red, 0; green, 0; blue, 0 }  ][fill={rgb, 255:red, 0; green, 0; blue, 0 }  ][line width=0.75]      (0, 0) circle [x radius= 1.34, y radius= 1.34]   ;
  %Straight Lines [id:da680174266702344] 
  \draw  [dash pattern={on 0.84pt off 2.51pt}]  (461.72,107.49) ;
  \draw [shift={(461.72,107.49)}, rotate = 0] [color={rgb, 255:red, 0; green, 0; blue, 0 }  ][fill={rgb, 255:red, 0; green, 0; blue, 0 }  ][line width=0.75]      (0, 0) circle [x radius= 1.34, y radius= 1.34]   ;
  \draw [shift={(461.72,107.49)}, rotate = 0] [color={rgb, 255:red, 0; green, 0; blue, 0 }  ][fill={rgb, 255:red, 0; green, 0; blue, 0 }  ][line width=0.75]      (0, 0) circle [x radius= 1.34, y radius= 1.34]   ;
  %Straight Lines [id:da5727304234448893] 
  \draw  [dash pattern={on 0.84pt off 2.51pt}]  (520.08,107.4) ;
  \draw [shift={(520.08,107.4)}, rotate = 0] [color={rgb, 255:red, 0; green, 0; blue, 0 }  ][fill={rgb, 255:red, 0; green, 0; blue, 0 }  ][line width=0.75]      (0, 0) circle [x radius= 1.34, y radius= 1.34]   ;
  \draw [shift={(520.08,107.4)}, rotate = 0] [color={rgb, 255:red, 0; green, 0; blue, 0 }  ][fill={rgb, 255:red, 0; green, 0; blue, 0 }  ][line width=0.75]      (0, 0) circle [x radius= 1.34, y radius= 1.34]   ;
  %Straight Lines [id:da39644996541790256] 
  \draw  [dash pattern={on 0.84pt off 2.51pt}]  (544.09,107.34) ;
  \draw [shift={(544.09,107.34)}, rotate = 0] [color={rgb, 255:red, 0; green, 0; blue, 0 }  ][fill={rgb, 255:red, 0; green, 0; blue, 0 }  ][line width=0.75]      (0, 0) circle [x radius= 1.34, y radius= 1.34]   ;
  \draw [shift={(544.09,107.34)}, rotate = 0] [color={rgb, 255:red, 0; green, 0; blue, 0 }  ][fill={rgb, 255:red, 0; green, 0; blue, 0 }  ][line width=0.75]      (0, 0) circle [x radius= 1.34, y radius= 1.34]   ;
  %Straight Lines [id:da6491622113622808] 
  \draw [color={rgb, 255:red, 0; green, 0; blue, 0 }  ,draw opacity=1 ]   (520.05,76.31) -- (520.07,103.55) ;
  \draw [shift={(520.07,106.55)}, rotate = 269.95] [fill={rgb, 255:red, 0; green, 0; blue, 0 }  ,fill opacity=1 ][line width=0.08]  [draw opacity=0] (7.14,-3.43) -- (0,0) -- (7.14,3.43) -- (4.74,0) -- cycle    ;
  %Straight Lines [id:da16547535781539757] 
  \draw [color={rgb, 255:red, 0; green, 0; blue, 0 }  ,draw opacity=1 ] [dash pattern={on 0.84pt off 2.51pt}]  (520.15,76.24) ;
  \draw [shift={(520.15,76.24)}, rotate = 0] [color={rgb, 255:red, 0; green, 0; blue, 0 }  ,draw opacity=1 ][fill={rgb, 255:red, 0; green, 0; blue, 0 }  ,fill opacity=1 ][line width=0.75]      (0, 0) circle [x radius= 1.34, y radius= 1.34]   ;
  \draw [shift={(520.15,76.24)}, rotate = 0] [color={rgb, 255:red, 0; green, 0; blue, 0 }  ,draw opacity=1 ][fill={rgb, 255:red, 0; green, 0; blue, 0 }  ,fill opacity=1 ][line width=0.75]      (0, 0) circle [x radius= 1.34, y radius= 1.34]   ;
  %Straight Lines [id:da8386396592299685] 
  \draw [color={rgb, 255:red, 0; green, 0; blue, 0 }  ,draw opacity=1 ]   (536.22,50.02) -- (521.98,73.25) ;
  \draw [shift={(520.41,75.81)}, rotate = 301.51] [fill={rgb, 255:red, 0; green, 0; blue, 0 }  ,fill opacity=1 ][line width=0.08]  [draw opacity=0] (7.14,-3.43) -- (0,0) -- (7.14,3.43) -- (4.74,0) -- cycle    ;
  %Straight Lines [id:da2753083382693] 
  \draw [color={rgb, 255:red, 0; green, 0; blue, 0 }  ,draw opacity=1 ] [dash pattern={on 0.84pt off 2.51pt}]  (536.35,50.02) ;
  \draw [shift={(536.35,50.02)}, rotate = 0] [color={rgb, 255:red, 0; green, 0; blue, 0 }  ,draw opacity=1 ][fill={rgb, 255:red, 0; green, 0; blue, 0 }  ,fill opacity=1 ][line width=0.75]      (0, 0) circle [x radius= 1.34, y radius= 1.34]   ;
  \draw [shift={(536.35,50.02)}, rotate = 0] [color={rgb, 255:red, 0; green, 0; blue, 0 }  ,draw opacity=1 ][fill={rgb, 255:red, 0; green, 0; blue, 0 }  ,fill opacity=1 ][line width=0.75]      (0, 0) circle [x radius= 1.34, y radius= 1.34]   ;
  
% Text Node
\draw (96.35,113.54) node [anchor=north west][inner sep=0.75pt]   [align=left] {$1$};
% Text Node
\draw (235.4,113.04) node [anchor=north west][inner sep=0.75pt]   [align=left] {$r$};
% Text Node
\draw (154.57,113.21) node [anchor=north west][inner sep=0.75pt]   [align=left] {$l_1$};
% Text Node
\draw (178.73,113.44) node [anchor=north west][inner sep=0.75pt]   [align=left] {$l_2$};
% Text Node
\draw (161.8,133.4) node [anchor=north west][inner sep=0.75pt]   [align=left] {(a)};
% Text Node
\draw (137.57,67.61) node [anchor=north west][inner sep=0.75pt]   [align=left] {$h_1$};
% Text Node
\draw (190.17,68.01) node [anchor=north west][inner sep=0.75pt]   [align=left] {$h_2$};
% Text Node
\draw (275.46,113.69) node [anchor=north west][inner sep=0.75pt]   [align=left] {$1$};
% Text Node
\draw (414.51,113.19) node [anchor=north west][inner sep=0.75pt]   [align=left] {$r$};
% Text Node
\draw (333.68,113.35) node [anchor=north west][inner sep=0.75pt]   [align=left] {$l_1$};
% Text Node
\draw (341.13,133.55) node [anchor=north west][inner sep=0.75pt]   [align=left] {(b)};
% Text Node
\draw (316.68,67.75) node [anchor=north west][inner sep=0.75pt]   [align=left] {$h_1$};
% Text Node
\draw (363.51,42.37) node [anchor=north west][inner sep=0.75pt]   [align=left] {$h_2$};
% Text Node
\draw (456.43,113.72) node [anchor=north west][inner sep=0.75pt]   [align=left] {$1$};
% Text Node
\draw (595.47,113.21) node [anchor=north west][inner sep=0.75pt]   [align=left] {$r$};
% Text Node
\draw (514.65,113.38) node [anchor=north west][inner sep=0.75pt]   [align=left] {$l_2$};
% Text Node
\draw (522.1,133.57) node [anchor=north west][inner sep=0.75pt]   [align=left] {(c)};
% Text Node
\draw (497.65,67.78) node [anchor=north west][inner sep=0.75pt]   [align=left] {$h_2$};
% Text Node
\draw (544.47,42.4) node [anchor=north west][inner sep=0.75pt]   [align=left] {$h_1$};
  
  \end{tikzpicture}
  
  \caption{Graphs for EYM amplitude with two gravitons. In (a) both gravitons $h_1$ and $h_2$ root at gluons. In (b), $h_1$ roots at a gluon while $h_2$ roots at $h_1$. The graph (c) is obtained by exchanging the roles of $h_1$ and $h_2$ in (b).}
  \label{single-t-fig2} 
\end{figure}
The one-graviton example is trivial, for the deformations are not presented in the lower-point amplitudes. On the contrary, when the single-trace amplitudes with more gravitons are reproduced according to ISL, the sub-amplitudes on the RHS of (\ref{single-t-isl}) generally contain the deformed spinors. Thus we need to relate the contributions of deformed sub-amplitudes to forests $\mathscr{G}$ that are introduced in \eqref{single-t-gra}. To clarify this point, we study the single-trace MHV amplitudes with two positive-helicity gravitons $h^+_1$ and $h^+_2$, as an example. Graphs for single-trace amplitude with two gravitons are given by \figref{single-t-fig2}. 

Using the formula (\ref{single-t-gra}), we can read off the contributions of these graphs directly:
\begin{equation}
\begin{aligned}
\text{\figref{single-t-fig2} (a)}&\quad\to\quad   A_1&=&      \sum_{l_1 \in \{1,\dots,r\}}  \mathcal{G} \left(h _ { 1 } , l_1 \right)  \sum_{l_2 \in \left\{1,\dots,r\right\} } \mathcal{G} \left(h _2 , l_2 \right)\,\textsc{PT},\\
\text{\figref{single-t-fig2} (b)}&\quad\to\quad    A_2&=&\quad \mathcal{G} \left(h _ { 2 } , h_1 \right)\,\sum_{l_1\in \left\{1,\dots,r\right\} } \mathcal{G} \left(h _ { 1 } , l_1 \right)\,\textsc{PT},\\
\text{\figref{single-t-fig2} (c)}&\quad\to\quad    A_3&=&\quad \mathcal{G} \left(h _ { 1 } , h_2 \right)\,\sum_{l_2\in \left\{1,\dots,r\right\} } \mathcal{G} \left(h _ { 2} , l_2 \right)\, \textsc{PT},
\end{aligned}
\label{2graviton-rep1}
\end{equation}
where PT denotes the "Parke-Taylor" factor $\frac{\langle ij\rangle ^4}{\langle 12\rangle \langle 23\rangle \cdots\langle r1\rangle }$. Here, all the edge factors $\mathcal{G}$'s take the same reference spinors $\xi$ and $\eta$. Thus, we omit $\xi$ in notation $\mathcal{G}$ for brevity. 

On the other hand, according to \eqref{single-t-isl}, the single-trace MHV amplitude with two gravitons is expanded as
\begin{equation}
A_{\text{MHV}}(1,\dots,r||h_1,h_2)=\sum_{l_2 \in T  \cup \{h_1\}} \frac{\langle l_2  ,\xi \rangle \langle l_2 ,\eta \rangle [l_2,h_2 ]  }{\langle h_2 , \xi\rangle \langle h_2,\eta \rangle \langle l_2,h_2  \rangle } A'_{\text{MHV}}(1,\dots,r||h_1)\Big|_{l'_2,\xi'},
\label{2graviton-isl}
\end{equation}
where we write $ \{1,\dots,r\}$ as $T$ for short. The amplitude $A_{\text{MHV}}(1,\dots,r||h_1)$ with one graviton $h_1$ has been shown to satisfy the expression of \eqref{1graviton-isl}. When the shifted spinors in \eqref{2.8} are substituted into $A_{\text{MHV}}(1,\dots,r||h_1)$, the Parke-Taylor factor in \eqref{1graviton-isl} keeps unchanged because it only contains spinors of such form $\langle...\rangle$. Thus the only factor that may be affected by the shifts in \eqref{2graviton-isl}  is $\mathcal{G} \left(h _ { 1 } ^ { + } , l_1 ,\xi\right)$. Particularly,
\begin{itemize}
\item If $l_2$ is a gluon in $ T\setminus\left\{l_1\right\}$, the factor $\mathcal{G} \left(h _ { 1 } ^ { + } , l_1 ,\xi\right)$ does not contain the shifted spinors $l'_2]$ and $\xi']$.
\item If $l_2$ is $l_1$ or $h_1$, the factor $\mathcal{G} \left(h _ { 1 } ^ { + } , l_1 ,\xi\right)$ must involve the shifted spinors $l'_1]$ or $h'_1]$, respectively.
\end{itemize}
Altogether, we expand the RHS of \eqref{2graviton-isl} as
\begin{equation}
\begin{aligned}
(\ref{2graviton-isl})&=\underbrace{\sum_{l_2\in T\setminus\left\{l_1\right\}}\mathcal{G}(h_2,l_2)\,\sum_{l_1\in T}\mathcal{G}(h_1,l_1)\,\textsc{PT}}_{{I_1}}+\underbrace{\sum_{l_1\in T}\mathcal{G}(h_2,l_1)\,\mathcal{G}'(h_1,l_1)\,\textsc{PT}}_{{I_2}}\\
&~~+\underbrace{\mathcal{G}(h_2,l_2)\Big|_{l_2=h_1}\,\sum_{l_1\in T}\mathcal{G}'(h_1,l_1)\,\textsc{PT}}_{{I_3}}.
\end{aligned}
\end{equation}
When the shifted spinor $l' ]$ in \eqref{2.8} is brought into the above expression, each of $I_2$ and $I_3$ further splits into two terms:
\begin{equation}
\begin{aligned}
{I_2}&=\underbrace{\sum_{l_2=l_1\in T}\frac{\langle l_1  , \xi \rangle \langle l_1 ,\eta \rangle [l_1,h_2 ]  }{\langle h_2 , \xi \rangle \langle h_2,\eta \rangle \langle l_1,h_2  \rangle }\,\frac{\langle l_1  , \xi \rangle \langle l_1 ,\eta \rangle [l_1,h_1 ]  }{\langle h_1 , \xi \rangle \langle h_1,\eta \rangle \langle l_1,h_1  \rangle }\,\textsc{PT}}_{{I_{2A}}}\\
&~~+\underbrace{\sum_{l_1\in T}\frac{\langle l_1  , \xi \rangle \langle l_1 ,\eta \rangle [l_1,h_2 ]  }{ \langle l_1,h_2  \rangle }\,\frac{\langle h_2,h_1\rangle\langle l_1,\eta\rangle}{\langle h_2,\eta\rangle\langle l_1,h_1\rangle}\,\frac{ [h_2,h_1 ]  }{\langle h_1 , \xi \rangle \langle h_1,\eta \rangle \langle h_2,h_1  \rangle }\,\textsc{PT}}_{{I_{2B}}},
\end{aligned}
\label{I2-split}
\end{equation}
\begin{equation}
\begin{aligned}
{I_3}&=\underbrace{\frac{\langle h_1  , \xi \rangle \langle h_1 ,\eta \rangle [h_1,h_2 ]  }{ \langle h_2,\xi\rangle \langle h_2,\eta\rangle\langle h_1,h_2\rangle}\,\sum_{l_1\in T}\frac{ \langle l_1,\xi\rangle\langle l_1,\eta\rangle[l_1,h_1] }{\langle h_1 , \xi \rangle \langle h_1,\eta \rangle \langle l_1,h_1  \rangle }\,\textsc{PT}}_{{I_{3A}}}\\
&~~+\underbrace{\sum_{l_1\in T}\frac{\langle l_1  , \xi \rangle \langle l_1 ,\eta \rangle [l_1,h_2 ]  }{ \langle l_1,h_2  \rangle }\,\frac{\langle h_1,\eta\rangle\langle l_1,h_2\rangle}{\langle h_2,\eta\rangle\langle l_1,h_1\rangle}\,\frac{ [h_2,h_1 ]  }{\langle h_1 , \xi \rangle \langle h_1,\eta \rangle \langle h_2,h_1  \rangle }\,\textsc{PT}}_{{I_{3B}}},
\end{aligned}
\label{I3-split}
\end{equation}
where $I_{2A}$ and $I_{3A}$ are resulted by the unchanged $l_2]$ part in $l'_2]$ expression, while $I_{2B}$ and $I_{3B}$ correspond to the shifted part $\frac{\langle h_2,\xi\rangle}{\langle l_2,\xi\rangle}h_2]$ of $l'_2]$. The correspondences between \eqref{2graviton-rep1} and \eqref{2graviton-isl} are then established as follows:
\begin{equation*}
\begin{aligned}
&{I_1}+{I_{2A}}&=&\quad A_1,\\
&\quad{I_{3A}}&=&\quad A_2,\\
&{I_{2B}}+{I_{3B}}&=&\quad A_3,
\end{aligned}
\end{equation*}
where the Schouten identity has been applied on the third line. 

So far, we have found the connection between the ISL approach (\ref{2graviton-isl}) and the spanning forest formula (\ref{2graviton-rep1}), for single-trace amplitude with two gravitons. We now summarize several key points here:
\begin{itemize}
\item Although there are two shifted spinors  $\xi']$ and $l'_2]$, only $ l'_2]$ shift has effects on amplitudes.
\item Amplitudes on the RHS of \eqref{2graviton-isl} with deformed momenta can be rearranged as the contributions of graphs in \figref{single-t-fig2}. 
\item Schouten identity is the core trick for reproducing the contribution of graphs where $h_2$ is not a leaf (i.e. an outermost node).
\end{itemize}
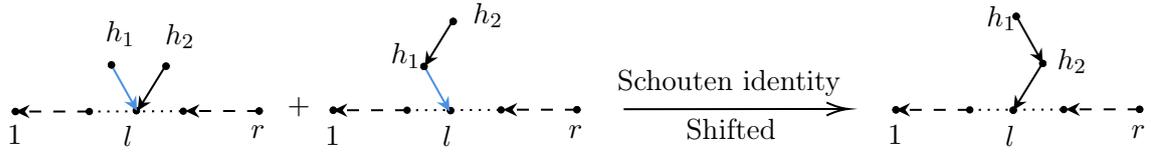
\begin{figure}
  \centering

  \tikzset{every picture/.style={line width=0.75pt}} %set default line width to 0.75pt        

  \begin{tikzpicture}[x=0.75pt,y=0.75pt,yscale=-1,xscale=1]
  %uncomment if require: \path (0,172); %set diagram left start at 0, and has height of 172
  
  %Straight Lines [id:da578845346029736] 
  \draw  [dash pattern={on 4.5pt off 4.5pt}]  (104.05,104.87) -- (70.18,104.87) ;
  \draw [shift={(67.18,104.87)}, rotate = 360] [fill={rgb, 255:red, 0; green, 0; blue, 0 }  ][line width=0.08]  [draw opacity=0] (7.14,-3.43) -- (0,0) -- (7.14,3.43) -- (4.74,0) -- cycle    ;
  %Straight Lines [id:da2512899257712078] 
  \draw  [dash pattern={on 0.84pt off 2.51pt}]  (104.05,104.87) -- (150.95,104.71) ;
  \draw [shift={(150.95,104.71)}, rotate = 359.8] [color={rgb, 255:red, 0; green, 0; blue, 0 }  ][fill={rgb, 255:red, 0; green, 0; blue, 0 }  ][line width=0.75]      (0, 0) circle [x radius= 1.34, y radius= 1.34]   ;
  \draw [shift={(104.05,104.87)}, rotate = 359.8] [color={rgb, 255:red, 0; green, 0; blue, 0 }  ][fill={rgb, 255:red, 0; green, 0; blue, 0 }  ][line width=0.75]      (0, 0) circle [x radius= 1.34, y radius= 1.34]   ;
  %Straight Lines [id:da16209685663667917] 
  \draw  [dash pattern={on 4.5pt off 4.5pt}]  (188.84,104.61) -- (154.97,104.61) ;
  \draw [shift={(151.97,104.61)}, rotate = 360] [fill={rgb, 255:red, 0; green, 0; blue, 0 }  ][line width=0.08]  [draw opacity=0] (7.14,-3.43) -- (0,0) -- (7.14,3.43) -- (4.74,0) -- cycle    ;
  %Straight Lines [id:da012671758460590654] 
  \draw  [dash pattern={on 0.84pt off 2.51pt}]  (188.84,104.61) ;
  \draw [shift={(188.84,104.61)}, rotate = 0] [color={rgb, 255:red, 0; green, 0; blue, 0 }  ][fill={rgb, 255:red, 0; green, 0; blue, 0 }  ][line width=0.75]      (0, 0) circle [x radius= 1.34, y radius= 1.34]   ;
  \draw [shift={(188.84,104.61)}, rotate = 0] [color={rgb, 255:red, 0; green, 0; blue, 0 }  ][fill={rgb, 255:red, 0; green, 0; blue, 0 }  ][line width=0.75]      (0, 0) circle [x radius= 1.34, y radius= 1.34]   ;
  %Straight Lines [id:da16068065324192582] 
  \draw  [dash pattern={on 0.84pt off 2.51pt}]  (67.18,104.87) ;
  \draw [shift={(67.18,104.87)}, rotate = 0] [color={rgb, 255:red, 0; green, 0; blue, 0 }  ][fill={rgb, 255:red, 0; green, 0; blue, 0 }  ][line width=0.75]      (0, 0) circle [x radius= 1.34, y radius= 1.34]   ;
  \draw [shift={(67.18,104.87)}, rotate = 0] [color={rgb, 255:red, 0; green, 0; blue, 0 }  ][fill={rgb, 255:red, 0; green, 0; blue, 0 }  ][line width=0.75]      (0, 0) circle [x radius= 1.34, y radius= 1.34]   ;
  %Straight Lines [id:da3648698156977901] 
  \draw  [dash pattern={on 0.84pt off 2.51pt}]  (127.54,104.42) ;
  \draw [shift={(127.54,104.42)}, rotate = 0] [color={rgb, 255:red, 0; green, 0; blue, 0 }  ][fill={rgb, 255:red, 0; green, 0; blue, 0 }  ][line width=0.75]      (0, 0) circle [x radius= 1.34, y radius= 1.34]   ;
  \draw [shift={(127.54,104.42)}, rotate = 0] [color={rgb, 255:red, 0; green, 0; blue, 0 }  ][fill={rgb, 255:red, 0; green, 0; blue, 0 }  ][line width=0.75]      (0, 0) circle [x radius= 1.34, y radius= 1.34]   ;
  %Straight Lines [id:da7800689192594783] 
  \draw [color={rgb, 255:red, 74; green, 144; blue, 226 }  ,draw opacity=1 ]   (114.95,81.31) -- (126.41,101.88) ;
  \draw [shift={(127.87,104.5)}, rotate = 240.87] [fill={rgb, 255:red, 74; green, 144; blue, 226 }  ,fill opacity=1 ][line width=0.08]  [draw opacity=0] (7.14,-3.43) -- (0,0) -- (7.14,3.43) -- (4.74,0) -- cycle    ;
  %Straight Lines [id:da5600399219198267] 
  \draw [color={rgb, 255:red, 0; green, 0; blue, 0 }  ,draw opacity=1 ] [dash pattern={on 0.84pt off 2.51pt}]  (115.07,81.43) ;
  \draw [shift={(115.07,81.43)}, rotate = 0] [color={rgb, 255:red, 0; green, 0; blue, 0 }  ,draw opacity=1 ][fill={rgb, 255:red, 0; green, 0; blue, 0 }  ,fill opacity=1 ][line width=0.75]      (0, 0) circle [x radius= 1.34, y radius= 1.34]   ;
  \draw [shift={(115.07,81.43)}, rotate = 0] [color={rgb, 255:red, 0; green, 0; blue, 0 }  ,draw opacity=1 ][fill={rgb, 255:red, 0; green, 0; blue, 0 }  ,fill opacity=1 ][line width=0.75]      (0, 0) circle [x radius= 1.34, y radius= 1.34]   ;
  %Straight Lines [id:da4547704692097485] 
  \draw [color={rgb, 255:red, 0; green, 0; blue, 0 }  ,draw opacity=1 ]   (142.19,81.7) -- (129.76,101.94) ;
  \draw [shift={(128.19,104.5)}, rotate = 301.56] [fill={rgb, 255:red, 0; green, 0; blue, 0 }  ,fill opacity=1 ][line width=0.08]  [draw opacity=0] (7.14,-3.43) -- (0,0) -- (7.14,3.43) -- (4.74,0) -- cycle    ;
  %Straight Lines [id:da7047262546656958] 
  \draw  [dash pattern={on 0.84pt off 2.51pt}]  (142.35,82.06) ;
  \draw [shift={(142.35,82.06)}, rotate = 0] [color={rgb, 255:red, 0; green, 0; blue, 0 }  ][fill={rgb, 255:red, 0; green, 0; blue, 0 }  ][line width=0.75]      (0, 0) circle [x radius= 1.34, y radius= 1.34]   ;
  \draw [shift={(142.35,82.06)}, rotate = 0] [color={rgb, 255:red, 0; green, 0; blue, 0 }  ][fill={rgb, 255:red, 0; green, 0; blue, 0 }  ][line width=0.75]      (0, 0) circle [x radius= 1.34, y radius= 1.34]   ;
  %Straight Lines [id:da6802940201623124] 
  \draw  [dash pattern={on 4.5pt off 4.5pt}]  (262.68,104.21) -- (228.81,104.21) ;
  \draw [shift={(225.81,104.21)}, rotate = 360] [fill={rgb, 255:red, 0; green, 0; blue, 0 }  ][line width=0.08]  [draw opacity=0] (7.14,-3.43) -- (0,0) -- (7.14,3.43) -- (4.74,0) -- cycle    ;
  %Straight Lines [id:da29327616648087984] 
  \draw  [dash pattern={on 0.84pt off 2.51pt}]  (262.68,104.21) -- (309.58,104.05) ;
  \draw [shift={(309.58,104.05)}, rotate = 359.8] [color={rgb, 255:red, 0; green, 0; blue, 0 }  ][fill={rgb, 255:red, 0; green, 0; blue, 0 }  ][line width=0.75]      (0, 0) circle [x radius= 1.34, y radius= 1.34]   ;
  \draw [shift={(262.68,104.21)}, rotate = 359.8] [color={rgb, 255:red, 0; green, 0; blue, 0 }  ][fill={rgb, 255:red, 0; green, 0; blue, 0 }  ][line width=0.75]      (0, 0) circle [x radius= 1.34, y radius= 1.34]   ;
  %Straight Lines [id:da13662121062670818] 
  \draw  [dash pattern={on 4.5pt off 4.5pt}]  (347.47,103.95) -- (313.59,103.95) ;
  \draw [shift={(310.59,103.95)}, rotate = 360] [fill={rgb, 255:red, 0; green, 0; blue, 0 }  ][line width=0.08]  [draw opacity=0] (7.14,-3.43) -- (0,0) -- (7.14,3.43) -- (4.74,0) -- cycle    ;
  %Straight Lines [id:da958047224527262] 
  \draw  [dash pattern={on 0.84pt off 2.51pt}]  (347.47,103.95) ;
  \draw [shift={(347.47,103.95)}, rotate = 0] [color={rgb, 255:red, 0; green, 0; blue, 0 }  ][fill={rgb, 255:red, 0; green, 0; blue, 0 }  ][line width=0.75]      (0, 0) circle [x radius= 1.34, y radius= 1.34]   ;
  \draw [shift={(347.47,103.95)}, rotate = 0] [color={rgb, 255:red, 0; green, 0; blue, 0 }  ][fill={rgb, 255:red, 0; green, 0; blue, 0 }  ][line width=0.75]      (0, 0) circle [x radius= 1.34, y radius= 1.34]   ;
  %Straight Lines [id:da562672529224385] 
  \draw  [dash pattern={on 0.84pt off 2.51pt}]  (225.81,104.21) ;
  \draw [shift={(225.81,104.21)}, rotate = 0] [color={rgb, 255:red, 0; green, 0; blue, 0 }  ][fill={rgb, 255:red, 0; green, 0; blue, 0 }  ][line width=0.75]      (0, 0) circle [x radius= 1.34, y radius= 1.34]   ;
  \draw [shift={(225.81,104.21)}, rotate = 0] [color={rgb, 255:red, 0; green, 0; blue, 0 }  ][fill={rgb, 255:red, 0; green, 0; blue, 0 }  ][line width=0.75]      (0, 0) circle [x radius= 1.34, y radius= 1.34]   ;
  %Straight Lines [id:da8007655934035283] 
  \draw  [dash pattern={on 0.84pt off 2.51pt}]  (284.38,104.39) ;
  \draw [shift={(284.38,104.39)}, rotate = 0] [color={rgb, 255:red, 0; green, 0; blue, 0 }  ][fill={rgb, 255:red, 0; green, 0; blue, 0 }  ][line width=0.75]      (0, 0) circle [x radius= 1.34, y radius= 1.34]   ;
  \draw [shift={(284.38,104.39)}, rotate = 0] [color={rgb, 255:red, 0; green, 0; blue, 0 }  ][fill={rgb, 255:red, 0; green, 0; blue, 0 }  ][line width=0.75]      (0, 0) circle [x radius= 1.34, y radius= 1.34]   ;
  %Straight Lines [id:da20516545329539504] 
  \draw [color={rgb, 255:red, 74; green, 144; blue, 226 }  ,draw opacity=1 ]   (270.95,81.6) -- (282.41,102.17) ;
  \draw [shift={(283.87,104.79)}, rotate = 240.87] [fill={rgb, 255:red, 74; green, 144; blue, 226 }  ,fill opacity=1 ][line width=0.08]  [draw opacity=0] (7.14,-3.43) -- (0,0) -- (7.14,3.43) -- (4.74,0) -- cycle    ;
  %Straight Lines [id:da3027659148970985] 
  \draw [color={rgb, 255:red, 0; green, 0; blue, 0 }  ,draw opacity=1 ] [dash pattern={on 0.84pt off 2.51pt}]  (271.13,82.2) ;
  \draw [shift={(271.13,82.2)}, rotate = 0] [color={rgb, 255:red, 0; green, 0; blue, 0 }  ,draw opacity=1 ][fill={rgb, 255:red, 0; green, 0; blue, 0 }  ,fill opacity=1 ][line width=0.75]      (0, 0) circle [x radius= 1.34, y radius= 1.34]   ;
  \draw [shift={(271.13,82.2)}, rotate = 0] [color={rgb, 255:red, 0; green, 0; blue, 0 }  ,draw opacity=1 ][fill={rgb, 255:red, 0; green, 0; blue, 0 }  ,fill opacity=1 ][line width=0.75]      (0, 0) circle [x radius= 1.34, y radius= 1.34]   ;
  %Straight Lines [id:da6073744159575036] 
  \draw [color={rgb, 255:red, 0; green, 0; blue, 0 }  ,draw opacity=1 ]   (285.53,59.19) -- (273.09,79.43) ;
  \draw [shift={(271.52,81.99)}, rotate = 301.56] [fill={rgb, 255:red, 0; green, 0; blue, 0 }  ,fill opacity=1 ][line width=0.08]  [draw opacity=0] (7.14,-3.43) -- (0,0) -- (7.14,3.43) -- (4.74,0) -- cycle    ;
  %Straight Lines [id:da7199450054627716] 
  \draw  [dash pattern={on 0.84pt off 2.51pt}]  (285.69,59.55) ;
  \draw [shift={(285.69,59.55)}, rotate = 0] [color={rgb, 255:red, 0; green, 0; blue, 0 }  ][fill={rgb, 255:red, 0; green, 0; blue, 0 }  ][line width=0.75]      (0, 0) circle [x radius= 1.34, y radius= 1.34]   ;
  \draw [shift={(285.69,59.55)}, rotate = 0] [color={rgb, 255:red, 0; green, 0; blue, 0 }  ][fill={rgb, 255:red, 0; green, 0; blue, 0 }  ][line width=0.75]      (0, 0) circle [x radius= 1.34, y radius= 1.34]   ;
  %Straight Lines [id:da8570180965853724] 
  \draw    (370.81,103.7) -- (480.99,103.58) ;
  \draw [shift={(482.99,103.58)}, rotate = 179.94] [color={rgb, 255:red, 0; green, 0; blue, 0 }  ][line width=0.75]    (10.93,-3.29) .. controls (6.95,-1.4) and (3.31,-0.3) .. (0,0) .. controls (3.31,0.3) and (6.95,1.4) .. (10.93,3.29)   ;
  %Straight Lines [id:da6287314858605337] 
  \draw  [dash pattern={on 4.5pt off 4.5pt}]  (543.54,104.03) -- (509.67,104.03) ;
  \draw [shift={(506.67,104.03)}, rotate = 360] [fill={rgb, 255:red, 0; green, 0; blue, 0 }  ][line width=0.08]  [draw opacity=0] (7.14,-3.43) -- (0,0) -- (7.14,3.43) -- (4.74,0) -- cycle    ;
  %Straight Lines [id:da8723769319163968] 
  \draw  [dash pattern={on 0.84pt off 2.51pt}]  (543.54,104.03) -- (590.44,103.87) ;
  \draw [shift={(590.44,103.87)}, rotate = 359.8] [color={rgb, 255:red, 0; green, 0; blue, 0 }  ][fill={rgb, 255:red, 0; green, 0; blue, 0 }  ][line width=0.75]      (0, 0) circle [x radius= 1.34, y radius= 1.34]   ;
  \draw [shift={(543.54,104.03)}, rotate = 359.8] [color={rgb, 255:red, 0; green, 0; blue, 0 }  ][fill={rgb, 255:red, 0; green, 0; blue, 0 }  ][line width=0.75]      (0, 0) circle [x radius= 1.34, y radius= 1.34]   ;
  %Straight Lines [id:da04587464971493738] 
  \draw  [dash pattern={on 4.5pt off 4.5pt}]  (628.33,103.77) -- (594.46,103.77) ;
  \draw [shift={(591.46,103.77)}, rotate = 360] [fill={rgb, 255:red, 0; green, 0; blue, 0 }  ][line width=0.08]  [draw opacity=0] (7.14,-3.43) -- (0,0) -- (7.14,3.43) -- (4.74,0) -- cycle    ;
  %Straight Lines [id:da3076425316606586] 
  \draw  [dash pattern={on 0.84pt off 2.51pt}]  (628.33,103.77) ;
  \draw [shift={(628.33,103.77)}, rotate = 0] [color={rgb, 255:red, 0; green, 0; blue, 0 }  ][fill={rgb, 255:red, 0; green, 0; blue, 0 }  ][line width=0.75]      (0, 0) circle [x radius= 1.34, y radius= 1.34]   ;
  \draw [shift={(628.33,103.77)}, rotate = 0] [color={rgb, 255:red, 0; green, 0; blue, 0 }  ][fill={rgb, 255:red, 0; green, 0; blue, 0 }  ][line width=0.75]      (0, 0) circle [x radius= 1.34, y radius= 1.34]   ;
  %Straight Lines [id:da7502082015605429] 
  \draw  [dash pattern={on 0.84pt off 2.51pt}]  (506.67,104.03) ;
  \draw [shift={(506.67,104.03)}, rotate = 0] [color={rgb, 255:red, 0; green, 0; blue, 0 }  ][fill={rgb, 255:red, 0; green, 0; blue, 0 }  ][line width=0.75]      (0, 0) circle [x radius= 1.34, y radius= 1.34]   ;
  \draw [shift={(506.67,104.03)}, rotate = 0] [color={rgb, 255:red, 0; green, 0; blue, 0 }  ][fill={rgb, 255:red, 0; green, 0; blue, 0 }  ][line width=0.75]      (0, 0) circle [x radius= 1.34, y radius= 1.34]   ;
  %Straight Lines [id:da5469255460468176] 
  \draw  [dash pattern={on 0.84pt off 2.51pt}]  (565.25,104.21) ;
  \draw [shift={(565.25,104.21)}, rotate = 0] [color={rgb, 255:red, 0; green, 0; blue, 0 }  ][fill={rgb, 255:red, 0; green, 0; blue, 0 }  ][line width=0.75]      (0, 0) circle [x radius= 1.34, y radius= 1.34]   ;
  \draw [shift={(565.25,104.21)}, rotate = 0] [color={rgb, 255:red, 0; green, 0; blue, 0 }  ][fill={rgb, 255:red, 0; green, 0; blue, 0 }  ][line width=0.75]      (0, 0) circle [x radius= 1.34, y radius= 1.34]   ;
  %Straight Lines [id:da6652646771506212] 
  \draw [color={rgb, 255:red, 0; green, 0; blue, 0 }  ,draw opacity=1 ]   (567.05,57.38) -- (578.51,77.95) ;
  \draw [shift={(579.97,80.57)}, rotate = 240.87] [fill={rgb, 255:red, 0; green, 0; blue, 0 }  ,fill opacity=1 ][line width=0.08]  [draw opacity=0] (7.14,-3.43) -- (0,0) -- (7.14,3.43) -- (4.74,0) -- cycle    ;
  %Straight Lines [id:da13485992717701634] 
  \draw [color={rgb, 255:red, 0; green, 0; blue, 0 }  ,draw opacity=1 ] [dash pattern={on 0.84pt off 2.51pt}]  (566.74,57) ;
  \draw [shift={(566.74,57)}, rotate = 0] [color={rgb, 255:red, 0; green, 0; blue, 0 }  ,draw opacity=1 ][fill={rgb, 255:red, 0; green, 0; blue, 0 }  ,fill opacity=1 ][line width=0.75]      (0, 0) circle [x radius= 1.34, y radius= 1.34]   ;
  \draw [shift={(566.74,57)}, rotate = 0] [color={rgb, 255:red, 0; green, 0; blue, 0 }  ,draw opacity=1 ][fill={rgb, 255:red, 0; green, 0; blue, 0 }  ,fill opacity=1 ][line width=0.75]      (0, 0) circle [x radius= 1.34, y radius= 1.34]   ;
  %Straight Lines [id:da23962482324110135] 
  \draw [color={rgb, 255:red, 0; green, 0; blue, 0 }  ,draw opacity=1 ]   (579.38,81.4) -- (566.95,101.65) ;
  \draw [shift={(565.38,104.2)}, rotate = 301.56] [fill={rgb, 255:red, 0; green, 0; blue, 0 }  ,fill opacity=1 ][line width=0.08]  [draw opacity=0] (7.14,-3.43) -- (0,0) -- (7.14,3.43) -- (4.74,0) -- cycle    ;
  %Straight Lines [id:da19969906506837942] 
  \draw  [dash pattern={on 0.84pt off 2.51pt}]  (579.99,80.71) ;
  \draw [shift={(579.99,80.71)}, rotate = 0] [color={rgb, 255:red, 0; green, 0; blue, 0 }  ][fill={rgb, 255:red, 0; green, 0; blue, 0 }  ][line width=0.75]      (0, 0) circle [x radius= 1.34, y radius= 1.34]   ;
  \draw [shift={(579.99,80.71)}, rotate = 0] [color={rgb, 255:red, 0; green, 0; blue, 0 }  ][fill={rgb, 255:red, 0; green, 0; blue, 0 }  ][line width=0.75]      (0, 0) circle [x radius= 1.34, y radius= 1.34]   ;
% Text Node
\draw (61.73,110.25) node [anchor=north west][inner sep=0.75pt]   [align=left] {$1$};
% Text Node
\draw (183.18,110.45) node [anchor=north west][inner sep=0.75pt]   [align=left] {$r$};
% Text Node
\draw (119.46,110.46) node [anchor=north west][inner sep=0.75pt]   [align=left] {$l$};
% Text Node
\draw (110.63,58.02) node [anchor=north west][inner sep=0.75pt]  [align=left] {$h_1$};
% Text Node
\draw (140.37,59.12) node [anchor=north west][inner sep=0.75pt]  [align=left] {$h_2$};
% Text Node
\draw (220.36,110.59) node [anchor=north west][inner sep=0.75pt]   [align=left] {$1$};
% Text Node
\draw (341.81,110.49) node [anchor=north west][inner sep=0.75pt]   [align=left] {$r$};
% Text Node
\draw (278.95,110.45) node [anchor=north west][inner sep=0.75pt]   [align=left] {$l$};
% Text Node
\draw (253.09,68.98) node [anchor=north west][inner sep=0.75pt]  [align=left] {$h_1$};
% Text Node
\draw (294.2,48.61) node [anchor=north west][inner sep=0.75pt]   [align=left] {$h_2$};
% Text Node
\draw (201.56,96.3) node [anchor=north west][inner sep=0.75pt]   [align=left] {$+$};
% Text Node
\draw (366.26,82.45) node [anchor=north west][inner sep=0.75pt]   [align=left] {Schouten identity};
% Text Node
\draw (400.52,107.24) node [anchor=north west][inner sep=0.75pt]  [color={rgb, 255:red, 0; green, 0; blue, 0 }  ,opacity=1 ] [align=left] {Shifted};
% Text Node
\draw (501.22,110.41) node [anchor=north west][inner sep=0.75pt]   [align=left] {$1$};
% Text Node
\draw (622.67,110.6) node [anchor=north west][inner sep=0.75pt]   [align=left] {$r$};
% Text Node
\draw (559.81,110.4) node [anchor=north west][inner sep=0.75pt]   [align=left] {$l$};
% Text Node
\draw (550.53,50.13) node [anchor=north west][inner sep=0.75pt]  [align=left] {$h_1$};
% Text Node
\draw (585.79,71.96) node [anchor=north west][inner sep=0.75pt]   [align=left] {$h_2$};

  \end{tikzpicture}
  
  \caption{A graphic representation for the summation of shifted parts in \eqref{2graviton-isl}. Here, the blue arrows represent the shifted part $I_{2B}$ and $I_{3B}$.}
  \label{2-graviton-shift-fig} 
\end{figure}

Noting that ISL is a recursive procedure, as shown in \eqref{2graviton-isl}, we multiply one more gravity soft factor $\mathcal{G}$ to a shifted lower-point amplitude which is supposed to satisfy the spanning forest formula. In the two-graviton example, when further separating the unshifted part from the full shifted amplitude, we find that (i). the unshifted parts give rise to \figref{single-t-fig2} (a) and (b) where the graviton $h_2$ plays  as a leaf, (ii). the shift results in those graphs with the structure \figref{single-t-fig2} (c) where the graviton $h_2$ is an internal node. The contribution of the shifted terms can be shown by \figref{2-graviton-shift-fig}, where the blue arrow lines are denoted as shifted parts. This observation inspires a general study of the ISL for single-trace EYM amplitudes.

\subsection{Terms with more shifted factors}
\begin{figure}
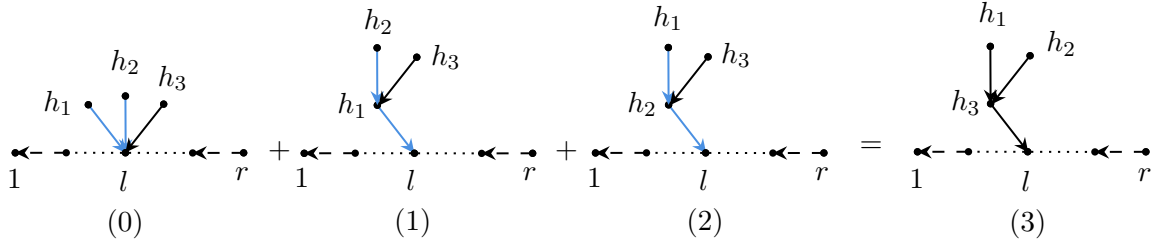

  \centering

  \tikzset{every picture/.style={line width=0.75pt}} %set default line width to 0.75pt        
  

  
  \caption{Graphic representation for the summation of terms containing two shifted factors derived from $A'_{\text{MHV}}(1,\dots,r||h_1,h_2)\big|_{l'_3}$.}
  \label{3-graviton-shift-fig}
  \end{figure}

  \begin{figure}
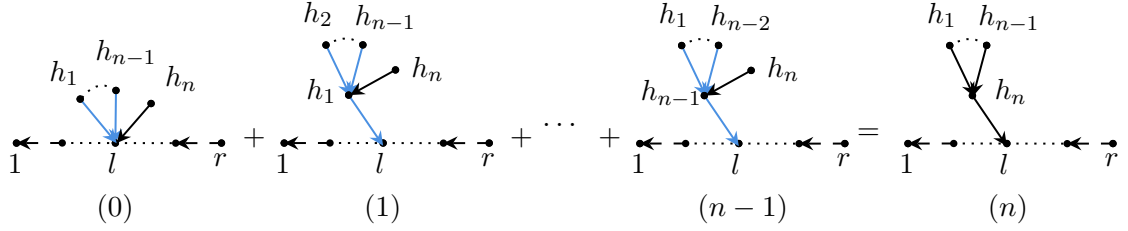

    \centering

  \tikzset{every picture/.style={line width=0.75pt}} %set default line width to 0.75pt        
  
  %
  
    \caption{Graphic representation for the summation of terms containing $n-1$ shifted factors in the $n$-graviton case.}
    \label{ngraviton-fig-rule}
    \end{figure}

In the two-graviton example, any term in the shifted amplitude contains one shifted factor $\frac{\langle h_k,\xi\rangle}{\langle l_k,\xi\rangle}h_k]$ at most. However, terms with more shifted factors have to be considered in amplitudes with more than two gravitons. We take the single-trace amplitude with three gravitons as an example to clarify this more complicated situation.  In the three-graviton case, when we insert the graviton $h_3$ to the spanning forests of the two-graviton amplitude, the momentum of the node which is adjacent to $h_3$ is shifted. Consequently, all edges attached to this node are affected by the shift. Considering this node can be attached by more than two edges, the graph will contribute to terms with up to two shifted factors $\left(\frac{\langle h_k,\xi\rangle}{\langle l_k,\xi\rangle}h_k]\right)^2$. In the single-trace amplitude with three gravitons, terms with two shifted factors are shown by the three graphs on the LHS of \figref{3-graviton-shift-fig}, where the blue arrow lines in each graph denote the shifted parts of the shifted spinors
\begin{equation}
l']=l]+\frac{\langle h_3,\xi\rangle}{\langle l,\xi\rangle}h_3],\quad h'_1]=h_1]+\frac{\langle h_3,\xi\rangle}{\langle h_1,\xi\rangle}h_3],\quad h'_2]=h_2]+\frac{\langle h_3,\xi\rangle}{\langle h_2,\xi\rangle}h_3],
\label{3-shift}
\end{equation}
respectively. The sum of the three graphs reproduces the RHS graph, where $h_3$ becomes an internal node and no shift appears in the edges. This can be verified by writing the contributions of LHS and the RHS explicitly.

To prove the identity shown in \figref{3-graviton-shift-fig}, we write down the contributions of graphs
\begin{equation}
\begin{aligned}
\text{\figref{3-graviton-shift-fig} (0)}\quad &=\quad\frac{[l,h_3]}{\langle l,h_3\rangle}\frac{[h_3,h_2]}{\langle l,h_2\rangle}\frac{[h_3,h_1]}{\langle l,h_1\rangle}\frac{\langle l,\xi\rangle\langle l,\eta\rangle^3\langle h_3,\xi\rangle}{\langle h_1,\xi\rangle\langle h_1,\eta\rangle\langle h_2,\xi\rangle\langle h_2,\eta\rangle\langle h_3,\eta\rangle}\,\textsc{PT},\\
\text{\figref{3-graviton-shift-fig} (1)} \quad &=\quad\frac{[l,h_3]}{\langle l,h_1\rangle}\frac{[h_3,h_2]}{\langle h_1,h_2\rangle}\frac{[h_3,h_1]}{\langle h_3,h_1\rangle}\frac{\langle l,\xi\rangle\langle l,\eta\rangle\langle h_1,\eta\rangle\langle h_3,\xi\rangle}{\langle h_2,\xi\rangle\langle h_2,\eta\rangle\langle h_3,\eta\rangle\langle h_1,\xi\rangle}\,\textsc{PT},\\
\text{\figref{3-graviton-shift-fig} (2)} \quad &=\quad\frac{[l,h_3]}{\langle l,h_2\rangle}\frac{[h_3,h_2]}{\langle h_3,h_2\rangle}\frac{[h_3,h_1]}{\langle h_2,h_1\rangle}\frac{\langle l,\xi\rangle\langle l,\eta\rangle\langle h_2,\eta\rangle\langle h_3,\xi\rangle}{\langle h_1,\xi\rangle\langle h_1,\eta\rangle\langle h_3,\eta\rangle\langle h_2,\xi\rangle}\,\textsc{PT},\\
\text{\figref{3-graviton-shift-fig} (3)} \quad &=\quad\frac{[l,h_3]}{\langle l,h_3\rangle}\frac{[h_3,h_2]}{\langle h_3,h_2\rangle}\frac{[h_3,h_1]}{\langle h_2,h_1\rangle}\frac{\langle l,\xi\rangle\langle l,\eta\rangle\langle h_2,\eta\rangle\langle h_3,\xi\rangle}{\langle h_1,\xi\rangle\langle h_1,\eta\rangle\langle h_3,\eta\rangle\langle h_2,\xi\rangle}\,\textsc{PT}.
\end{aligned}
\label{3-id-list}
\end{equation}
Once the identity of \figref{3-graviton-shift-fig}, $(0)+(1)+(2)=(3)$ holds, we can extract the common factors and get 
\bea
\langle h_1,\eta\rangle^2\langle l,h_3\rangle\langle l,h_2\rangle\langle h_2,h_3\rangle&+&\langle h_2,\eta\rangle^2\langle l,h_3\rangle\langle l,h_1\rangle\langle h_3,h_1\rangle\nn
&+&\langle h_3,\eta\rangle^2\langle l,h_2\rangle\langle l,h_1\rangle\langle h_1,h_2\rangle+\langle l,\eta\rangle^2\langle h_3,h_2\rangle\langle h_3,h_1\rangle\langle h_2,h_1\rangle=0,\label{3graviton-id}
\eea
which is just the aforementioned identity (\ref{4-id}).

So far, we have concluded two rules shown by  \figref{2-graviton-shift-fig} and \figref{3-graviton-shift-fig}, which are encountered in the study of two- and three-graviton cases. These identities can further be extended to a more general identity which is shown by  \figref{ngraviton-fig-rule} and will be encountered in our general proof.

To prove it, we write down the LHS of \figref{ngraviton-fig-rule} explicitly. The graph \figref{ngraviton-fig-rule} ($0$) contributes
\begin{equation}
  \text{\figref{ngraviton-fig-rule} (0)}=\frac{[l,h_n]\langle l,\xi\rangle\langle l,\eta\rangle^{n}\langle h_n,\xi\rangle^{n-2}}{\langle l,h_n\rangle\langle h_n,\eta\rangle}\prod_{i\in\{1,\dots,n-1\}}\frac{[h_n,h_i]}{\langle l,h_i\rangle\langle h_i,\xi\rangle\langle h_i,\eta\rangle}\text{PT}.
  \label{figure70}
  \end{equation}
Each of \figref{ngraviton-fig-rule} ($1$)-($n-1$) on the LHS corresponds to
\begin{equation}
\text{\figref{ngraviton-fig-rule} }(k)=\frac{[l,h_n][h_n,h_k]\langle l,\xi\rangle\langle l,\eta\rangle(\langle h_n,\xi\rangle\langle h_k,\eta\rangle)^{n-2}}{\langle l,h_k\rangle\langle h_n,h_k\rangle\langle h_k,\xi\rangle\langle h_n,\eta\rangle}\prod_{i\in\{1,\dots,n-1\}\setminus\{k\}}\frac{[h_n,h_i]}{\langle h_k,h_i\rangle\langle h_i,\xi\rangle\langle h_i,\eta\rangle}\text{PT},
\label{figure7k}
\end{equation}
where $k=1,...,n-1$. In the above expression, we have already incorporated the shifted factor $\left(\frac{\langle h_n,\xi\rangle}{\langle h_k,\xi\rangle}h_i]\right)^{n-1}$ into the entire expression. The graph \figref{ngraviton-fig-rule} ($n$) contributes
\begin{equation}
  \text{\figref{ngraviton-fig-rule} }(n)=\frac{[l,h_n]\langle l,\xi\rangle\langle l,\eta\rangle\left(\langle h_n,\xi\rangle\langle h_n,\eta\rangle\right)^{n-2}}{\langle l,h_n\rangle}\prod_{i\in\{1,\dots,n-1\}}\frac{[h_n,h_i]}{\langle h_n,h_i\rangle\langle h_i,\xi\rangle\langle h_i,\eta\rangle}\text{PT}.
  \label{figure7n}
\end{equation}
When a factor (\ref{figure7n}) is extracted out,  \eqref{figure70} and \eqref{figure7k} can further be arranged as
\bea
\text{\figref{ngraviton-fig-rule} }(0)&=&\left(\frac{\langle l,\eta\rangle}{\langle h_n,\eta\rangle}\right)^{n-1}\prod_{i\in\{1,\dots,n-1\}}\frac{\langle h_n,h_i\rangle}{-\langle h_i,l\rangle}\times\text{\figref{ngraviton-fig-rule} }(n),
\label{S-id-0}\\
\text{\figref{ngraviton-fig-rule} }(k)&=&\left(\frac{\langle h_k,\eta\rangle}{\langle h_n,\eta\rangle}\right)^{n-1}\frac{\langle h_n, l\rangle}{\langle h_k,l \rangle}\nn
&&~~~~~~~~~\times(-1)^{n-k+1}\prod_{i\in\{1,\dots,k-1\}}\frac{\langle h_n,h_i\rangle}{\langle h_k,h_i\rangle}\prod_{i\in\{k+1,\dots,n-1\}}\frac{\langle h_n,h_i\rangle}{\langle h_i,h_k\rangle}\times\text{\figref{ngraviton-fig-rule} }(n).
\label{S-id-k}
\eea
Thus the identity encoded in \figref{ngraviton-fig-rule} is equivalent to 
\begin{equation}
\left[(\ref{S-id-0})+\sum_{k\in\{1,\dots,n-1\}}(\ref{S-id-k})\right](\ref{figure7n})^{-1}=1.
\label{ngraviton-shift-id}
\end{equation}
To further simplify \eqref{ngraviton-shift-id}, we need to find out the common denominator of $\frac{(\ref{S-id-0})}{(\ref{figure7n})}$ and $\sum\limits_{k\in\{1,\dots,n-1\}}\frac{(\ref{S-id-k})}{(\ref{figure7n})}$. It's not hard to see that the common denominator takes the form of
\bea
\langle h_n,\eta\rangle^{n-1}\prod\limits_{0\leq j<i\leq n-1}\langle h_i,h_j\rangle,
\label{st-id-n}
\eea
where we mark $l$ as $h_0$. Multiplying this common factor by \eqref{ngraviton-shift-id}, it simplifies to 
\begin{equation}
\begin{aligned}
\langle l,\eta\rangle^{n-1}(-1)^{n+1}\prod\limits_{1\leq j<i\leq n}\langle h_i,h_j\rangle+\sum\limits_{k\in\{1,\dots,n-1\}}&\langle h_k,\eta\rangle^{n-1}(-1)^{n-k+1}\prod\limits_{\substack{0\leq j<i\leq n\\i,j\neq k}}\langle h_i,h_j\rangle \\
&~~~~~~~~~-\langle h_n,\eta\rangle^{n-1}\prod\limits_{0\leq j<i\leq n-1}\langle h_i,h_j\rangle=0.
\end{aligned}
\end{equation}
This identity must hold because of \eqref{S-id-ab}.

\subsection{General proof of the single-trace $(g_i^-,g_j^-)$ case}

In this section, we prove that the ISL expression of single-trace amplitudes (\ref{single-t-isl}) in general reproduces the contribution of spanning forests as shown by \figref{single-t-fig1}, and thus is equivalent to \eqref{single-t-gra}. We have already proved the one-, two-graviton cases and evaluated the effect of shifts in the $n$-graviton cases in previous subsections. To complete the final step of the inductive proof, we suppose the ISL has been proven to be effective for the amplitude with $n-1$ gravitons, then the following formula must hold
\bea
   A_{\text{MHV}}(1,\dots,r||\textsc{H}\setminus\left\{h_n\right\})&=&\mathcal{G}(h_{n-1},l_{n-1})\, A'_{\text{MHV}}(1,\dots,r||\textsc{H}\setminus\{h_n,h_{n-1}\})\nn
&=&\text{PT}\,
   \left[\sum_{\mathscr{G}^{n-1}}\prod_{e(x,y) \in \mathcal{E}(\mathscr{G}^{n-1})}\frac{\langle y, \xi \rangle \langle y, \eta \rangle [y,x]}{\langle x, \xi \rangle \langle x, \eta \rangle \langle y,x \rangle}
\right],
\label{n-1-ISL}
\eea
where we have summed over forests $\mathscr{G}^{n-1}$ generated by graviton set $\textsc{H}\setminus\left\{h_n\right\}$. As \eqref{n-1-ISL} holds, we insert the graviton $h_n$ to \eqref{n-1-ISL}, then we obtain
\bea
   A_{\text{MHV}}(1,\dots,r||\textsc{H}) &=&\sum_{l_{n}\in T \cup \text{H}\setminus\left\{h_n\right\}}\mathcal{G}(h_n,l_n)\, A'_{\text{MHV}}(1,\dots,r||\textsc{H}\setminus\left\{h_n\right\})\nn
&=&\sum_{l_{n}\in T \cup \text{H}\setminus\left\{h_n\right\}}\mathcal{G}(h_n,l_n)\,
   \left[\sum_{\mathscr{G}^{n-1}}\prod_{e(x,y) \in \mathcal{E}(\mathscr{G}^{n-1})}\frac{\langle y, \xi \rangle \langle y, \eta \rangle [y,x]}{\langle x, \xi\rangle \langle x, \eta \rangle \langle y,x \rangle}
\right]'\Bigg|_{l'_n,\xi'}\,\textsc{PT},
\label{n-ISL}
\eea
where the Parke-Taylor factor is shift-independent and the expression in the square brackets involves shifted spinors. Terms in the above summation can be classified according to distinct positions of $l_n$ that is the neighbor of the newly inserted graviton $h_n$: 
\begin{itemize}
\item  If $l_n\in \mathscr{G}^{n-1}$ is a gluon with no tree structure planted on it, the shift is applied on the Parke-Taylor factor, and thus has no effect on the expression above.
\item If $l_n\in \mathscr{G}^{n-1}$ belongs to a tree structure (either the root or other nodes), the shift has nontrivial effects on the expression inside the square brackets. One can split the shifted spinor ${l'_n}]$ into an unshifted one ${l_n}]$ and a shifted part $\frac{\langle h_n,\xi\rangle}{\langle l_n,\xi\rangle}h_n]$. Then the expression is expanded according to different powers of the shifted part $\frac{\langle h_n,\xi\rangle}{\langle l_n,\xi\rangle}h_n]$. 

\end{itemize}
\begin{figure}
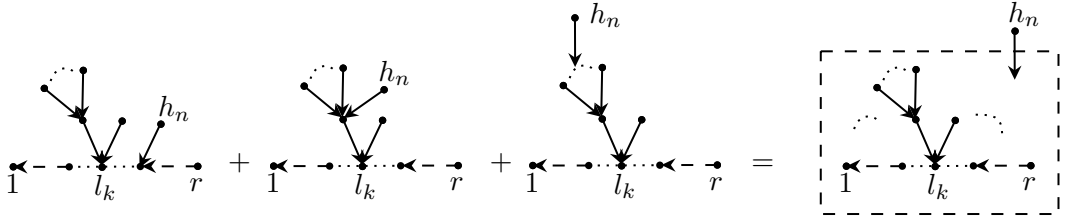

  \centering

  \tikzset{every picture/.style={line width=0.75pt}} %set default line width to 0.75pt        

  
  \caption{In the summation of all unshifted terms, $h_n$ can root at a gluon, an internal graviton, or an outermost graviton. In all these graphs $h_n$ plays as a leaf (an outermost node). All such graphs are packaged into the RHS.}
  \label{un-shifted-contribution} 
\end{figure}
The contribution of all unshifted terms is represented in \figref{un-shifted-contribution}, where $h_n$ is simply inserted as a new leaf into the original graphs. The contribution involving the $f$-th ($f>0$) power of the shifted parts can be collected together, as shown by the LHS of \figref{ngraviton-fig-gen}. As the identity shown in \figref{ngraviton-fig-rule} is applied, we immediately get the RHS of \figref{ngraviton-fig-gen}, where the graviton $h_n$ is an internal node and the structures attached to $h_1,\dots, h_f$ are rooted at $h_n$. Conversely, for any graph on the RHS, one can always find a set of graphs uniquely shown by the LHS, each of them containing the $f$-th power of the shifted part.
Therefore, we have established a one-to-one correspondence between a set of graphs with shifted parts and a graph where $h_n$ plays as an internal node. Consequently, all the graphs on the RHS of \eqref{single-t-gra} are produced.  Hence the equivalence between the ISL and the spanning forest form for single-trace amplitudes has been proven.

\begin{figure}
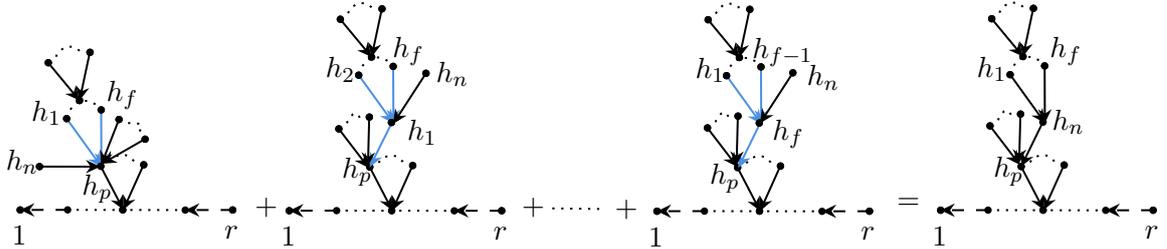

\centering

\tikzset{every picture/.style={line width=0.75pt}} %set default line width to 0.75pt        



\caption{Graphic representation for the summation of an arbitrary $f$-th power shifted terms which contributes to the new configuration on the RHS of the identity.}
\label{ngraviton-fig-gen}
\end{figure}

\subsection{Single-trace amplitudes with the ($h^-_i$, $g^-_j$) and the ($h^-_i$, $h^-_j$) configurations}
\label{sec4.5}

Other two configurations ($h^-_i$, $g^-_j$) and ($h^-_i$, $h^-_j$) with two negative-helicity particles have not been discussed yet. When considering the ($h^-_i$, $g^-_j$) configuration, one may try to insert all gravitons one by one into an $r$-gluon amplitude with only one negative-helicity gluon. This approach fails when $r>3$ because the $r$ ($r$>3) gluon amplitude with only one negative helicity particle must vanish. Thus, we begin with a three-gluon amplitude which contains one negative-helicity gluon and two positive-helicity gluons. Then we construct the full amplitude by inserting other particles in the following way:
\begin{itemize}
\item {\bf Step-1}  Supposing the negative-helicity gluon is the gluon $1$, we begin with $A_3(1^-,2^+,3^+)$ (where $2^+$ and $3^+$ are positive-helicity gluons) and insert the negative-helicity graviton, say $h^-_i$ according to ISL\cite{Nandan:2012rk}: 
\begin{equation}
  \begin{aligned}
    A(1^-,2^+,3^+||h^-_i)&=\sum_{l\in\{2,3\}}\mathcal{G}(h^-_i,l,1)A'_3(1^-,2^+,3^+)\\
  &=\sum_{l\in\{2,3\}}\frac{[l,1][l,\eta]\langle l,h_i\rangle}{[h_i,1][h_i,\eta][l,h_i]}\frac{[2,3]^4}{[1,2][2,3][3,1]}\\
  &=\left\{\frac{[2,1][2,\eta]\langle2,3\rangle[h_i,3]}{[h_i,1][h_i,\eta][2,3]\langle h_i,3\rangle}+\frac{[3,1][3,\eta]\langle3,1\rangle[h_i,1]}{[h_i,1][h_i,\eta][3,1]\langle h_i,1\rangle}\right\}\frac{\langle 1,h_i\rangle^4}{\langle1,2\rangle\langle2,3\rangle\langle 3,1\rangle}\\
 % &=\left\{\frac{[2,1][2,\eta]\langle2|3|h_i]}{[h_i,1][h_i,\eta]\langle h_i|3|2]}+\frac{[3,\eta]\langle3,1\rangle}{[h_i,\eta]\langle h_i,1\rangle}\right\}\frac{\langle 1,h_i\rangle^4}{\langle1,2\rangle\langle2,3\rangle\langle 3,1\rangle}\\
  &=\sum_{l\in\{1,2,3\}}\frac{[l,\eta]\langle l,1\rangle}{[h_i,\eta]\langle h_i,1\rangle}\frac{\langle 1,h_i\rangle^4}{\langle1,2\rangle\langle2,3\rangle\langle 3,1\rangle}\\
  &=(-1)\frac{\langle 1,h_i\rangle^4}{\langle1,2\rangle\langle2,3\rangle\langle 3,1\rangle}.
  \end{aligned}
\label{hg-1}
\end{equation}
In the above equation, we have followed the same discussion as that in the $(g^-_i,g^-_j)$ case, but exchanged the roles of the two types of spinors $\rangle$ and $]$, momentum conservation was  applied. This is reasonable because our starting point is an $\overline{\text{MHV}}$ amplitude and the graviton carries negative helicity now.

\item {\bf Step-2} After step-1, we obtained an EYM amplitude with two negative-helicity particles. Now we insert positive-helicity gluons into this amplitude according to the ISL for gluons \cite{Boucher-Veronneau:2011rwd}, then get the amplitude with one negative-helicity graviton, of the form
\bea
 A(1^-,2^+,...,r^+||h^-_i)=(-1)\frac{\langle 1,h_i\rangle^4}{\langle1,2\rangle\langle2,3\rangle...\langle r,1\rangle},\label{hg-2}
\eea
which agrees with the known result \cite{Tian:2021dzf}. 

\item {\bf Step-3}  Based on the expression (\ref{hg-2}), we insert the positive-helicity gravitons (as we have done in the ($g^-_i$, $g^-_j$) case) in turn, but fix the gauge $\xi=h_i$ in each soft factor of the form $\frac{\langle h_i,\xi \rangle \langle h_i,\eta \rangle [h_i,x]  }{\langle x , \xi \rangle \langle x ,\eta \rangle \langle y,x \rangle }$ this time, which means that every configuration with positive-helicity gravitons growing on the negative-helicity graviton $h_i$ will vanish. It turns out that the final expression of amplitude with ($h^-_i$, $g^-_j$) configuration can be obtained from the  ($g^-_i$, $g^-_j$) case via (i). replacing the Parke-Taylor factor $\frac{\langle i,j\rangle^4}{\langle1,2\rangle\langle2,3\rangle...\langle r,1\rangle}$ in (\ref{single-t-gra}) by (\ref{hg-2}) and (ii). adding the edges factors associated with positive-helicity gravitons in (\ref{single-t-gra}). Then the ($h^-_i$, $g^-_j$)-amplitude is expressed by the spanning forest formula in which all gluons and positive-helicity gravitons are nodes, and all gluons play as roots. This result precisely agrees with earlier studies in \cite{Bern:1999bx,Du:2016wkt}.

\end{itemize}
One may also try to apply an analogous discussion to the $(h_i^-,h_j^-)$-configuration, where all gluons carry positive helicity, starting from a nonvanishing three-point amplitude. Nevertheless,  such a nonvanishing starting point does not exist \cite{Benincasa:2007xk, Chen:2010ct}. This the $(h_i^-,h_j^-)$-amplitude must vanish.

\section{Constructing double-trace MHV amplitudes}
\label{Sec5}
\begin{figure}
  \centering

  \tikzset{every picture/.style={line width=0.75pt}} %set default line width to 0.75pt        

  \begin{tikzpicture}[x=0.75pt,y=0.75pt,yscale=-1,xscale=1]
  %uncomment if require: \path (0,175); %set diagram left start at 0, and has height of 175
  
  %Straight Lines [id:da47109549816252305] 
  \draw [color={rgb, 255:red, 0; green, 0; blue, 0 }  ,draw opacity=1 ] [dash pattern={on 4.5pt off 4.5pt}]  (228.1,126.15) -- (197.35,126.15) ;
  \draw [shift={(194.35,126.15)}, rotate = 360] [fill={rgb, 255:red, 0; green, 0; blue, 0 }  ,fill opacity=1 ][line width=0.08]  [draw opacity=0] (7.14,-3.43) -- (0,0) -- (7.14,3.43) -- (4.74,0) -- cycle    ;
  %Straight Lines [id:da5347664691050218] 
  \draw [color={rgb, 255:red, 0; green, 0; blue, 0 }  ,draw opacity=1 ] [dash pattern={on 0.84pt off 2.51pt}]  (228.1,126.15) -- (271.03,125.96) ;
  \draw [shift={(271.03,125.96)}, rotate = 359.75] [color={rgb, 255:red, 0; green, 0; blue, 0 }  ,draw opacity=1 ][fill={rgb, 255:red, 0; green, 0; blue, 0 }  ,fill opacity=1 ][line width=0.75]      (0, 0) circle [x radius= 1.34, y radius= 1.34]   ;
  \draw [shift={(228.1,126.15)}, rotate = 359.75] [color={rgb, 255:red, 0; green, 0; blue, 0 }  ,draw opacity=1 ][fill={rgb, 255:red, 0; green, 0; blue, 0 }  ,fill opacity=1 ][line width=0.75]      (0, 0) circle [x radius= 1.34, y radius= 1.34]   ;
  %Straight Lines [id:da29547044442058046] 
  \draw [color={rgb, 255:red, 0; green, 0; blue, 0 }  ,draw opacity=1 ] [dash pattern={on 4.5pt off 4.5pt}]  (304.79,125.96) -- (274.03,125.96) ;
  \draw [shift={(271.03,125.96)}, rotate = 360] [fill={rgb, 255:red, 0; green, 0; blue, 0 }  ,fill opacity=1 ][line width=0.08]  [draw opacity=0] (7.14,-3.43) -- (0,0) -- (7.14,3.43) -- (4.74,0) -- cycle    ;
  %Straight Lines [id:da6673138633213014] 
  \draw [color={rgb, 255:red, 0; green, 0; blue, 0 }  ,draw opacity=1 ] [dash pattern={on 0.84pt off 2.51pt}]  (305.72,125.85) ;
  \draw [shift={(305.72,125.85)}, rotate = 0] [color={rgb, 255:red, 0; green, 0; blue, 0 }  ,draw opacity=1 ][fill={rgb, 255:red, 0; green, 0; blue, 0 }  ,fill opacity=1 ][line width=0.75]      (0, 0) circle [x radius= 1.34, y radius= 1.34]   ;
  \draw [shift={(305.72,125.85)}, rotate = 0] [color={rgb, 255:red, 0; green, 0; blue, 0 }  ,draw opacity=1 ][fill={rgb, 255:red, 0; green, 0; blue, 0 }  ,fill opacity=1 ][line width=0.75]      (0, 0) circle [x radius= 1.34, y radius= 1.34]   ;
  %Straight Lines [id:da32663365588965343] 
  \draw [color={rgb, 255:red, 0; green, 0; blue, 0 }  ,draw opacity=1 ] [dash pattern={on 0.84pt off 2.51pt}]  (194.35,126.15) ;
  \draw [shift={(194.35,126.15)}, rotate = 0] [color={rgb, 255:red, 0; green, 0; blue, 0 }  ,draw opacity=1 ][fill={rgb, 255:red, 0; green, 0; blue, 0 }  ,fill opacity=1 ][line width=0.75]      (0, 0) circle [x radius= 1.34, y radius= 1.34]   ;
  \draw [shift={(194.35,126.15)}, rotate = 0] [color={rgb, 255:red, 0; green, 0; blue, 0 }  ,draw opacity=1 ][fill={rgb, 255:red, 0; green, 0; blue, 0 }  ,fill opacity=1 ][line width=0.75]      (0, 0) circle [x radius= 1.34, y radius= 1.34]   ;
  %Straight Lines [id:da8491251618460129] 
  \draw [color={rgb, 255:red, 0; green, 0; blue, 0 }  ,draw opacity=1 ] [dash pattern={on 0.84pt off 2.51pt}]  (247.97,126.36) ;
  \draw [shift={(247.97,126.36)}, rotate = 0] [color={rgb, 255:red, 0; green, 0; blue, 0 }  ,draw opacity=1 ][fill={rgb, 255:red, 0; green, 0; blue, 0 }  ,fill opacity=1 ][line width=0.75]      (0, 0) circle [x radius= 1.34, y radius= 1.34]   ;
  \draw [shift={(247.97,126.36)}, rotate = 0] [color={rgb, 255:red, 0; green, 0; blue, 0 }  ,draw opacity=1 ][fill={rgb, 255:red, 0; green, 0; blue, 0 }  ,fill opacity=1 ][line width=0.75]      (0, 0) circle [x radius= 1.34, y radius= 1.34]   ;
  %Straight Lines [id:da17156941422218375] 
  \draw [color={rgb, 255:red, 0; green, 0; blue, 0 }  ,draw opacity=1 ]   (260.09,99.36) -- (249.2,123.62) ;
  \draw [shift={(247.97,126.36)}, rotate = 294.18] [fill={rgb, 255:red, 0; green, 0; blue, 0 }  ,fill opacity=1 ][line width=0.08]  [draw opacity=0] (7.14,-3.43) -- (0,0) -- (7.14,3.43) -- (4.74,0) -- cycle    ;
  %Straight Lines [id:da6935019283489505] 
  \draw [color={rgb, 255:red, 0; green, 0; blue, 0 }  ,draw opacity=1 ] [dash pattern={on 0.84pt off 2.51pt}]  (260.09,99.36) ;
  \draw [shift={(260.09,99.36)}, rotate = 0] [color={rgb, 255:red, 0; green, 0; blue, 0 }  ,draw opacity=1 ][fill={rgb, 255:red, 0; green, 0; blue, 0 }  ,fill opacity=1 ][line width=0.75]      (0, 0) circle [x radius= 1.34, y radius= 1.34]   ;
  \draw [shift={(260.09,99.36)}, rotate = 0] [color={rgb, 255:red, 0; green, 0; blue, 0 }  ,draw opacity=1 ][fill={rgb, 255:red, 0; green, 0; blue, 0 }  ,fill opacity=1 ][line width=0.75]      (0, 0) circle [x radius= 1.34, y radius= 1.34]   ;
  %Straight Lines [id:da8810051238165255] 
  \draw [color={rgb, 255:red, 0; green, 0; blue, 0 }  ,draw opacity=1 ] [dash pattern={on 4.5pt off 4.5pt}]  (427.86,127.14) -- (397.1,127.14) ;
  \draw [shift={(394.1,127.14)}, rotate = 360] [fill={rgb, 255:red, 0; green, 0; blue, 0 }  ,fill opacity=1 ][line width=0.08]  [draw opacity=0] (7.14,-3.43) -- (0,0) -- (7.14,3.43) -- (4.74,0) -- cycle    ;
  %Straight Lines [id:da4309833640836176] 
  \draw [color={rgb, 255:red, 0; green, 0; blue, 0 }  ,draw opacity=1 ] [dash pattern={on 0.84pt off 2.51pt}]  (427.86,127.14) -- (470.79,126.95) ;
  \draw [shift={(470.79,126.95)}, rotate = 359.75] [color={rgb, 255:red, 0; green, 0; blue, 0 }  ,draw opacity=1 ][fill={rgb, 255:red, 0; green, 0; blue, 0 }  ,fill opacity=1 ][line width=0.75]      (0, 0) circle [x radius= 1.34, y radius= 1.34]   ;
  \draw [shift={(427.86,127.14)}, rotate = 359.75] [color={rgb, 255:red, 0; green, 0; blue, 0 }  ,draw opacity=1 ][fill={rgb, 255:red, 0; green, 0; blue, 0 }  ,fill opacity=1 ][line width=0.75]      (0, 0) circle [x radius= 1.34, y radius= 1.34]   ;
  %Straight Lines [id:da39777873923951] 
  \draw [color={rgb, 255:red, 0; green, 0; blue, 0 }  ,draw opacity=1 ] [dash pattern={on 4.5pt off 4.5pt}]  (505.47,126.84) -- (474.72,126.84) ;
  \draw [shift={(471.72,126.84)}, rotate = 360] [fill={rgb, 255:red, 0; green, 0; blue, 0 }  ,fill opacity=1 ][line width=0.08]  [draw opacity=0] (7.14,-3.43) -- (0,0) -- (7.14,3.43) -- (4.74,0) -- cycle    ;
  %Straight Lines [id:da5364772690639508] 
  \draw [color={rgb, 255:red, 0; green, 0; blue, 0 }  ,draw opacity=1 ] [dash pattern={on 0.84pt off 2.51pt}]  (505.47,126.84) ;
  \draw [shift={(505.47,126.84)}, rotate = 0] [color={rgb, 255:red, 0; green, 0; blue, 0 }  ,draw opacity=1 ][fill={rgb, 255:red, 0; green, 0; blue, 0 }  ,fill opacity=1 ][line width=0.75]      (0, 0) circle [x radius= 1.34, y radius= 1.34]   ;
  \draw [shift={(505.47,126.84)}, rotate = 0] [color={rgb, 255:red, 0; green, 0; blue, 0 }  ,draw opacity=1 ][fill={rgb, 255:red, 0; green, 0; blue, 0 }  ,fill opacity=1 ][line width=0.75]      (0, 0) circle [x radius= 1.34, y radius= 1.34]   ;
  %Straight Lines [id:da9315578951366397] 
  \draw [color={rgb, 255:red, 0; green, 0; blue, 0 }  ,draw opacity=1 ] [dash pattern={on 0.84pt off 2.51pt}]  (394.1,127.14) ;
  \draw [shift={(394.1,127.14)}, rotate = 0] [color={rgb, 255:red, 0; green, 0; blue, 0 }  ,draw opacity=1 ][fill={rgb, 255:red, 0; green, 0; blue, 0 }  ,fill opacity=1 ][line width=0.75]      (0, 0) circle [x radius= 1.34, y radius= 1.34]   ;
  \draw [shift={(394.1,127.14)}, rotate = 0] [color={rgb, 255:red, 0; green, 0; blue, 0 }  ,draw opacity=1 ][fill={rgb, 255:red, 0; green, 0; blue, 0 }  ,fill opacity=1 ][line width=0.75]      (0, 0) circle [x radius= 1.34, y radius= 1.34]   ;
  %Straight Lines [id:da4200099765620471] 
  \draw [color={rgb, 255:red, 0; green, 0; blue, 0 }  ,draw opacity=1 ] [dash pattern={on 0.84pt off 2.51pt}]  (447.72,127.34) ;
  \draw [shift={(447.72,127.34)}, rotate = 0] [color={rgb, 255:red, 0; green, 0; blue, 0 }  ,draw opacity=1 ][fill={rgb, 255:red, 0; green, 0; blue, 0 }  ,fill opacity=1 ][line width=0.75]      (0, 0) circle [x radius= 1.34, y radius= 1.34]   ;
  \draw [shift={(447.72,127.34)}, rotate = 0] [color={rgb, 255:red, 0; green, 0; blue, 0 }  ,draw opacity=1 ][fill={rgb, 255:red, 0; green, 0; blue, 0 }  ,fill opacity=1 ][line width=0.75]      (0, 0) circle [x radius= 1.34, y radius= 1.34]   ;
  %Straight Lines [id:da7544212419776233] 
  \draw [color={rgb, 255:red, 0; green, 0; blue, 0 }  ,draw opacity=1 ] [dash pattern={on 0.84pt off 2.51pt}]  (280.62,82.65) -- (260.09,99.36) ;
  %Straight Lines [id:da2253896057396818] 
  \draw [color={rgb, 255:red, 0; green, 0; blue, 0 }  ,draw opacity=1 ] [dash pattern={on 0.84pt off 2.51pt}]  (280.62,82.65) ;
  \draw [shift={(280.62,82.65)}, rotate = 0] [color={rgb, 255:red, 0; green, 0; blue, 0 }  ,draw opacity=1 ][fill={rgb, 255:red, 0; green, 0; blue, 0 }  ,fill opacity=1 ][line width=0.75]      (0, 0) circle [x radius= 1.34, y radius= 1.34]   ;
  \draw [shift={(280.62,82.65)}, rotate = 0] [color={rgb, 255:red, 0; green, 0; blue, 0 }  ,draw opacity=1 ][fill={rgb, 255:red, 0; green, 0; blue, 0 }  ,fill opacity=1 ][line width=0.75]      (0, 0) circle [x radius= 1.34, y radius= 1.34]   ;
  %Straight Lines [id:da5111774940180422] 
  \draw [color={rgb, 255:red, 0; green, 0; blue, 0 }  ,draw opacity=1 ]   (306.05,70.65) -- (283.33,81.37) ;
  \draw [shift={(280.62,82.65)}, rotate = 334.74] [fill={rgb, 255:red, 0; green, 0; blue, 0 }  ,fill opacity=1 ][line width=0.08]  [draw opacity=0] (7.14,-3.43) -- (0,0) -- (7.14,3.43) -- (4.74,0) -- cycle    ;
  %Straight Lines [id:da025048019982318204] 
  \draw [color={rgb, 255:red, 0; green, 0; blue, 0 }  ,draw opacity=1 ] [dash pattern={on 0.84pt off 2.51pt}]  (306.05,70.65) ;
  \draw [shift={(306.05,70.65)}, rotate = 0] [color={rgb, 255:red, 0; green, 0; blue, 0 }  ,draw opacity=1 ][fill={rgb, 255:red, 0; green, 0; blue, 0 }  ,fill opacity=1 ][line width=0.75]      (0, 0) circle [x radius= 1.34, y radius= 1.34]   ;
  \draw [shift={(306.05,70.65)}, rotate = 0] [color={rgb, 255:red, 0; green, 0; blue, 0 }  ,draw opacity=1 ][fill={rgb, 255:red, 0; green, 0; blue, 0 }  ,fill opacity=1 ][line width=0.75]      (0, 0) circle [x radius= 1.34, y radius= 1.34]   ;
  %Straight Lines [id:da6811980683190562] 
  \draw [color={rgb, 255:red, 0; green, 0; blue, 0 }  ,draw opacity=1 ]   (434.02,99.79) -- (446.39,124.66) ;
  \draw [shift={(447.72,127.34)}, rotate = 243.56] [fill={rgb, 255:red, 0; green, 0; blue, 0 }  ,fill opacity=1 ][line width=0.08]  [draw opacity=0] (7.14,-3.43) -- (0,0) -- (7.14,3.43) -- (4.74,0) -- cycle    ;
  %Straight Lines [id:da32267012472109347] 
  \draw [color={rgb, 255:red, 0; green, 0; blue, 0 }  ,draw opacity=1 ] [dash pattern={on 0.84pt off 2.51pt}]  (434.02,99.79) ;
  \draw [shift={(434.02,99.79)}, rotate = 0] [color={rgb, 255:red, 0; green, 0; blue, 0 }  ,draw opacity=1 ][fill={rgb, 255:red, 0; green, 0; blue, 0 }  ,fill opacity=1 ][line width=0.75]      (0, 0) circle [x radius= 1.34, y radius= 1.34]   ;
  \draw [shift={(434.02,99.79)}, rotate = 0] [color={rgb, 255:red, 0; green, 0; blue, 0 }  ,draw opacity=1 ][fill={rgb, 255:red, 0; green, 0; blue, 0 }  ,fill opacity=1 ][line width=0.75]      (0, 0) circle [x radius= 1.34, y radius= 1.34]   ;
  %Straight Lines [id:da8537335895590181] 
  \draw [color={rgb, 255:red, 0; green, 0; blue, 0 }  ,draw opacity=1 ] [dash pattern={on 0.84pt off 2.51pt}]  (407.45,82.36) -- (434.02,99.79) ;
  %Straight Lines [id:da19884460189758646] 
  \draw [color={rgb, 255:red, 0; green, 0; blue, 0 }  ,draw opacity=1 ] [dash pattern={on 0.84pt off 2.51pt}]  (407.45,82.36) ;
  \draw [shift={(407.45,82.36)}, rotate = 0] [color={rgb, 255:red, 0; green, 0; blue, 0 }  ,draw opacity=1 ][fill={rgb, 255:red, 0; green, 0; blue, 0 }  ,fill opacity=1 ][line width=0.75]      (0, 0) circle [x radius= 1.34, y radius= 1.34]   ;
  \draw [shift={(407.45,82.36)}, rotate = 0] [color={rgb, 255:red, 0; green, 0; blue, 0 }  ,draw opacity=1 ][fill={rgb, 255:red, 0; green, 0; blue, 0 }  ,fill opacity=1 ][line width=0.75]      (0, 0) circle [x radius= 1.34, y radius= 1.34]   ;
  %Straight Lines [id:da7147349858166712] 
  \draw [color={rgb, 255:red, 0; green, 0; blue, 0 }  ,draw opacity=1 ]   (382.63,70.34) -- (405.33,81.09) ;
  \draw [shift={(408.04,82.38)}, rotate = 205.35] [fill={rgb, 255:red, 0; green, 0; blue, 0 }  ,fill opacity=1 ][line width=0.08]  [draw opacity=0] (7.14,-3.43) -- (0,0) -- (7.14,3.43) -- (4.74,0) -- cycle    ;
  %Straight Lines [id:da9193694317446872] 
  \draw [color={rgb, 255:red, 0; green, 0; blue, 0 }  ,draw opacity=1 ] [dash pattern={on 0.84pt off 2.51pt}]  (382.63,70.34) ;
  \draw [shift={(382.63,70.34)}, rotate = 0] [color={rgb, 255:red, 0; green, 0; blue, 0 }  ,draw opacity=1 ][fill={rgb, 255:red, 0; green, 0; blue, 0 }  ,fill opacity=1 ][line width=0.75]      (0, 0) circle [x radius= 1.34, y radius= 1.34]   ;
  \draw [shift={(382.63,70.34)}, rotate = 0] [color={rgb, 255:red, 0; green, 0; blue, 0 }  ,draw opacity=1 ][fill={rgb, 255:red, 0; green, 0; blue, 0 }  ,fill opacity=1 ][line width=0.75]      (0, 0) circle [x radius= 1.34, y radius= 1.34]   ;
  %Straight Lines [id:da5431845048921826] 
  \draw [color={rgb, 255:red, 0; green, 0; blue, 0 }  ,draw opacity=1 ]   (309.05,70.63) -- (379.63,70.35) ;
  \draw [shift={(382.63,70.34)}, rotate = 179.77] [fill={rgb, 255:red, 0; green, 0; blue, 0 }  ,fill opacity=1 ][line width=0.08]  [draw opacity=0] (7.14,-3.43) -- (0,0) -- (7.14,3.43) -- (4.74,0) -- cycle    ;
  \draw [shift={(306.05,70.65)}, rotate = 359.77] [fill={rgb, 255:red, 0; green, 0; blue, 0 }  ,fill opacity=1 ][line width=0.08]  [draw opacity=0] (7.14,-3.43) -- (0,0) -- (7.14,3.43) -- (4.74,0) -- cycle    ;
  %Straight Lines [id:da620034735615216] 
  \draw [color={rgb, 255:red, 0; green, 0; blue, 0 }  ,draw opacity=1 ]   (214.76,101.01) -- (227.46,123.65) ;
  \draw [shift={(228.93,126.27)}, rotate = 240.7] [fill={rgb, 255:red, 0; green, 0; blue, 0 }  ,fill opacity=1 ][line width=0.08]  [draw opacity=0] (7.14,-3.43) -- (0,0) -- (7.14,3.43) -- (4.74,0) -- cycle    ;
  %Straight Lines [id:da9223393490482599] 
  \draw [color={rgb, 255:red, 0; green, 0; blue, 0 }  ,draw opacity=1 ] [dash pattern={on 0.84pt off 2.51pt}]  (214.88,101.15) ;
  \draw [shift={(214.88,101.15)}, rotate = 0] [color={rgb, 255:red, 0; green, 0; blue, 0 }  ,draw opacity=1 ][fill={rgb, 255:red, 0; green, 0; blue, 0 }  ,fill opacity=1 ][line width=0.75]      (0, 0) circle [x radius= 1.34, y radius= 1.34]   ;
  \draw [shift={(214.88,101.15)}, rotate = 0] [color={rgb, 255:red, 0; green, 0; blue, 0 }  ,draw opacity=1 ][fill={rgb, 255:red, 0; green, 0; blue, 0 }  ,fill opacity=1 ][line width=0.75]      (0, 0) circle [x radius= 1.34, y radius= 1.34]   ;
  %Straight Lines [id:da41816140415982894] 
  \draw [color={rgb, 255:red, 0; green, 0; blue, 0 }  ,draw opacity=1 ]   (239.18,99.87) -- (229.27,123.39) ;
  \draw [shift={(228.1,126.15)}, rotate = 292.85] [fill={rgb, 255:red, 0; green, 0; blue, 0 }  ,fill opacity=1 ][line width=0.08]  [draw opacity=0] (7.14,-3.43) -- (0,0) -- (7.14,3.43) -- (4.74,0) -- cycle    ;
  %Straight Lines [id:da9305031789262517] 
  \draw [color={rgb, 255:red, 0; green, 0; blue, 0 }  ,draw opacity=1 ] [dash pattern={on 0.84pt off 2.51pt}]  (239.18,99.87) ;
  \draw [shift={(239.18,99.87)}, rotate = 0] [color={rgb, 255:red, 0; green, 0; blue, 0 }  ,draw opacity=1 ][fill={rgb, 255:red, 0; green, 0; blue, 0 }  ,fill opacity=1 ][line width=0.75]      (0, 0) circle [x radius= 1.34, y radius= 1.34]   ;
  \draw [shift={(239.18,99.87)}, rotate = 0] [color={rgb, 255:red, 0; green, 0; blue, 0 }  ,draw opacity=1 ][fill={rgb, 255:red, 0; green, 0; blue, 0 }  ,fill opacity=1 ][line width=0.75]      (0, 0) circle [x radius= 1.34, y radius= 1.34]   ;
  %Curve Lines [id:da3891982913990655] 
  \draw [color={rgb, 255:red, 0; green, 0; blue, 0 }  ,draw opacity=1 ] [dash pattern={on 0.84pt off 2.51pt}]  (219.93,98.56) .. controls (225.9,91.01) and (233.45,94.43) .. (235.63,100.67) ;
  %Straight Lines [id:da5466222132091556] 
  \draw [color={rgb, 255:red, 0; green, 0; blue, 0 }  ,draw opacity=1 ]   (456.91,101.81) -- (469.61,124.45) ;
  \draw [shift={(471.08,127.07)}, rotate = 240.7] [fill={rgb, 255:red, 0; green, 0; blue, 0 }  ,fill opacity=1 ][line width=0.08]  [draw opacity=0] (7.14,-3.43) -- (0,0) -- (7.14,3.43) -- (4.74,0) -- cycle    ;
  %Straight Lines [id:da43988741389275754] 
  \draw [color={rgb, 255:red, 0; green, 0; blue, 0 }  ,draw opacity=1 ] [dash pattern={on 0.84pt off 2.51pt}]  (457.03,101.95) ;
  \draw [shift={(457.03,101.95)}, rotate = 0] [color={rgb, 255:red, 0; green, 0; blue, 0 }  ,draw opacity=1 ][fill={rgb, 255:red, 0; green, 0; blue, 0 }  ,fill opacity=1 ][line width=0.75]      (0, 0) circle [x radius= 1.34, y radius= 1.34]   ;
  \draw [shift={(457.03,101.95)}, rotate = 0] [color={rgb, 255:red, 0; green, 0; blue, 0 }  ,draw opacity=1 ][fill={rgb, 255:red, 0; green, 0; blue, 0 }  ,fill opacity=1 ][line width=0.75]      (0, 0) circle [x radius= 1.34, y radius= 1.34]   ;
  %Straight Lines [id:da6003766519737448] 
  \draw [color={rgb, 255:red, 0; green, 0; blue, 0 }  ,draw opacity=1 ]   (481.33,100.67) -- (471.42,124.19) ;
  \draw [shift={(470.25,126.95)}, rotate = 292.85] [fill={rgb, 255:red, 0; green, 0; blue, 0 }  ,fill opacity=1 ][line width=0.08]  [draw opacity=0] (7.14,-3.43) -- (0,0) -- (7.14,3.43) -- (4.74,0) -- cycle    ;
  %Straight Lines [id:da7186018670181573] 
  \draw [color={rgb, 255:red, 0; green, 0; blue, 0 }  ,draw opacity=1 ] [dash pattern={on 0.84pt off 2.51pt}]  (481.33,100.67) ;
  \draw [shift={(481.33,100.67)}, rotate = 0] [color={rgb, 255:red, 0; green, 0; blue, 0 }  ,draw opacity=1 ][fill={rgb, 255:red, 0; green, 0; blue, 0 }  ,fill opacity=1 ][line width=0.75]      (0, 0) circle [x radius= 1.34, y radius= 1.34]   ;
  \draw [shift={(481.33,100.67)}, rotate = 0] [color={rgb, 255:red, 0; green, 0; blue, 0 }  ,draw opacity=1 ][fill={rgb, 255:red, 0; green, 0; blue, 0 }  ,fill opacity=1 ][line width=0.75]      (0, 0) circle [x radius= 1.34, y radius= 1.34]   ;
  %Curve Lines [id:da41757086135108046] 
  \draw [color={rgb, 255:red, 0; green, 0; blue, 0 }  ,draw opacity=1 ] [dash pattern={on 0.84pt off 2.51pt}]  (462.08,99.36) .. controls (468.05,91.81) and (475.6,95.23) .. (477.78,101.47) ;
  
% Text Node
\draw (188.34,133.5) node [anchor=north west][inner sep=0.75pt]   [align=left] {$x_1$};
% Text Node
\draw (241.94,133.5) node [anchor=north west][inner sep=0.75pt]   [align=left] {$x_a$};
% Text Node
\draw (237.93,67.98) node [anchor=north west][inner sep=0.75pt]   [align=left] {$\mathscr{G}_1$};
% Text Node
\draw (433.93,67.18) node [anchor=north west][inner sep=0.75pt]   [align=left] {$\mathscr{G}_2$};
% Text Node
\draw (299.72,133.5) node [anchor=north west][inner sep=0.75pt]   [align=left] {$x_r$};
% Text Node
\draw (387.09,134.5) node [anchor=north west][inner sep=0.75pt]   [align=left] {$y_1$};
% Text Node
\draw (440.69,134.5) node [anchor=north west][inner sep=0.75pt]   [align=left] {$y_b$};
% Text Node
\draw (499.47,134.5) node [anchor=north west][inner sep=0.75pt]   [align=left] {$y_s$};
% Text Node
\draw (300.15,55) node [anchor=north west][inner sep=0.75pt]   [align=left] {$a$};
% Text Node
\draw (377.65,55) node [anchor=north west][inner sep=0.75pt]   [align=left] {$b$};
% Text Node
\draw (339.09,77.97) node [anchor=north west][inner sep=0.75pt]   [align=left] {$\mathcal{B}$};

  \end{tikzpicture}
  
\caption{A typical graph for a double-trace MHV amplitude $A^{(g^-_i,g^-_j)}(x_1,\dots, x_r|y_1,\dots,y_s||\textsc{H})$, where a bridge inside two forests connects two traces.}
\label{double-t-fig1}
\end{figure}
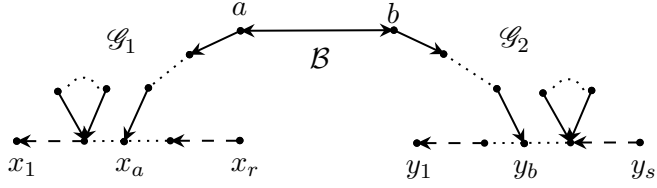
In this section, we turn to double-trace MHV amplitude with $(g_i^-,g_j^-)$-configuration, which was shown to have the following form \cite{Tian:2021dzf}:
\bea
A(x_1,\dots,x_r|y_1,\dots,y_s||\textsc{H})
=\frac{\langle g_i,g_j\rangle^4}{(x_1,\dots,x_r)(y_1,\dots,y_s)}\left[\sum\limits_{\mathscr{G}=\mathscr{G}_1\oplus\mathscr{G}_2}\sum\limits_{\substack{a\in\mathscr{G}_1\\b\in\mathscr{G}_2}}\mathcal{B}(a,b)\prod\limits_{e(x,y)\in\mathcal{E}(\mathscr{G})}\mathcal{G}(x,y,\xi)\right].\nn
\label{double-t-graph}
\eea
In the above expression, the two gluon traces correspond to Parke-Taylor factor $\frac{\langle g_i,g_j\rangle^4}{(x_1,\dots,x_r)(y_1,\dots,y_s)}$ and we use $(x_1,\dots,x_r)$ to denote $\Spaa{x_1,x_2}\dots\Spaa{x_r,x_1}$. Contents in the square brackets are explained as follows:
\begin{itemize}
\item $\mathscr{G}=\mathscr{G}_1\oplus\mathscr{G}_2$ denotes spanning forests as shown by \figref{double-t-fig1}, where gluons in both traces are considered as roots.
\item For a given $\mathscr{G}=\mathscr{G}_1\oplus\mathscr{G}_2$, we pick out two nodes $a\in \mathscr{G}_1$ and $b\in\mathscr{G}_2$ and connect these nodes by a line with two arrows pointing to opposite directions, as shown by \figref{double-t-fig1}. This line is associated with the factor 
\begin{equation}
    \mathcal{B}(a,b)=\left( - k _ { a } \cdot k _ { b } \right) \frac { \langle a , \zeta \rangle \langle b , \chi \rangle } { \langle a , b \rangle \langle \zeta , \chi \rangle }
    \xrightarrow[\text{constant factor}]{\text{absorbing}} [a,b]\frac { \langle a, \zeta \rangle \langle b , \chi \rangle } {  \langle \zeta , \chi \rangle }.\label{bridge-factor}
\end{equation}
With this edge, we have a path between the two traces, called the bridge.
\item Each edge in $\mathscr{G}=\mathscr{G}_1\oplus\mathscr{G}_2$ is accompanied by a factor
\begin{equation}
 \mathcal{G}(x,y)=\frac { \langle y ,\xi  \rangle \left\langle y ,\lambda _ { e } \right\rangle [ x , y ] } 
 { \langle x , \xi \rangle \left\langle x , \lambda _ { e } \right\rangle \langle x , y \rangle },
\end{equation}
where the reference spinor $\lambda_e\rangle$ can be chosen separately according to distinct positions of the forests in the graph
\begin{equation}\label{lambda}
 \lambda _ { e }\rangle = \left\{ 
\begin{array} 
    { l l } \eta\rangle & ( \text { if } e ( x , y ) \text { is an outer edge } ) \\ 
    \zeta\rangle & \left( \text { if } e ( x , y ) \text { is spanned on bridge and } e ( x , y ) \in \mathscr { G } _ { 1 } \right) \\ 
    \chi\rangle & \left( \text { if } e ( x , y ) \text { is spanned on bridge and } e ( x , y ) \in \mathscr { G } _ { 2 } \right) .
  \end{array}\right.
\end{equation}
\end{itemize}
Summing over all possible $\mathscr{G}=\mathscr{G}_1\oplus\mathscr{G}_2$ and all possible choices of $a,b$, we get the expression inside the square brackets. In the following subsections, we reconstruct \eqref{double-t-graph} by ISL.

\subsection{Double-trace MHV amplitudes with one graviton}

\begin{figure}
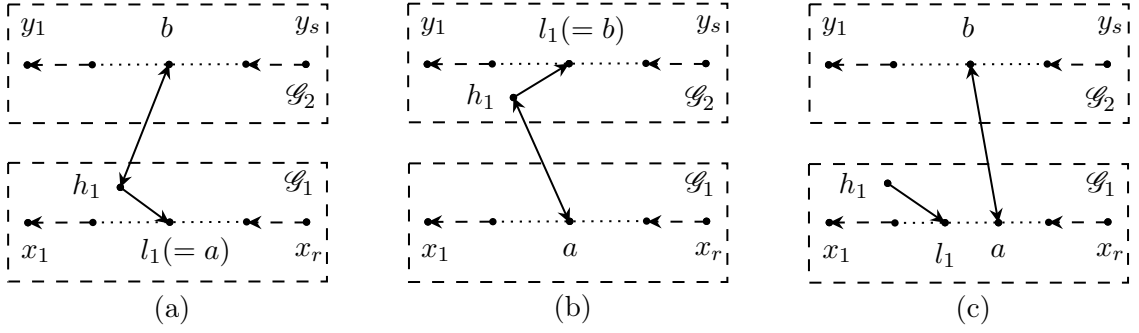

  \centering

  \tikzset{every picture/.style={line width=0.75pt}} %set default line width to 0.75pt        

  %
  
\caption{ Graphs for double-trace amplitudes with one graviton. The double arrow lines represent the $k\cdot k$ structure. In graphs (a) and (b), $l_1=a \in T_1$, $l_1= b \in T_2$, respectively, and $h_1$ belongs to the bridge. In graph (c), $l_1$ can be a gluon in both trace $T_1$ and $T_2$.}
%\label{double 1grav}
\label{double-t-1graviton}
\end{figure}

Double-trace amplitude with one graviton requires contributions from the three types of graphs in \figref{double-t-1graviton}, which correspond to (a). $h_1\in \mathscr{G}_1$, $h_1$ lives on the bridge between the two traces, (b). $h_1\in \mathscr{G}_2$ and $h_1$ lives on the bridge, (c). $h_1$ root at a gluon and it does not belong to the bridge. Terms corresponding to these graphs are given by 
\begin{equation}
\begin{aligned}
\text{\figref{double-t-1graviton} (a)}&\to\quad   A_1&=&    \sum_{a\in T_1,b\in T_2}\frac{\langle a,\xi\rangle[a,h_1]}{\langle h_1,\xi\rangle\langle a,h_1\rangle }[h_1,b]\frac{\langle a,\zeta\rangle\langle b,\chi\rangle}{\langle\zeta,\chi\rangle}\,\frac{\langle i,j\rangle^4}{(x_1,\dots,x_r)(y_1,\dots,y_s)},\\
\text{\figref{double-t-1graviton} (b)}&\to\quad    A_2&=&   \sum_{a\in T_1,b\in T_2}\frac{\langle b,\xi\rangle[b,h_1]}{\langle h_1,\xi\rangle\langle b,h_1\rangle }[a,h_1]\frac{\langle a,\zeta\rangle\langle b,\chi\rangle}{\langle\zeta,\chi\rangle}\,\frac{\langle i,j\rangle^4}{(x_1,\dots,x_r)(y_1,\dots,y_s)},\\
\text{\figref{double-t-1graviton} (c)}&\to\quad    A_3&=&\sum_{l_1\in T_1\cup T_2}\mathcal{G}(h_1,l_1)\,\sum_{a\in T_1,b\in T_2}\mathcal{B}(a,b)\,\frac{\langle i,j\rangle^4}{(x_1,\dots,x_r)(y_1,\dots,y_s)},
\end{aligned}
\label{double-1-graph}
\end{equation}
where $T_1$ and $T_2$ represent the two gluon traces respectively. To reconstruct the double-trace amplitude with one graviton, we write down the ISL formula
\begin{equation}
    A_{\text{MHV}}\left(  x _ { 1 },\dots, x _ { r } | y _ { 1 }, \dots, y _ { s } || h _ { 1 } \right) = \sum_{l\in T_1\cup T_2}\mathcal{G}(h_1,l_1)\,\left[\mathcal{B}(a,b)\,\textsc{PT}\right]'\big|_{l'},
\label{double-1-ISL}
\end{equation}

\begin{figure}
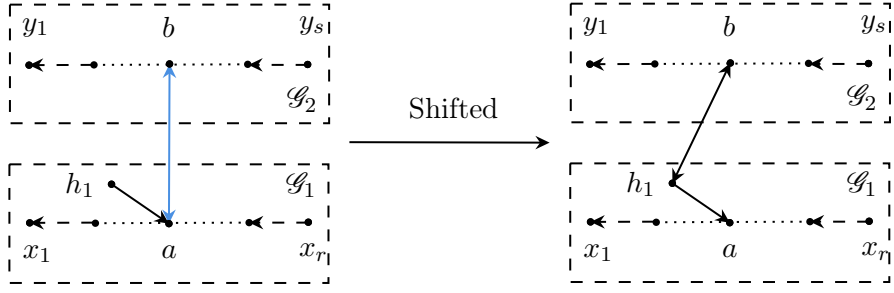

  \centering
\tikzset{every picture/.style={line width=0.75pt}} %set default line width to 0.75pt        



  \caption{The contribution from the term with shifted part of $\mathcal{B}(a,b)$  in \eqref{double-1-ISL}.}
  \label{double-1graviton-rule} 
\end{figure}

Now the $\text{PT}$ denotes the factor $\frac{\langle i,j\rangle^4}{(x_1,\dots,x_r)(y_1,\dots,y_s)}$ in the double-trace discussion. The expression inside the square brackets is just the double-trace amplitude with no external graviton. When the explicit expression of $\mathcal{G}(h_1,l_1)$ and the shifted spinor  $l'_1]$ are inserted, one gets
\bea
   (\ref{double-1-ISL}) &=&\sum_{l_1\in T_1 \cup T_2\setminus\left\{a,b\right\}}\underbrace{\mathcal{G}(h_1,l_1)\, \mathcal{B}(a,b)\,\textsc{PT}}_{{I_1}}\nn
&&+\quad\sum_{a\in T_1,b\in T_2}\frac{\langle a,\xi\rangle\langle a,\eta\rangle[a,h_1]}{\langle h_1,\xi\rangle\langle h_1,\eta\rangle\langle a,h_1\rangle}\left\{\underbrace{[a,b]\frac{\langle a,\zeta\rangle\langle b,\chi\rangle}{\langle \zeta,\chi\rangle}}_{I_{2A}}+\underbrace{\frac{\langle h_1,\xi\rangle}{\langle a,\xi\rangle}[h_1,b]\frac{\langle a,\zeta\rangle\langle b,\chi\rangle}{\langle \zeta,\chi\rangle}}_{I_{2B}}\right\}\,\textsc{PT}\nn
&&+\quad\sum_{a\in T_1,b\in T_2}\frac{\langle b,\xi\rangle\langle b,\eta\rangle[b,h_1]}{\langle h_1,\xi\rangle\langle h_1,\eta\rangle\langle b,h_1\rangle}\left\{\underbrace{[a,b]\frac{\langle a,\zeta\rangle\langle b,\chi\rangle}{\langle \zeta,\chi\rangle}}_{I_{3A}}+\underbrace{\frac{\langle h_1,\xi\rangle}{\langle b,\xi\rangle}[a,h_1]\frac{\langle a,\zeta\rangle\langle b,\chi\rangle}{\langle \zeta,\chi\rangle}}_{I_{3B}}\right\}\,\textsc{PT}.
\label{double-1-split}
\eea
In the above expression, the total contribution of $I_{2B}$ is given by $\sum\limits_{a\in T_1,b\in T_2}\frac{\langle a,\eta\rangle[a,h_1]}{\langle h_1,\eta\rangle\langle a,h_1\rangle}[h_1,b]\frac{\langle a,\zeta\rangle\langle b,\chi\rangle}{\langle \zeta,\chi\rangle}\,\text{PT}$ which is the same with $A_1$ in \eqref{double-1-graph}, up to a replacement of the reference spinors $\eta\rangle\to \xi\rangle$. One can verify that this term precisely equals $A_1$. Similarly, the total contribution of $I_{3B}$ matches with $A_2$ in \eqref{double-1-graph}. This process is shown by \figref{double-1graviton-rule}.

All the remaining terms $I_1$, $I_{2A}$ and $I_{3A}$ together reproduce the $A_3$ in \eqref{double-1-graph}.
 Now we can give the following summary:
\begin{equation*}
\begin{aligned}
&I_{2B}&\to&\quad A_1&=&\text{\figref{double-t-1graviton} (a)},\\
&I_{3B}&\to&\quad A_2&=&\text{\figref{double-t-1graviton} (b)},\\
I_1+&I_{2A}+I_{3A}&\to&\quad A_3&=&\text{\figref{double-t-1graviton} (c)},
\end{aligned}
\end{equation*}
which are precisely all contributions of the spanning forests for the double-trace MHV amplitude with one graviton.

\subsection{Double-trace MHV amplitudes with two gravitons}

\begin{figure}
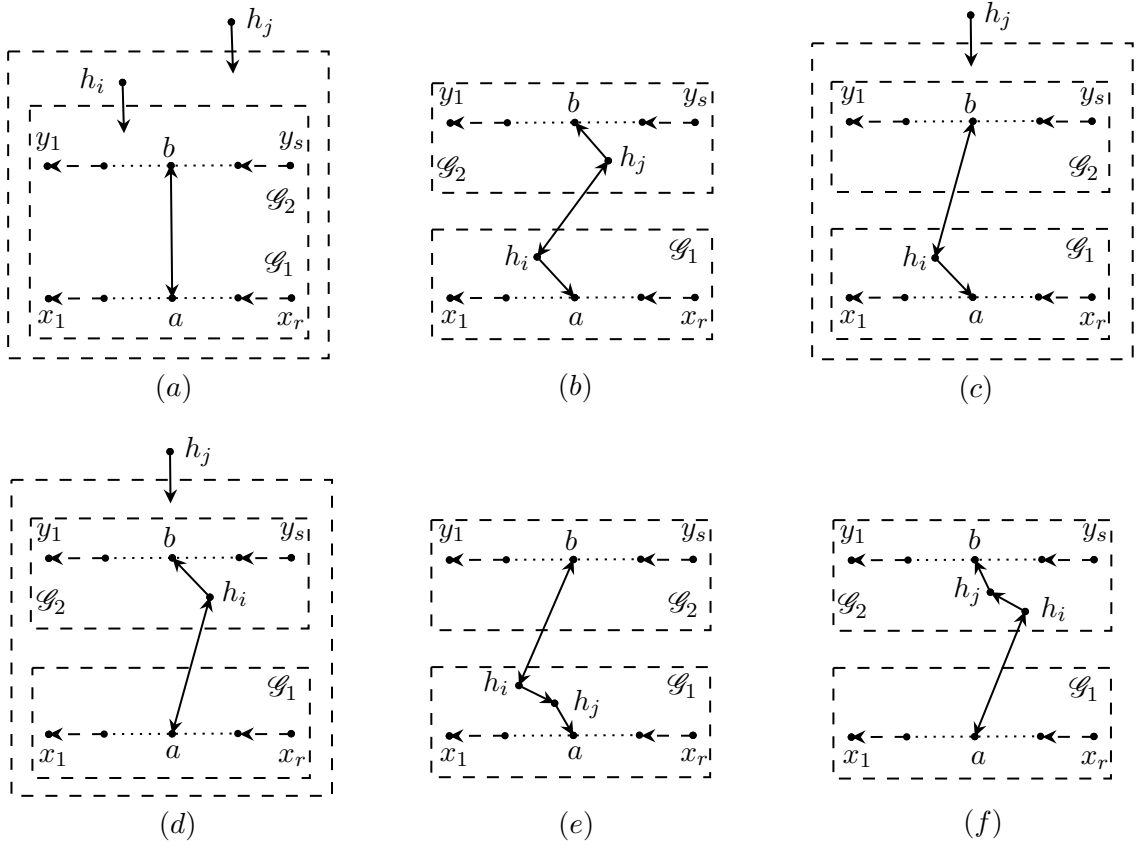

  \centering

  \tikzset{every picture/.style={line width=0.75pt}} %set default line width to 0.75pt        


  
\caption{ Graphs for double trace MHV amplitude with two gravitons. In graph (a), $h_i$ is inserted before $h_j$, thus $h_j$ can root at two traces and $h_i$. In graph (b), $h_i$ and $h_j$ root at $\mathscr{G}_1$ and $\mathscr{G}_2$ respectively, and both of them belong to the bridge structure. In graphs (c) and (d), $h_j$ can root at all nodes in two traces and $h_i$. As for graph (e) (or (f)), both $h_i$ and $h_j$ live on the bridge and they both belong to $\mathscr{G}_1$ (or $\mathscr{G}_2$). For all of the graphs above, $i$, $j$ can be $1$, $2$ or $2$, $1$. }
\label{double-t-2graviton}
\end{figure}

Double-trace MHV amplitudes with two gravitons get contributions from the graphs in \figref{double-t-2graviton}, where each graviton can either be inserted on the bridge between the two traces or not. The explicit expressions of these graphs are correspondingly displayed as:
\begin{equation}
 \begin{aligned}
&\text{\figref{double-t-2graviton} (a)}&\to &\sum\limits_{l_j\in T\cup\{h_i\}}\mathcal{G}(h_j,l_j)\sum\limits_{l_i\in T}\mathcal{G}(h_i,l_i)\sum\limits_{a\in T_1,b\in T_2}\mathcal{B}(a,b)\, \textsc{PT},\\
&\text{\figref{double-t-2graviton} (b)}&\to &\sum\limits_{a\in T_1,b\in T_2}\mathcal{G}(h_i,a)\mathcal{G}(h_j,b)\,\mathcal{B}(h_i,h_j)\, \textsc{PT},\\
&\text{\figref{double-t-2graviton} (c)}&\to &\sum\limits_{l_j\in T\cup\{h_i\}}\mathcal{G}(h_j,l_j)\sum\limits_{a\in T_1,b\in T_2}\mathcal{G}(h_i,a)\mathcal{B}(h_i,b)\, \textsc{PT},\\
&\text{\figref{double-t-2graviton} (d)}&\to & \sum\limits_{l_j\in T\cup\{h_i\}}\mathcal{G}(h_j,l_j)\sum\limits_{a\in T_1,b\in T_2}\mathcal{G}(hi,b)\mathcal{B}(a,h_i)\, \textsc{PT},\\
&\text{\figref{double-t-2graviton} (e)}&\to & \sum\limits_{a\in T_1,b\in T_2}\mathcal{G}(h_i,h_j)\mathcal{G}(h_j,a)\mathcal{B}(h_i,b)\, \textsc{PT},\\
&\text{\figref{double-t-2graviton} (f)}&\to & \sum\limits_{a\in T_1,b\in T_2}\mathcal{G}(h_i,h_j)\mathcal{G}(h_j,b)\mathcal{B}(a,h_i)\, \textsc{PT}.
\end{aligned}
\label{double-t-2gra-graphs}
\end{equation}

\begin{figure}
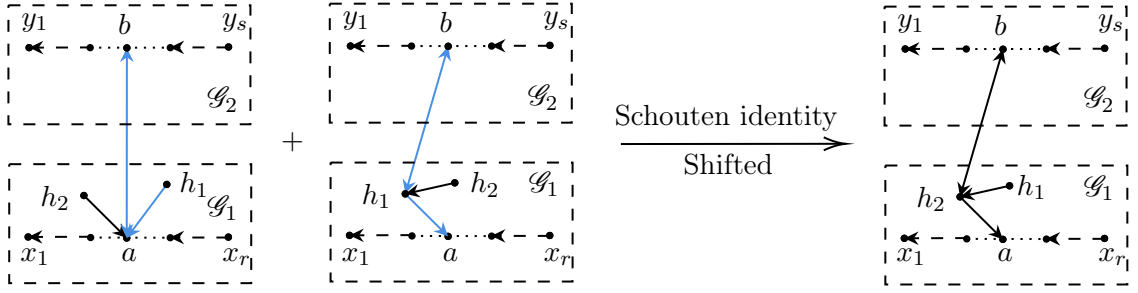

  \centering

  \tikzset{every picture/.style={line width=0.75pt}} %set default line width to 0.75pt        


    
  \caption{ Summation of terms with two shifted factors: the first and the second graphs on the LHS is accompanied by factors $\mathcal{G}'(h_1,a')\mathcal{B}'(a',b)$ and $\mathcal{G}'(h'_1,a)\mathcal{B}'(h'_1,b)$, respectively.}
  \label{double-trace-2shift-rule}
  \end{figure}
We expect that they can be reconstructed through the ISL below:
\begin{equation}
   A_{\text{MHV}}\left(  x _ { 1 },\dots, x _ { r } | y _ { 1 }, \dots, y _ { s } || h _ { 1 } ,h_2\right) = \sum_{l_2\in T\cup\{h_1\}}\mathcal{G}(h_2, l_2)A'(x_1,\dots,x_r|y_1,\dots,y_s||h_1)\Big|_{l'_2},
\label{double-t-2gra-ISL}
\end{equation}
where the amplitude with one graviton in \eqref{double-t-2gra-ISL} is expressed by graphs in \figref{double-t-1graviton}. The expansion of \eqref{double-t-2gra-ISL} contains plenty of terms that can be classified into three categories corresponding to distinct powers of the shifted factor: (i). unshifted terms; (ii). terms with one shifted factor ($l'_1]$, $h'_1]$, $a']$ or $b']$); (iii). terms with two shifted factors ($a']=h'_1]$, $a']=l'_1]$ and $b']=h'_1]$, $b']=l'_1]$). As the first two cases have already been discussed in previous subsections, we list several credible conclusions directly.
\begin{itemize}
\item Terms in category (i) contribute to part of the graph (a), part of the graph (c) with $h_j=h_2$, $h_i=h_1$ and part of the graph (d) with $h_j=h_2$, $h_i=h_1$ in \figref{double-t-2graviton}.
\item Terms in category (ii) cover the contributions of the rest part of the graph (a) (where $h_2$ is the root of $h_1$), the whole graphs (b), (e), (f), and part of graphs (c), (d) with $h_j=h_1$, $h_i=h_2$, $l_1\neq h_2$.
\end{itemize}
Terms in category (iii) get contribution from graphs shown by the LHS of \figref{double-trace-2shift-rule}, which have the following expressions
\begin{equation}
\sum_{a\in T_1,b\in T_2}\mathcal{G}(h_2,a)\mathcal{G}'(h_1,a')\mathcal{B}'(a',b)\, \textsc{PT}=\frac{[a,h_2]}{\langle a,h_2\rangle}\frac{[h_2,h_1]}{\langle a,h_1\rangle}\frac{\langle a,\xi\rangle^2}{\langle h_1,\xi\rangle\langle h_1,\eta\rangle\langle h_2,\xi\rangle}\tilde{\mathcal{B}}(h_2,b),
\label{d-t-rule2-lhs1}
\end{equation}
\begin{equation}
\mathcal{G}(h_2,h_1)\sum_{a\in T_1,b\in T_2}\mathcal{G}'(h'_1,a)\mathcal{B}'(h'_1,b)\, \textsc{PT}=\frac{[a,h_2]}{\langle a,h_1\rangle}\frac{[h_2,h_1]}{\langle h_2,h_1\rangle}\frac{\langle a,\xi\rangle\langle h_2,\eta\rangle}{\langle h_1,\eta\rangle\langle h_2,\xi\rangle}\tilde{\mathcal{B}}(h_2,b),
\label{d-t-rule2-lhs2}
\end{equation}
where we define $\tilde{\mathcal{B}}(i,j)\equiv[i,j]\frac{\langle a,\zeta\rangle\langle b,\chi\rangle}{\langle\zeta,\chi\rangle}\text{PT}$ for convenience. summation of these two graphs produces the RHS of \figref{double-trace-2shift-rule} which are explicitly displayed as 
\begin{equation}
\begin{aligned}
\mathcal{G}(h_1,h_2)\sum_{a\in T_1,b\in T_2}\mathcal{G}(h_2,a)\mathcal{B}(h_2,b)\, \textsc{PT}
&=\frac{[a,h_2]}{\langle a,h_2\rangle}\frac{[h_2,h_1]}{\langle h_2,h_1\rangle}\frac{\langle a,\xi\rangle\langle h_2,\eta\rangle}{\langle h_1,\xi\rangle\langle h_1,\eta\rangle}[h_2,b] \frac{\langle a,\zeta\rangle\langle b,\chi\rangle}{\langle\zeta,\chi\rangle}\text{PT}\\
&=\frac{[a,h_2]}{\langle a,h_2\rangle}\frac{[h_2,h_1]}{\langle h_2,h_1\rangle}\frac{\langle a,\xi\rangle\langle h_2,\eta\rangle}{\langle h_1,\xi\rangle\langle h_1,\eta\rangle}\tilde{\mathcal{B}}(h_2,b).
\end{aligned}
\label{d-t-rule2-rhs}
\end{equation}
It's not hard to check that:
\begin{equation}
(\ref{d-t-rule2-lhs1})+(\ref{d-t-rule2-lhs2})=(\ref{d-t-rule2-rhs})\times\left(\frac{\langle h_2,h_1\rangle\langle a,\xi\rangle}{\langle a,h_1\rangle\langle h_2,\xi\rangle}+\frac{\langle a,h_2\rangle\langle h_1,\xi\rangle}{\langle a,h_1\rangle\langle h_2,\xi\rangle}\right)=(\ref{d-t-rule2-rhs}).
\label{double-t-id-2}
\end{equation}
where we have extracted the common factor here and applied the Schouten identity (\ref{Sc-id}). The case that $h_1$ root at $h_2 \in \mathscr{G}_2$ which belongs to the bridge follows from a similar discussion. Therefore, all the spanning forests with two gravitons can be reproduced through ISL.

\subsection{Terms with more shifted factors}

\begin{figure}
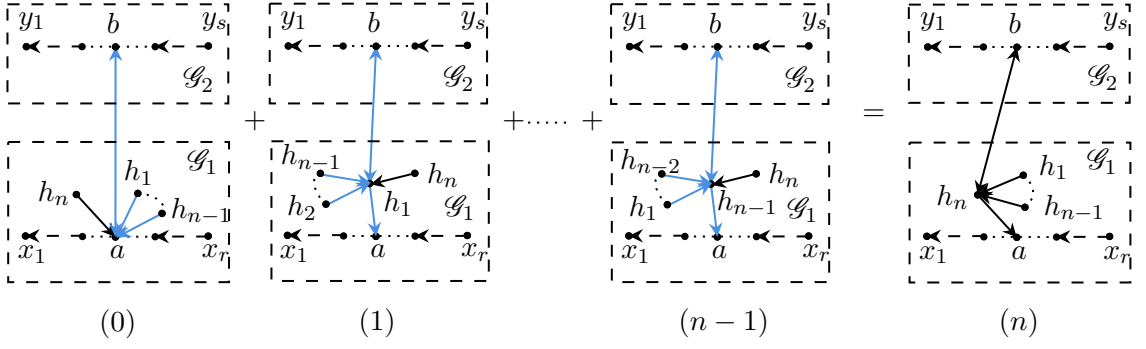

\centering

\tikzset{every picture/.style={line width=0.75pt}} %set default line width to 0.75pt        



\caption{ Summation of terms with $n$ shifted factors: for graphs on the LHS, the first one takes the shift $a']=l'_n]$ and the rest take $h'_i]=l'_n]$ ($i=1,\dots, n-1$).}
\label{double-trace-nshift-rule}
\end{figure}

The observation in the two-graviton example is naturally generalized to amplitudes with more gravitons: when reconstructing amplitudes containing more gravitons, we have to encounter terms with higher powers of shifted factors. In general, these terms with the $n$-th power of shifted factors are shown by the LHS of \figref{double-trace-nshift-rule} which can be explicitly written as
\bea
\text{\figref{double-trace-nshift-rule} }(0)&=&\left[\,\prod\limits_{i\in\{1,\dots,n-1\}}\frac{[h_n,h_i]}{\langle a,h_i\rangle\langle h_i,\xi\rangle\langle h_i,\eta\rangle}\right]\frac{[a,h_n]}{\langle a,h_n\rangle}\frac{\langle a,\xi\rangle^n\langle h_n,\eta\rangle^{n-1}}{\langle h_n,\xi\rangle}\,\tilde{\mathcal{B}}(h_n,b),
\label{dt-n-shift-fig0}\\
\text{\figref{double-trace-nshift-rule} }(k)&=&\left[\,\prod\limits_{i\in\{1,\dots,n-1\}\setminus\{k\}}\frac{[h_n,h_i]}{\langle h_k,h_i\rangle\langle h_i,\xi\rangle\langle h_i,\eta\rangle}\right]\label{dt-n-shift-figk}\\
&&~~~~~~~~~~~~~~~~~~~~~\times\frac{[a,h_n]}{\langle a,h_k\rangle}\frac{[h_n,h_k]}{\langle h_n,h_k\rangle}\frac{\langle a,\xi\rangle\langle h_k,\xi\rangle^{n-2}\langle h_n,\eta\rangle^{n-1}}{\langle h_n,\xi\rangle\langle h_k,\eta\rangle}\,\tilde{\mathcal{B}}(h_n,b)~(k=1,...,n-1).
\nonumber
\eea
We now show that the sum of the LHS of \figref{double-trace-nshift-rule} equals to the RHS  which has the following contribution
\begin{equation}
\text{\figref{double-trace-nshift-rule} }(n)=\left[\,\prod\limits_{i\in\{1,\dots,n-1\}}\frac{[h_n,h_i]}{\langle h_n,h_i\rangle\langle h_i,\xi\rangle\langle h_i,\eta\rangle}\right]\frac{[a,h_n]\langle a,\xi\rangle\langle h_n,\xi\rangle^{n-2}\langle h_n,\eta\rangle^{n-1}}{\langle a,h_n\rangle}\,\tilde{\mathcal{B}}(h_n,b).
\label{dt-n-shift-fign}
\end{equation}
To see this, we rewrite \eqref{dt-n-shift-fig0} and \eqref{dt-n-shift-figk} by extracting out the common factor (\ref{dt-n-shift-fign})
\bea
\text{\figref{double-trace-nshift-rule} }(0)&=&\text{\figref{double-trace-nshift-rule} }(n)\times\frac{\langle a,\xi\rangle^{n-1}}{\langle h_n,\xi\rangle^{n-1}}(-1)^{n-1}\prod\limits_{i\in\{1,\dots,n-1\}}\frac{\langle h_n,h_i\rangle}{\langle h_i,a\rangle},
\label{dt-n-shift-fig0n}\\
\text{\figref{double-trace-nshift-rule} }(k)&=&\text{\figref{double-trace-nshift-rule} }(n)\times\frac{\langle h_k,\xi\rangle^{n-1}\langle h_n,a\rangle}{\langle h_n,\xi\rangle^{n-1}\langle h_k,a\rangle}\prod\limits_{i\in\{1,\dots,n-1\}\setminus\{k\}}\frac{\langle h_n,h_i\rangle}{\langle h_k,h_i\rangle}.
\label{dt-n-shift-figkn}
\eea
Then the identity shown by \figref{double-trace-nshift-rule} is equivalent to the following identity
%We express \eqref{dt-n-shift-fig0} and \eqref{dt-n-shift-figk} with \eqref{dt-n-shift-fign}. For the convenience of notation, we take “$h_0=a$”, though $a$ label a gluon node. 
%Thus the identity encoded in \figref{double-trace-nshift-rule} can be expressed as
\bea
(-1)^{n-1}\frac{\langle h_0,\xi\rangle^{n-1}}{\langle h_n,\xi\rangle^{n-1}}\prod\limits_{i\in\{1,\dots,n-1\}}\frac{\langle h_n,h_i\rangle}{\langle h_i,h_0\rangle}+\sum\limits_{k\in\{1,\dots,n-1\}}\frac{\langle h_k,\xi\rangle^{n-1}\langle h_n,h_0\rangle}{\langle h_n,\xi\rangle^{n-1}\langle h_k,h_0\rangle}\prod\limits_{i\in\{1,\dots,n-1\}\setminus\{k\}}\frac{\langle h_n,h_i\rangle}{\langle h_k,h_i\rangle}-1=0,\nn
\label{dt-n-shift-id}
\eea
where we have renamed the gluon $a$ by $h_0$ . When a common factor  $\left[\langle h_n,\xi\rangle^{n-1}\prod\limits_{0\leq j<i\leq n-1}\langle h_i,h_j\rangle\right]^{-1}$ is extracted out, \eqref{dt-n-shift-id} turns into:
\bea
\langle h_0,\xi\rangle^{n-1}(-1)^{n+1}\prod\limits_{1\leq j<i\leq n}\langle h_i,h_j\rangle+\sum\limits_{k\in\{1,\dots,n-1\}}\langle h_k,\xi\rangle^{n-1}(-1)^{n-k+1}\prod\limits_{\substack{0\leq j<i\leq n\\i,j\neq k}}\langle h_i,h_j\rangle&& \nn
-\langle h_n,\xi\rangle^{n-1}\prod\limits_{0\leq j<i\leq n-1}\langle h_i,h_j\rangle&=&0,
\eea
which has the same form as \eqref{S-id-ab}. Therefore, the identity shown by \figref{double-trace-nshift-rule} has been proven.

According to the symmetry between the two traces of the formula (\ref{double-t-graph}), the corresponding situation where the graviton $h_n$ is connected to some node $b$ which plays as an end of the $k_a\cdot k_b$ line and belongs to the component involving the other trace has also been proven. For now, we can see that the identity in \figref{double-trace-nshift-rule} which is relevant to the factor $k_a\cdot k_b$ on the bridge is essentially the same as the identity in \figref{ngraviton-fig-rule} which has been applied in the single-trace discussion. Nevertheless, there is a subtle difference between the two identities shown by \figref{double-trace-nshift-rule} and \figref{ngraviton-fig-rule}: In the identity in \figref{ngraviton-fig-rule}, the graphs ($1$)-($n-1$) include all possible graphs where one edge $e(h_i,l)$ in ($0$) is kept and all other edges end at the node $h_i$.  But, the LHS of the identity in \figref{double-trace-nshift-rule}, does not contain the graph where all edges end at the node $b$ that belongs to the $k_a\cdot k_b$ edge. This is because of the difference between the $k\cdot k$ edge and the other graviton edges.

\subsection{General proof of the double-trace case}
Now let us prove that the general formula (\ref{double-t-graph}) (which are characterized by graphs of the pattern \figref{double-t-fig1}) can be derived from ISL, i.e., 
\begin{equation}
A(x_1,\dots,x_r|y_1,\dots,y_s||\textsc{H})=\sum_{l_n\in T\oplus \textsc{H}\setminus\{h_n\}}\mathcal{G}(h_n,l_n) A'(T_1|T_2||\textsc{H}\setminus\{h_n\})\big|_{l'_n}.
\label{double-t-n-1}
\end{equation}
The cases of $n=1,2$ have already been proved. Assuming that the amplitude with $n-1$ gravitons already satisfy \eqref{double-t-graph}, the RHS of the above expression then reads
\bea
\sum_{l_n\in T\oplus \textsc{H}\setminus\{h_n\}}\mathcal{G}(h_n,l_n)\left[\sum\limits_{\mathscr{G}^{n-1}}\sum\limits_{\substack{a\in\mathscr{G}_1\setminus\{h_n\}\\b\in\mathscr{G}_2\setminus\{h_n\}}}\mathcal{B}(a,b)\prod\limits_{e(x,y)\in\mathcal{E}(\mathscr{G}^{n-1})}\mathcal{G}(x,y)\right]'\Bigg|_{l'_n}\, \text{PT},
\label{double-t-n-2}
\eea
where $\mathscr{G}^{n-1}=\mathscr{G}_1\oplus\mathscr{G}_2\setminus\{h_n\}$.

\begin{figure}
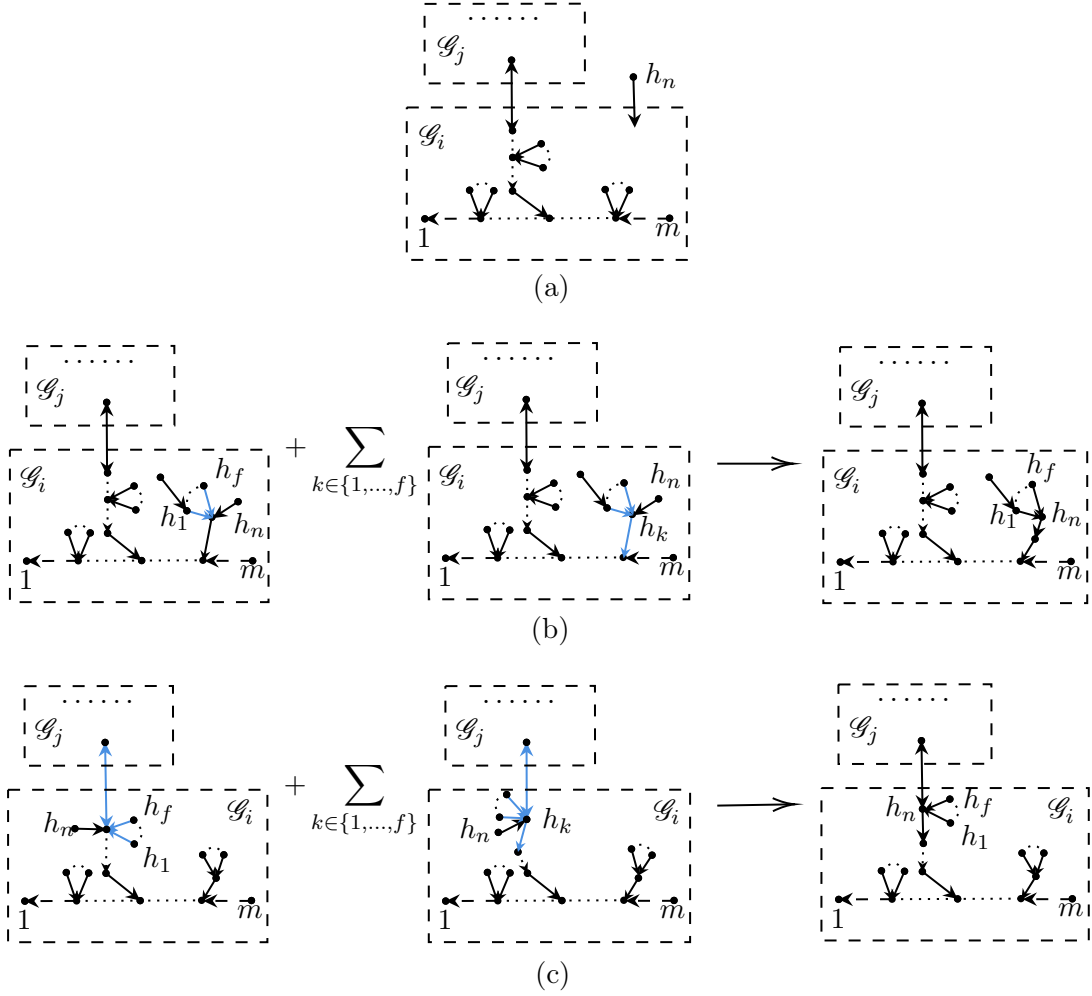

\centering

\tikzset{every picture/.style={line width=0.75pt}} %set default line width to 0.75pt        



\caption{Contributions of terms in different classes, where $\mathscr{G}_i$, $\mathscr{G}_j$ represent two different forests belong to the correspond traces. The blue arrows in graphs (b) and (c) represent the shifted part of these edges (and bridge) fators, and all the terms they shift only are summed up.}
\label{dt-general-proof}
\end{figure}
Terms in the above expression can be classified into three parts: (i). unshifted terms, (ii). terms with shifted factors only accompanying to edges of the form in \eqref{G-factor}, (iii).  terms with shifted factors accompanying the $k_a\cdot k_b$ edge. Contributions of these three types of terms are separately considered as follows:
\begin{itemize}
\item Terms in (i)  result in all forests where $h_n$ plays as a leaf (in other words an outermost node), as shown by \figref{dt-general-proof} (a).
\item For terms in (ii), the graviton $h_n$ can be connected to a node belonging to either  $\mathscr{G}_1$ or $\mathscr{G}_2$. In each case, one can apply the identity shown by \figref{ngraviton-fig-gen} and then insert $h_n$ into either $\mathscr{G}_1$ or $\mathscr{G}_2$ as an internal node (see \figref{dt-general-proof} (b)). More concretely, $h_n$ may become an internal node that does not belong to the bridge, or become an internal node on the bridge but does not play as an end node of the $k_a\cdot k_b$ edge. 
\item As shown by \figref{dt-general-proof} (c), terms in (iii) reproduce those graphs, in which $h_n$ plays as an end node of the $k_a\cdot k_b$ edge, according to the identity shown by \figref{double-trace-nshift-rule}.
\end{itemize}
Altogether, we have proven that the ISL formula (\ref{double-t-2gra-ISL}) indeed reproduces all the graphs in the formula (\ref{double-t-graph}).

\subsection{ Comments on the vanishing configurations}

For completeness, other configurations with two negative-helicity particles (i.e. the ($h^-_i$, $g^-_j$) and ($h^-_i$, $h^-_j$) configurations) for amplitudes with two traces, and all amplitudes with two negative-helicity particles and more than two traces shall also be considered. Such amplitudes have already been shown to vanish, from the angle of the Cachazo-He-Yuan \cite{Cachazo:2013gna, Cachazo:2013hca, Cachazo:2013iea, Cachazo:2014nsa, Cachazo:2014xea} formalism \cite{Cachazo:2014xea, Xie:2022nfu}, the graphic expansion rule \cite{Tian:2021dzf}, as well as the BCFW recursion (where internal gravitons, B fields, and dilaton should all be taken into account) \cite{Cachazo:2014xea} which can be considered as the starting point for deriving the ISL method.

\section{Conclusion}\label{Sec6}

In this paper, we verified that the spanning forest formulas of single- and double-trace MHV amplitudes in EYM can be reconstructed precisely through the inverse soft limit, all possible MHV configurations have been discussed and are consistent with previous conclusions in \cite{Cachazo:2014xea,Du:2016wkt,Tian:2021dzf}. A generalized identity of spinnor products, obtained by braiding Schouten identities and interpreted by graphs, was proposed and proved. It is worth highlighting again that such an ISL approach to amplitudes is not unique and should not be limited to MHV sector. In a similar vein, there may be some other generalizations of Schouten identities that may help to uncover symmetric formulas for amplitudes beyond MHV.  In addition, starting from a nonvanishing three-point amplitude and making an analogous discussion with those in  \secref{sec4.5}, (as pointed in \cite{Nandan:2012rk}), one can evaluate single-trace EYM amplitudes with more complicated helicity configurations, and even multi-trace amplitude when the B fields and dilatons are also taken into account. Nevertheless, a more symmetric form of such amplitudes still deserves some further study.

%%%%%%%%%%%%%%%%%%%%%%%
\section*{Acknowledgments}
%%%%%%%%%%%%%%%%%%%%%%
This work is supported by NSFC under Grant No. 11875206.

%%%-------------- Appendix, If needed-----------%%%

%\appendix
%\section{Some title}
%Please always give a title also for appendices. 
%heading.

%\paragraph{Note added.} This is also a good position for notes added
%after the paper has been written.

% query inspire, trying to verify the data (this will probably spot
% in short all the information that clearly identify a document.

%%%--------Current Bibliography---------%%%

%\bibliographystyle{unsrt}
%\bibliography{reference}

%%%--------The Original References Template---------%%%
%\begin{thebibliography}{99}

%\bibitem{a}
%Author, \emph{Title}, \emph{J. Abbrev.} {\bf vol} (year) pg.

%\bibitem{b}
%Author, \emph{Title},
%arxiv:1234.5678.

%\bibitem{c}
%Author, \emph{Title},
%Publisher (year).

% Please avoid comments such as "For a review'', "For some examples",
% "and references therein" or move them in the text. In general,
% please leave only references in the bibliography and move all
% accessory text in footnotes.

% Also, please have only one work for each \bibitem.

%\end{thebibliography}

%%%--------The Original References Template---------%%%

\bibliographystyle{JHEP}
\bibliography{reference}
\end{document}